\documentclass[review,fleqn,sort&compress]{elsarticle}
\usepackage{amssymb,amsbsy,amsthm,amsmath,amsfonts,amssymb,amscd}
\usepackage{geometry}
\usepackage{bm} % for \bm macro
\usepackage{graphicx}
\usepackage{float}
\usepackage{multirow}
\usepackage{subfigure}
\usepackage{lineno}
\usepackage{hyperref}
\hypersetup{colorlinks, citecolor=black, filecolor=black, linkcolor=black, urlcolor=black}
\usepackage{booktabs}
\usepackage[ruled,linesnumbered]{algorithm2e}
\usepackage{appendix}
\usepackage{algpseudocode}
\usepackage{mathtools,nccmath}
\usepackage{adjustbox}
\usepackage{cleveref}
\journal{Computer Methods in Applied Mechanics and Engineering}
\graphicspath{{fig/}}
\theoremstyle{definition}

\usepackage{color}

\begin{document}

\begin{frontmatter}
\title{Integrated Finite Element Neural Network (I-FENN) for non-local continuum damage mechanics}
\author{Panos Pantidis\corref{cor1}}
\author{Mostafa E. Mobasher}\corref{cor2}
\cortext[cor1]{Corresponding author. \emph{E-mail address:} \texttt{pp2624@nyu.edu} (Panos Pantidis)}
\cortext[cor2]{Corresponding author. \emph{E-mail address:} \texttt{mostafa.mobasher@nyu.edu} (Mostafa Mobasher)}
\address{Civil and Urban Engineering Department, New York University Abu Dhabi, Abu Dhabi, P.O. Box 129188, UAE}

\begin{abstract}

We present a new Integrated Finite Element Neural Network framework (I-FENN), with the objective to accelerate the numerical solution of nonlinear computational mechanics problems. We leverage the swift predictive capability of neural networks (NNs) and we embed them inside the finite element stiffness function, to compute element-level state variables and their derivatives within a nonlinear, iterative numerical solution. This process is conducted jointly with conventional finite element methods that involve shape functions: the NN receives input data that resembles the material point deformation and its output is used to construct element-level field variables such as the element Jacobian matrix and residual vector. Here we introduce I-FENN to the continuum damage analysis of quasi-brittle materials, and we establish a new non-local gradient-based damage framework which operates at the cost of a local damage approach. First, we develop a physics informed neural network (PINN) to resemble the non-local gradient model and then we train the neural network offline. The network learns to predict the non-local equivalent strain at each material point, as well as its derivative with respect to the local strain. Then, the PINN is integrated in the element stiffness definition and conducts the local to non-local strain transformation, whereas the two PINN outputs are used to construct the element Jacobian matrix and residual vector. This process is carried out within the nonlinear solver, until numerical convergence is achieved. The resulting method bears the computational cost of the conventional local damage approach, but ensures mesh-independent results and a diffused non-local strain and damage profile. As a result, the proposed method tackles the vital drawbacks of both the local and non-local gradient method, respectively being the mesh-dependence and additional computational cost. We showcase through a series of numerical examples the computational efficiency and generalization capability of I-FENN in the context of non-local continuum damage, and we discuss the future outlook. {The PINN training code and associated training data files have been made publicly available online, to enable reproducibility of our results.}

\end{abstract}

\begin{keyword}
\texttt I-FENN \sep finite element method \sep neural networks \sep continuum damage mechanics \sep non-local gradient 
\end{keyword}

\end{frontmatter}
%\linenumbers

%\newpage

\section{Introduction}
\label{Sec:Introduction}

\subsection{Literature review: machine-learning in computational mechanics}

Maturing over the course of several decades, the Finite Element Method (FEM) has played a fundamental role in the understanding and prediction of complex mechanical phenomena and processes, constituting arguably one of the most robust, reliable and versatile numerical approaches which are currently available. The FEM's capability to model systems of arbitrary geometry and loading/boundary conditions, as well as the capability of handling non-linear problems, have established the role of FEM in several fields of computational mechanics \cite{reddy2014introduction, belytschko2014nonlinear, wriggers2008nonlinear}. This has led to the development and use of FEM to the solution of many problems of immense practical importance such as brittle and ductile fracture \cite{mcauliffe2016173, duda2015269, duarte2001generalized}, time-dependent response of materials \cite{barlat1991693, barlat20111309}, non-linear multi-physics problems \cite{mobasher2017non, mobasher2018thermodynamic}, multi-scale \cite{ozturk2021phcms} and multiple length-scale \cite{mobasher2022duallength} problems, as well as other applications \cite{pantidis3DEuler, zhangcardiovascular}. Despite its proven strengths, the computational cost of non-linear FEM models presents a challenge as the level of detail in a model increases, since it requires the iterative solution of significantly larger non-linear systems of equations. Therefore, continuous research efforts aim at reducing the cost of non-linear FEM models \cite{farhat2001feti,mobasher2016adaptive,white2019two,castelletto2015accuracy,waisman2013adaptive,fish2012staggered}.

Machine-learning (ML) strategies have recently emerged as promising candidates to overcome these shortcomings, by improving the accuracy \cite{tancogne2021recurrent} and lowering the computational expense of computational mechanics simulations \cite{logarzo2021smartlaws}. One potential pathway is the development of data-driven surrogate models, which can be implemented in multi-scale methods in order to bypass the computationally exhaustive representative volume element (RVE) simulations. These pre-trained constitutive models represent the homogenized stress-strain response at the micro-scale, thus allowing for more numerically efficient multi-scale analyses \cite{fascetti2019multiscale, ozturk2021phcms}. A few representative examples of this approach include micromechanical modeling of polycrystalline materials \cite{reimann2019micromodel, vlassis2020geometric}, path-dependent plasticity in aluminum-rubber composite materials \cite{mozaffar2019pathplasticity} and history-dependent hyperelasticity models \cite{huang2020proporthdecom}. Alternatively, purely data-driven approaches attempt to avoid simulation biases and offer an alternative approach by calculating the material point equilibrium states based purely on experimentally available data. The work of Kirchdoerfer and Ortiz \cite{kirchdoerfer2016data} introduced the principles of this framework, and extensions of the original elasticity-based work include applications in computational plasticity \cite{chinesta2017datadrivenplasticity} and fracture \cite{carrara2020datadrivenfracture}. However, data-driven frameworks have two major limitations. The first is the lack of physical bounds to the solution, which can lead to difficulties of the physical interpretation of their results \cite{rao2020physics}. The second is that the data-driven nature limits the extensibility of these approaches towards a more generic solution method \cite{kochman2021machinelearning,liphysicsguided2021, abdulla2021fracprop}. Consequently data-driven methods still suffer fundamentally from their dependency on the training datasets \cite{chen2021learning}, whether these originate from expensive nonlinear RVE-level simulations or from the available experimental data points. 

To overcome these issues, several researchers have implemented more physics-consistent efforts, such as the incorporation of the thermodynamics relationships in the learning process \cite{masi2022tann}. Several variational methods were also introduced to enhance data-driven approaches by relating energy-based formulations to the problem setup \cite{yu2018deep, abueidda2022hypervisco, kharazmi2021113547, liphysicsguided2021}. Additionally, Samaniego et al \cite{samaniego2020energy} developed the Deep Energy Method (DEM) which utilized the neural networks relationships as the discretization functions for the potential energy minimization statement, and more recently DEM was extended to modeling finite strain hyperelasticity \cite{fuhg2022DEMhyperelast}. However, the non-convex nature of the optimization problems still poses significant challenges \cite{samaniego2020energy} and the presence of multiple local optima often hinders the efficiency of these methods, necessitating further work along this frontier \cite{yu2018deep, abueidda2022hypervisco}. 

Other approaches have also attempted to enrich the finite element method using neural networks in a conceptually different way than the aforementioned. The Finite Element Network Analysis framework \cite{jokarFENA2021, jokarFENA2022}mimics the conventional finite element assembly process by utilizing pre-trained neural networks as surrogate building-blocks to formulate the physical system. Despite the generalization potential of this method to more complicated cases, this approach has been thus far implemented only to linear elastic 1D bars and 2D thin plates. More recently, Mitusch et al \cite{mitusch2021hybridFEMNN} demonstrated the ability of their hybrid FEM-NN network to recover unknown PDE terms by augmenting the differential equations and subsequently discretize them in space using FEM. 

A conceptually different approach to minimize the cost of computational mechanics models is centered around the capability of machine-learning methods to predict the solution of the governing partial differential equations (PDEs), which were otherwise numerically solved using FEM and other numerical approaches. Along this pathway, physics-informed neural networks (PINNs) have gained significant attention over the last years. PINNs are tailored to the discovery of PDEs or the prediction of their solution. The idea behind them is to convert the objective of the network learning process from the minimization of the error between data points into a minimization process of the PDE residual error. This goal is achieved during the neural network training stage, by equating the expression of the network cost function to the PDE residual while also accounting for the imposed boundary conditions. The origins of PINNs can be traced back to almost three decades ago \cite{psichogios1992hybrid,lagaris1998artificial}, but the major developments were proposed more recently in the seminal work of Raissi et al \cite{raissi2017physics, raissi2019physics}. Since then, interest in this field has emerged and PINNs have found applications in several fields, such as quantum physics \cite{stiller2020large}, flow problems \cite{mathews2021uncovering}, uncertainty models \cite{yang2020physics}, optics and electromagnetics \cite{fang2021high}, and more. However, they have found only limited applications in the field of computational solid mechanics, particularly with respect to nonlinear problems \cite{haghighat2021physics, henkes2022pinnsmicromech, rao2020physics}.

\subsection{Overview of proposed methodology and implementation}
\begin{figure}[H]
	\centering
	\includegraphics[ width=0.9\textwidth]{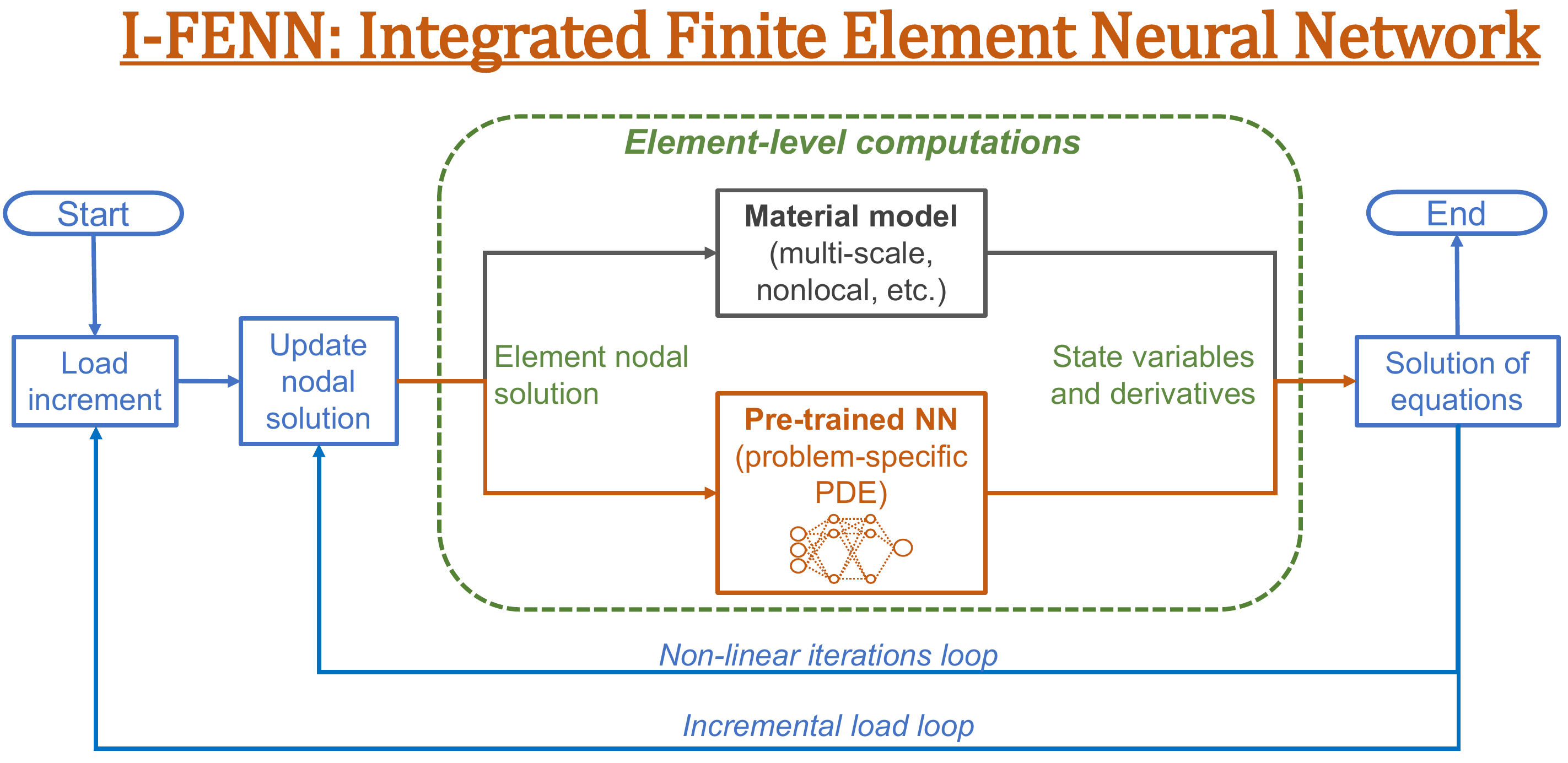}
	\caption{Flowchart of the proposed Integrated Finite Element Neural Network (I-FENN) methodology. A pre-trained physics-informed neural network is embedded inside the finite element stiffness function, and it is used to compute element-level state variables and their derivatives jointly with finite element shape functions. The entire scheme is encapsulated within a nonlinear, iterative numerical solver.}
	\label{Fig_Methodology_Flowchart_Intro}
\end{figure}

% The PINN purpose is to target a problem-specific partial differential equation and solve it at a fraction of time compared to the conventional FEM discretization. 
In this paper, we develop a methodology with the overarching goal of reducing the computational expense of nonlinear mechanics problems by leveraging the swift predictive capability of pre-trained physics-informed neural networks. The pivotal point from other methods in the literature is that we utilize the pre-trained Neural Network (NN) to compute both state variables and their derivatives on an element-level basis, jointly with conventional finite element procedures that involve shape functions and their derivatives. These two methods act in a synergistic fashion: the NN receives input data that resembles the material point deformation, and then its output data is used to construct element-level variables such as the element Jacobian matrix and the element residual vector. One of the key characteristics of the proposed framework is that the NN is directly embedded inside the element stiffness function, and therefore its utilization within a nonlinear, iterative procedure becomes straightforward once the network has been trained. More importantly this approach preserves the framework of the solution of non-linear problems within FEM, and assures that the state variables and their derivatives which are derived as the NN output are re-computed in each iteration. The continuous update of the NN input and output within the iterative procedure is the key feature which allows for the convergence of the nonlinear finite element solution. A descriptive flowchart of our approach is shown in Figure \ref{Fig_Methodology_Flowchart_Intro}. 

In this work we demonstrate the overall strategy of the proposed framework and we develop a specific implementation which targets the continuum damage mechanics (CDM) analysis of quasi-brittle materials. Specifically, we develop a CDM framework which has a non-local nature and utilizes PINNs as the key computational tool to predict the non-local strain field based on the gradient non-local PDE \cite{peerlings1996gradient}. Generally, the main motivation behind using a non-local damage is to regularize the mesh dependence and the lack of solution uniqueness associated with local damage models \cite{pijaudier1987nonlocal,de1995comparison,peerlings2001critical}. However, the use of non-local damage models is often associated with additional computational cost and implementation complexities \cite{peerlings1996gradient, kiefer2018gradient}. The specific objective of our framework is to decrease the computational expense by developing a non-local damage model that resembles the gradient non-local relationship, while reducing the size of the system of equations. This is achieved by predicting the solution of the non-local gradient equation using a trained network, and hence eliminating the need for its numerical solution which includes solving for additional degrees of freedom and the need to implement a non-symmetric Jacobian matrix \cite{peerlings1996gradient}. In other words, we eliminate the need for doing any geometric search as in the non-local integral approach \cite{pijaudier1987nonlocal}, and we eliminate the need for the expanded system of equations exhibited by the non-local gradient model \cite{peerlings1996gradient}. Thus, the output framework presents a non-local damage model that works at the computational cost of a local damage model. 

The road-map of the implementation is briefly presented below, and a more extensive discussion is provided in the later sections. First, a PINN is developed and trained to predict the local to non-local strain transformation based on the gradient non-local strain equation \cite{peerlings1996gradient}. Adopting a mathematical approach which is similar in spirit to the conventional local damage framework, we perform the following steps at the element-level basis: a) using the nodal displacements, we compute the local equivalent strain at each integration point, b) using the pre-trained PINN, we transform the latter to a non-local equivalent strain and extract their differential relationship, c) using the constitutive damage law, we compute a non-local damage variable based on the non-local strain PINN output, and d) we utilize the non-local damage from step (c), and the partial derivative of non-local to local equivalent strain from step (b) to construct the element Jacobian matrix and residual vector which are used for the non-linear FEM solution. Ultimately, non-locality is embedded in the FEM solution via the pre-trained PINN, whereas the size of the equations system matrix is in the order of the local damage method. As a result, we tackle the drawbacks of both the local and non-local gradient method, respectively being the mesh-dependence and computational cost. {The PINN training code and associated training data files which are used in this study have been made publicly available online and can be found at: https://drive.google.com/drive/folders/1mM35pXk7gzyx27NbIF3flqQxehr1tebh?usp=sharing}.

This study presents several novel contributions. From the integration of NNs in non-linear finite element point of view, the paper a) establishes the theoretical principles of the I-FENN framework, b) presents the details of a problem-specific implementation, which is targeted in the continuum damage analysis of quasi-brittle materials, and c) showcases through a series of numerical examples the feasibility and potential of the proposed methodology. In addition, from a damage modeling point of view, this paper presents a new non-local damage model in which the non-local damage variable is computed from the output of a neural network without the need for spatial averaging or gradient non-local PDE solution. 

The paper is structured as follows: Section \ref{Sec:TheoreticalOverview} presents a theoretical overview and a brief mathematical description of the local and non-local gradient theory. Section \ref{Sec:Methodology} details the principles and background of the proposed framework. Section \ref{Sec:ConstitutiveModeling} discusses the constitutive models which are adopted in this study, while Section \ref{Sec:NumericalExamples} presents the numerical examples and elaborates on the observations and limitations of this work. Finally, Section \ref{Sec:Conclusions} presents the overall conclusions of this paper and discusses the future outlook.

%%%%%%%%%%%%%%%%%%%%%%%%%
\section{Theoretical Overview: local and non-local damage methods}
\label{Sec:TheoreticalOverview}

\subsection{Continuum damage mechanics methods}
\label{Sec:CDMtheories}

Material damage is the physical process during which degradation or complete loss of the mechanical properties occurs \cite{lemaitrebook,kachanovbook}. In the case of quasi-brittle materials, such as ceramics and cementitious ones, damage evolves through crack initiation and subsequent propagation. From a representation standpoint, one can use either \textit{discrete} and \textit{smeared} numerical methods to describe the crack formulation. The first family of methods is based on the concepts of Linear Elastic Fracture Mechanics \cite{griffiththeory}, where the crack is explicitly represented as a material discontinuity. The second approach adopts the fundamentals of Continuum Damage Mechanics (CDM) \cite{lemaitrebook,kachanovbook}, where the failure process is described through the use of strain-softening constitutive relationships. In this case, damage is mathematically treated as a gradual stiffness loss which is calculated at each material point.

The definition of the damage variable at a local material point leads to loss of ellipticity of the PDE, which yields an ill-posed mathematical problem and consequently causes lack of uniqueness of the numerical solution \cite{geersstrainbased}. From a numerical perspective, during the material degradation phase, this leads to the phenomenon of strain localization over an ever-decreasing area upon refinement of the mesh resolution. As a result, numerical methods suffer from mesh dependency, not only with regards to the element size but also to the mesh orientation. To overcome this deficiency, non-local constitutive models have been developed, aiming to model damage as a diffused property over a material region. In CDM, the most common frameworks are the a) \textit{non-local integral} \cite{pijaudier1987nonlocal} and b) \textit{non-local gradient} method \cite{peerlings1996gradient}. {Several notable attempts have been made towards eliminating the need for mathematically imposing the non-local field by leveraging Thick Level Set and Lip-Field approaches} {\cite{moes2011level,moes2021lip,parrillagomez201775,moeslip2}}. Phase-field approaches have also been developed and used to describe diffused material damage \cite{phasefieldtheory}, and several studies indicate their analogy to the non-local gradient methods \cite{deborstcomparison}.

In the non-local integral approach, the non-local damage variable at a material point is computed as a weighted average of the local damage values at points within a characteristic length scale. This integral-type definition results to mesh-independence, but it is associated with several numerical challenges: a) the need to perform an algorithmic search for all integration points that lie within the neighbourhood of each material point, b) performance issues at the domain boundary, where natural discontinuities such as external boundaries and holes exist, and c) the difficulty of the derivation of the Jacobian matrix \cite{jirasek2002consistent,chen2022dynamic}. Alternatively, the non-local gradient approach replaces the integral definition of the macroscopic deformation with a gradient-based formulation. This method introduces another governing PDE in the problem setup, coupling the local and non-local deformation measures. The major advantage of this method is that it retains a non-local character in an implicit manner, without the integral requirements of search and averaging continuously throughout the finite element analysis. Nevertheless, the inclusion of an additional PDE increases the computational cost. This will become apparent in Section \ref{Sec:Methodology}, where we explicitly report the size of the Jacobian matrix for the local and the non-local gradient method. In addition, the resulting system of equations is not symmetric, which further increases the cost of the solution of the system of equations \cite{peerlings1996gradient}.  

\subsection{Mathematical formulation of the local and non-local gradient damage theories}
\label{Sec:Nonlocaldamageformulation}

\begin{figure}[H]
	\centering
	\includegraphics[scale=0.45]{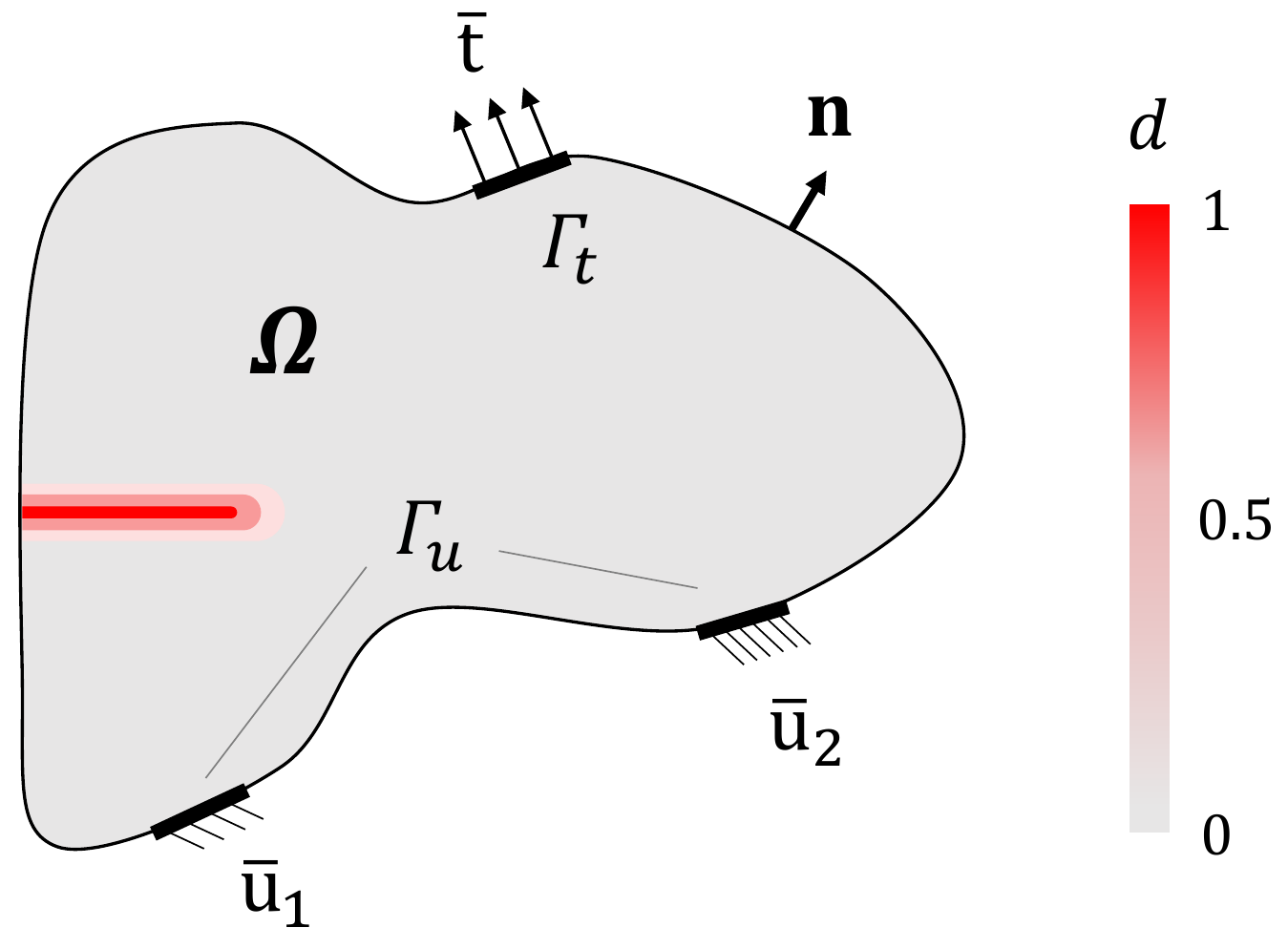}
	\caption{Schematic representation of a body with prescribed displacement and traction boundary conditions}
	\label{Fig_Elastic_Domain}
\end{figure}

Consider the elastic domain $\Omega$ shown in Figure \ref{Fig_Elastic_Domain}, where $\Gamma$ denotes its boundary. Displacements $\bar{\bf{u}}$ are prescribed in the Dirichlet boundary part $\Gamma_{u}$ and traction forces ${\bar{t}}_{j}$ are applied in the Neumann boundary part $\Gamma_{t}$ ($\Gamma = \Gamma_{u} \cup \Gamma_{t}$). The analysis is governed by the equilibrium equation \eqref{Equilibrium_Condition} and appropriate boundary conditions \eqref{Dirichlet_BCs} and \eqref{Neumann_BCs}:

\begin{equation}
\begin{split}
    \sigma_{ij,j} = 0 \; \; \; in \; \; \Omega
\end{split}
\label{Equilibrium_Condition}
\end{equation}

\begin{equation}
\begin{split}
    u_{i} = {\bar{u}_{i}} \; \; \; in \; \; \Gamma_{u}
\end{split}
\label{Dirichlet_BCs}
\end{equation}

\begin{equation}
\begin{split}
    \sigma_{ij} \cdot n_{i} = {\bar{t}_{j}} \; \; \; in \; \; \Gamma_{t}
\end{split}
\label{Neumann_BCs}
\end{equation}

\noindent where $\sigma_{ij}$ is the Cauchy stress tensor and $n_{i}$ denotes the outward unit normal vector on $\Gamma$. Assuming isotropic behavior, a scalar variable $d$ is used to represent the stiffness loss magnitude. This state variable takes values between 0 and 1, with $d = 0$ indicating the intact material and $d = 1$ denoting the fully damaged state. The total stress $\sigma_{ij}$ is expressed as follows \cite{kachanovbook}:

% Cauchy stress-strain relationship
\begin{equation}
\begin{split}
    \sigma_{ij} = (1 - d) \: C_{ijkl} \varepsilon_{kl} 
\end{split}
\label{Cauchy_stress}
\end{equation}

\noindent where $C_{ijkl}$ is the fourth-order elasticity tensor and $\varepsilon_{kl}$ is the strain tensor. The elasticity tensor is given by:

\begin{equation}
	C_{ijkl} = \left[K \delta_{ij}\delta_{kl}+\mu \left[\delta_{ik}\delta_{jl}+\delta_{il}\delta_{jk}-\frac{2}{3}\delta_{ij}\delta_{kl}\right]\right]
\label{C_tensor}
\end{equation}

\noindent where $K$ is the bulk modulus, $\mu$ is the shear modulus, and $\delta_{ij}$ is the Kronecker delta. Assuming small deformations, the strain-displacement relationship reads: 

\begin{equation}
	\varepsilon_{ij} = \frac{1}{2}\left[u_{i,j}+u_{j,i}\right] 
\label{e_tensor}
\end{equation}

A scalar invariant measure of deformation, termed $\textit{local equivalent strain}$, can be defined as a function of the strain tensor: ${\varepsilon}_{eq} = {\varepsilon}_{eq}(\varepsilon_{ij})$. The local damage variable $d$ is a function of the local equivalent strain, $d = d({\varepsilon}_{eq})$, and their relationship is given by the appropriate damage evolution law.

In the non-local gradient damage model \cite{peerlings1996gradient} the damage variable $d$ evolves as a diffusive property over a finite material zone which is defined by a characteristic length scale measure $\l_{c}$ \cite{pijaudier1987nonlocal}. This is achieved by evolving the damage variable as a function of a non-local equivalent strain $d = d(\bar{\varepsilon}_{eq})$. The non-local equivalent strain $\bar{\varepsilon}_{eq}$ is calculated based on the following gradient-based PDE:

% Local-nonlocal strain PDE
\begin{equation}
\begin{split}
    {\bar{\varepsilon}}_{eq} - g \nabla^{2} {\bar{\varepsilon}}_{eq} = \varepsilon_{eq}
\label{NonlocalGradientPDE}
\end{split}
\end{equation}

In the expression above, $\nabla^{2}$ is the Laplacian operator and $g$ is defined as $g = l_{c}^{2}/2$. The PDE definition in Equation \eqref{NonlocalGradientPDE} is usually complimented by the following boundary condition:

% Local-nonlocal strain PDE
\begin{equation}
\begin{split}
    \nabla {\bar{\varepsilon}}_{eq} \cdot n_{i} = 0 \; \; \; in \; \; \Gamma
\end{split}
\label{NonlocalGradientPDE_BCs}
\end{equation}

Here we adopt equation \eqref{NonlocalGradientPDE_BCs} for the boundary conditions expression, a choice which is in consistence with previous works \cite{peerlings1996gradient, peerlings2001critical}. This condition does not interfere with the physical interpretation of the non-local gradient model and allows for consistent damage diffusion in areas close to the boundaries.

%%%%%%%%%%%%%%%%%%%%%%%%%%%%%%%%%%%%%%%%
\section{Methodology}
\label{Sec:Methodology}

\begin{figure}[H]
	\centering
	\includegraphics[width=0.9\textwidth]{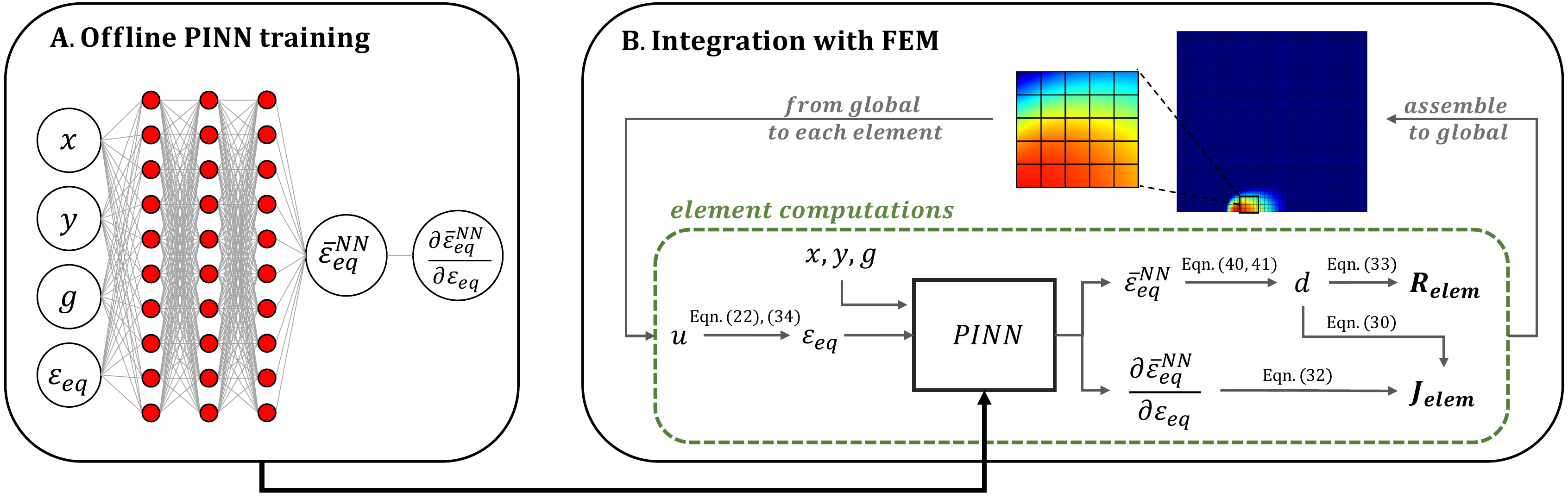}
	\caption{A schematic of the overall methodology of integrating the PINN with FEM towards and I-FENN aimed at the efficient solution of non-local damage. Box {\bf{A}}. Schematic formulation of the adopted PINN. The input variables are the material point coordinates $x$ and $y$, the local equivalent strain $\varepsilon_{eq}$ and the internal length scale measure $g$. The output variables which will be used inside FEM are the non-local equivalent strain ${\bar{\varepsilon}}_{eq}$ and its partial derivative with respect to the local strain, $\frac{\partial \bar{\varepsilon}^{NN}_{eq}}{\partial \varepsilon_{eq}}$. Box {\bf{B}}. Integration of the pre-trained PINN within the element-level computations. The PINN input is computed using the nodal displacements and the shape function derivative matrix. The PINN output is utilized in the calculation of the element Jacobian matrix ${\bf{J_{elem}}}$ and residual force vector ${\bf{R_{elem}}}$.}
	\label{Fig_Methodology_Flowchart_Method}
\end{figure}

This section illustrates the implementation of the proposed Integrated Finite Element Neural Network I-FENN framework for the solution of the non-local damage model based on the gradient representation \cite{peerlings1996gradient}. This process is carried out in two steps and it is graphically represented in Figure \ref{Fig_Methodology_Flowchart_Method}. First, we do an offline training of the neural network on the governing PDE with the objective to learn the local to non-local strain transformation. Subsection \ref{SubSec:MethodPINNs} discusses the formulation and underlying assumptions of the proposed PINN, and presents its design and training principles. The second step involves the integration of the PINN inside the element-level stiffness function, and for this purpose we devise the following strategy: 

\begin{enumerate}

\item Using the appropriate nodal displacements $\hat{\bf{u}}$ and shape function derivatives $\bf{B}$, we calculate the local equivalent strain ${\varepsilon}_{eq}$ at each integration point (IP).

\item Using the pre-trained PINN, we predict the non-local equivalent strain $\bar{\varepsilon}^{NN}_{eq}$ and its derivative with respect to the local equivalent strain $\frac{\partial \bar{\varepsilon}^{NN}_{eq}}{\partial \varepsilon_{eq}}$ at each IP.

\item Using the governing damage law, we compute the damage state variable $d$ as a function of the non-local equivalent strain, $d = d(\bar{\varepsilon}^{NN}_{eq})$.

\item Using the conventional finite element procedures and utilizing $d$ (step 3) and $\frac{\partial \bar{\varepsilon}^{NN}_{eq}}{\partial \varepsilon_{eq}}$ (step 2), we construct the element Jacobian matrix $\bf{J}_{elem}$ and residual vector $\bf{R}_{elem}$. 

\item Using the finite element connectivity matrix, we assemble the system-level Jacobian matrix $\bf{J}$ and residual vector $\bf{R}$; which is followed by solution of the equations.

\item Steps 1-5 are repeated until $\bf{R}$ converges within the nonlinear iterative solver.

\end{enumerate} 

Subsection \ref{SubSec:FEM_implementation} presents the theoretical derivation of the new non-local damage model, and provides more details on the strategy of trained PINN integration into the framework of a nonlinear finite element analysis as a complementary component on the element-level computations.

\subsection{Physics-informed Neural Network formulation for the non-local gradient model}
\label{SubSec:MethodPINNs}

{In the next paragraphs we detail the generic setup of Physics-Informed Neural Networks, while commenting on our choice of all the network hyperparameters. For a broader discussion on PINNs, the interested reader is directed to} \cite{raissi2017physics, raissi2019physics, cuomo2022scientific}. {PINNs are a scientific machine-learning technique which is used to approximate the solution of partial differential equations (PDEs). Compared to the traditional multi-layer perceptrons, they are different in the sense that they a) compute partial derivatives of the output variable with respect to the input variables, and b) utilize these derivatives in the definition of the objective cost function.} PINNs construct approximate numerical solutions to PDEs by minimizing the PDE residual error at selected interior points of the domain, termed \textit{collocation points}, as well as enforcing the initial/boundary conditions along the spatio-temporal domain boundary. In the case of supervised learning several labeled value pairs may also be available, and therefore some of the predictions can be directly compared to the corresponding labeled values. 

A generic mathematic formulation of the differential expressions solved by PINNs is given below \cite{cuomo2022scientific}:

\begin{equation}
\begin{split}
    \bf{F} (\bf{h(z)}); \gamma) = \bf{f(z)} \; \; \; \; \;  \bf{z} \; \; on \; \; \Omega 
\end{split}
\label{General_PINN_equation1}
\end{equation}

\begin{equation}
\begin{split}
    \bf{I} (\bf{h(z)})) = \bf{b(z)} \; \; \; \; \;  \bf{z} \; \; on \; \; \Gamma
\end{split}
\label{General_PINN_equation2}
\end{equation}

\noindent where $\bf{z}$ contains the spatio-temporal input variables, $\bf{h}$ is the unknown quantity we aim to solve for, $\gamma$ are physics-related parameters, $\bf{F}$ is the differential expression, $\bf{f}$ is a function relevant to the problem data, $\bf{I}$ denotes initial or boundary conditions which constrain the solution space, $\bf{b}$ is a boundary-data related function, and $\bf{\Omega}$ is the domain with $\bf{\Gamma}$ indicating its boundary. PINNs approximate $\bf{h}$ with a surrogate solution ${\bf{h}}_{\theta}$, where $\theta$ indicates the network parameters. {The training process begins by initialization of these parameters using any of the existing methods which exist in the literature with regards to this aspect} {\cite{glorot2010understanding, he2015delving}}. {In our case, we utilize the Xavier initialization {\cite{glorot2010understanding}}, which is a common practice in the PINN literature} {\cite{rao2020physics, markidis2021old, wang2021learning}}. The network learns the parameter vector $\theta$ by minimizing a cost function which is defined as the weighted sum of the PDE residual error $J_{PDE}$, the initial/boundary conditions error $J_{BCs}$ and the labeled data mismatch $J_{Data}$. Each term in this expression is multiplied with a corresponding weight factor, namely $w_{PDE}$, $w_{BCs}$ and $w_{Data}$ respectively, and therefore the cost function $J$ reads:

\begin{equation}
\begin{split}
    J(\theta) = w_{PDE} J_{PDE}(\theta) + w_{BCs} J_{BCs}(\theta)  + w_{Data} J_{Data}(\theta)
\end{split}
\label{General_Cost_function}
\end{equation}

The optimum parameter vector $\theta^{*}$ is computed as:

\begin{equation}
\begin{split}
    \theta^{*} = \arg \min_{\theta} (J)
\end{split}
\label{General_PINN_equation3}
\end{equation}

\noindent by the solution of the optimization problem involving the minimization of the cost function $J$. Minimization of the cost function $J$ is carried out by performing sequential forward and backward propagation passes across the layers. This process continuously updates the network parameters, typically with gradient-based algorithms \cite{lecun2015deep} such as the stochastic gradient descent or its more advanced successors (AdaGrad \cite{duchi2011adaptive}, Adam \cite{kingma2014adam}), until a predefined convergence criterion is satisfied. The calculation of the partial derivatives between the input and output quantities in the PDE and BCs residuals is enabled via a powerful technique which is available on many platforms platforms, namely the $\textit{automatic differentiation}$ \cite{rumelhart1986learning, baydin2018automatic}.
In this study, we make additional use of the automatic differentiation to compute the derivative of the network output with respect to an input variable, which allows us to compute the Jacobian matrix and residual for the FEM computations as discussed in the next section.

\begin{figure}
	\centering
	\includegraphics[scale=0.48]{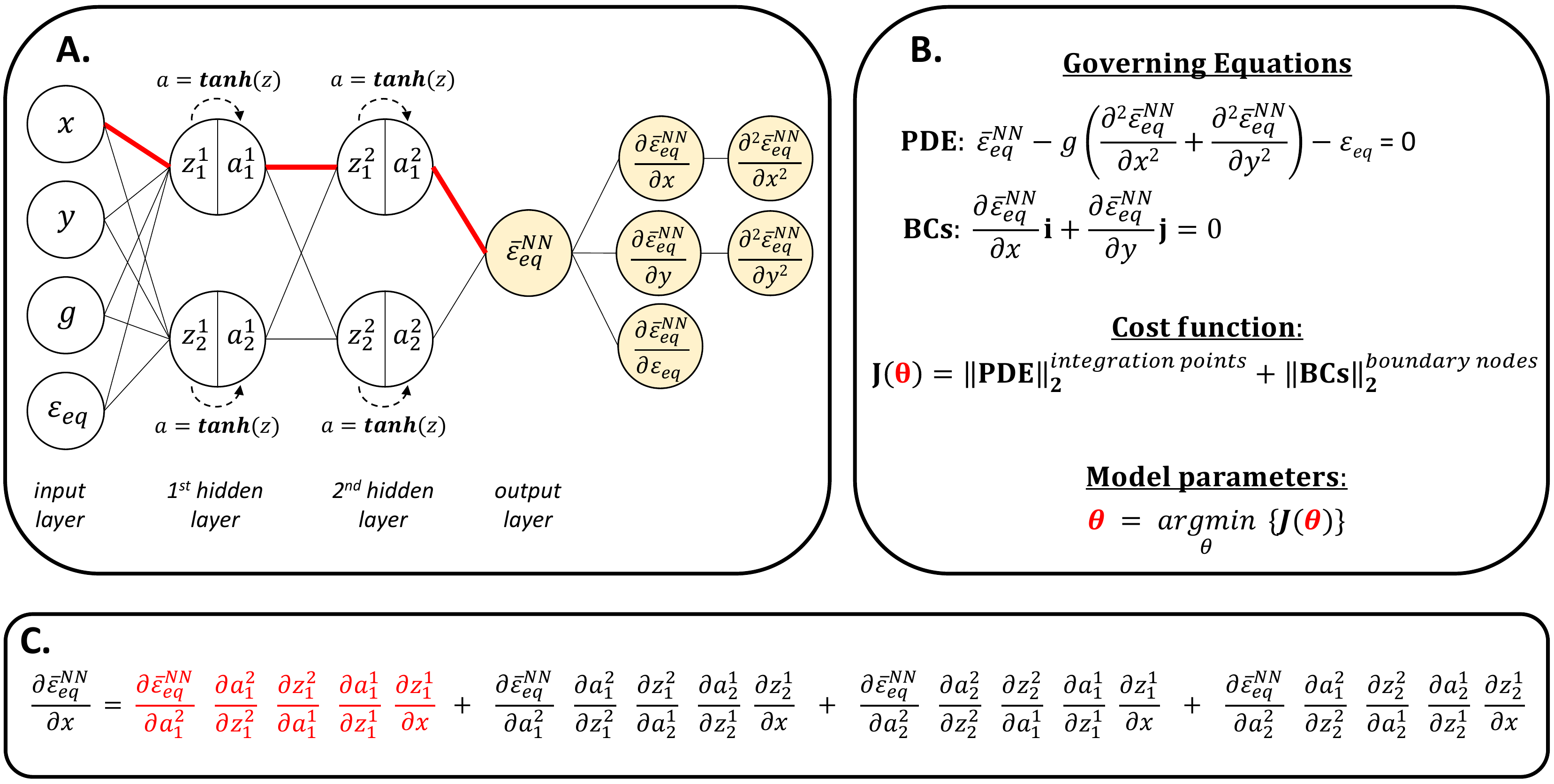}
	\caption{An overview of the developed PINN to represent the non-local gradient model. Box {\bf{A}}. Detailed representation of our PINN shown with only two hidden layers and two neurons for clarity. Box {\bf{B}}. The mathematical backbone of our framework: the partial differential equations which govern the non-local gradient formulation, the cost function definition and the expression for the PINN parameters $\theta$. Box {\bf{C}}. The analytical expression of the first-order partial derivative of $\bar{\varepsilon}^{NN}_{eq}$ with respect to the $x$ coordinate. The red-highlighted term corresponds to the path shown with the red lines in (A).}
	\label{Figure_Methodology_PINNs}
\end{figure}

The PINN we develop in this study follows the layout and principles displayed in Figure \ref{Figure_Methodology_PINNs}. The first sub-plot of this figure (Box A) shows a schematic representation of the network architecture, with just two hidden layers with two neurons each for the sake of clarity. The proposed PINN has a fully-connected deep neural network structure. The input data in our case are the $x$ and $y$ coordinate of the material points, the internal length scale measure $g$, and the local equivalent strain $\varepsilon_{eq}$. The PINN output variables are the:

\begin{itemize}

\item non-local equivalent strain $\bar{\varepsilon}_{eq}^{NN}$

\item second order derivatives of $\bar{\varepsilon}_{eq}^{NN}$ with respect to the $x$ and and $y$ coordinates, $\frac{\partial^{2}{{\bar{\varepsilon}^{NN}_{eq}}}}{\partial{x}^{2}}$ and $\frac{\partial^{2}{{\bar{\varepsilon}^{NN}_{eq}}}}{\partial{y}^{2}}$, computed at the integration points

\item first order derivatives of $\bar{\varepsilon}^{NN}_{eq}$ with respect to the $x$ and and $y$ coordinates, $\frac{\partial{{\bar{\varepsilon}^{NN}_{eq}}}}{\partial{x}}$ and $\frac{\partial{{\bar{\varepsilon}^{NN}_{eq}}}}{\partial{y}}$, computed at the boundary nodes

\item first order derivatives of $\bar{\varepsilon}^{NN}_{eq}$ with respect to the local equivalent strain, $\frac{\partial{{\bar{\varepsilon}^{NN}_{eq}}}}{\partial{\varepsilon}_{eq}}$, computed at the integration points

\end{itemize}

The first four partial derivatives $\left(\frac{\partial^{2}{{\bar{\varepsilon}^{NN}_{eq}}}}{\partial{x}^{2}},\frac{\partial^{2}{{\bar{\varepsilon}^{NN}_{eq}}}}{\partial{y}^{2}},\frac{\partial{{\bar{\varepsilon}^{NN}_{eq}}}}{\partial{x}},\frac{\partial{{\bar{\varepsilon}^{NN}_{eq}}}}{\partial{y}}\right)$ are used in the cost function definition, and more details are given in the next paragraphs. The partial derivative $\frac{\partial{{\bar{\varepsilon}^{NN}_{eq}}}}{\partial{\varepsilon}_{eq}}$ is incorporated in the nonlinear FEM solver and it is used for the Jacobian matrix calculation; the details of this implementation are given in Section \ref{SubSec:FEM_implementation}.

Regarding the cost function, we first clarify that in the proposed model we assume a completely unsupervised learning context, meaning that there are no labeled pairs for the output predictions $\bar{\varepsilon}^{NN}_{eq}$. Therefore, $w_{data}=0$ and the cost function degrades to the weighted sum of the governing differential equation and the boundary condition cost terms. Assuming a value of unity for each of the other weight terms $\left(w_{PDE}=1,w_{BCs}=1\right)$, the proposed cost function reads:

\begin{equation}
\begin{split}
    J(\theta) = J_{PDE}(\theta) + J_{BCs}(\theta)
\end{split}
\label{Our_Cost_function_s}
\end{equation}

It is important to clarify the material point locations where each of the cost function terms is evaluated. In FEM, the unknown displacements are computed at the nodes, whereas using the shape functions derivatives the strains are calculated at the Integration Points (IPs). Since strains lie at the core of the problem-specific PDE, we treat the IPs as the PINN collocation points and we minimize the PDE residual in these locations. As a result, the input dataset used for the evaluation of the governing PDE is tied to the IPs, and so do the output variables $\bar{\varepsilon}^{NN}_{eq}$, $\frac{\partial^{2}{{\bar{\varepsilon}^{NN}_{eq}}}}{\partial{x}^{2}}$, and  $\frac{\partial^{2}{{\bar{\varepsilon}^{NN}_{eq}}}}{\partial{y}^{2}}$. The output partial derivative $\frac{\partial{{\bar{\varepsilon}^{NN}_{eq}}}}{\partial{\varepsilon}_{eq}}$ is also calculated at the IPs, and it will be used to calculate the Jacobian matrix based on the I-FENN approach postulated in section \ref{SubSec:FEM_implementation}. On the other hand, the boundary conditions are applied on the domain edges. Therefore, the input dataset used for the evaluation of the boundary condition differential expression is related to the boundary nodes, and so do the output variables $\frac{\partial{{\bar{\varepsilon}^{NN}_{eq}}}}{\partial{x}}$ and $\frac{\partial{{\bar{\varepsilon}^{NN}_{eq}}}}{\partial{y}}$. Ultimately, both $J_{PDE}$ and $J_{BCs}$ are vectors of dimension equal to the total number of IPs and boundary nodes respectively, and each entry contains the residual value at the relevant IP or boundary node. The error metric used for both vectors is the L2-norm, and its formula reads:

\begin{equation}
\begin{split}
    ||.||_{2} = \sqrt{\sum {(.)^{2}}}
\end{split}
\label{L2_norm_formula}
\end{equation}

The reasoning behind adopting the L2-norm, a metric which has been used in other studies as well \cite{gao2021phygeonet}, is the following. The appropriateness and accuracy of the selected error metric depends on the output range values \cite{jokarFENA2021}. For the specific type of quasi-brittle materials under consideration, strain values are in the order of $10^{-3}$ and below. Error metrics such as the mean squared error and its relatives tend to yield lower values for this range than the L2-norm, and therefore the latter is more critical and it is used as our cost function metric. 

To summarize, the expressions for the governing differential equations which dictate the PINN training, as well as the cost function equation, are given below:

\begin{equation}
    {\bf{PDE}}: {\bar{\varepsilon}}^{NN}_{eq} - g \Big(\frac{\partial^{2} \bar{\varepsilon}^{NN}_{eq}}{\partial {x^{2}}} + \frac{\partial^{2} \bar{\varepsilon}^{NN}_{eq}}{\partial {y^{2}}}\Big) -\varepsilon_{eq} = 0 
\label{NonlocalGradientPDE_PINN}
\end{equation}

\begin{equation}
{\bf{BCs}}: \frac{\partial \bar{\varepsilon}^{NN}_{eq}}{\partial x} {\bf{i}} + \frac{\partial \bar{\varepsilon}^{NN}_{eq}}{\partial y} {\bf{j}} = 0
\label{NonlocalGradientBCs_PINN}
\end{equation}

\begin{equation}
\begin{split}
    J = ||{\bf{PDE}}||^{integration \ points}_{2} + ||{\bf{BCs}}||^{boundary \ nodes}_{2}
\end{split}
\label{Our_Cost_function}
\end{equation}

\noindent where {\bf{i}} and {\bf{j}} are the outward unit normal vectors in the x and y directions, respectively.

The presence of second-order derivatives in the PDE residual dictates the use of activation functions which are at least twice differentiable; otherwise, these derivatives will trivially degrade to zero. Therefore we select the \textit{hyperbolic tangent} ($tanh()$) function, which is always an admissible solution for PINNs \cite{mishra2020}, and we denote that piece-wise linear activation functions such as ReLU or leaky-ReLU are not appropriate choices for this specific problem. The search for the optimum hyperparameters, such as the number of neurons, layers, epochs and learning rate value, still lies as an open question of active research in many areas \cite{lye2020109339}, and this is true for the subject problem as well. {The network performance is a priori sensitive to these choices, and the later still relies heavily on trial-and-error methods, empirical observations and literature guidance}. Each of the numerical examples which are demonstrated in Section \ref{Sec:NumericalExamples} utilizes a different value for these hyperparameters, which are btained through trial-and-error {and they are reported in Table {\ref{Table_PINNs_hyperparameters}}}. As for the chosen optimizers, at the beginning of the training we use the Adam \cite{kingma2014adam} algorithm for a predefined number of epochs, followed by the L-BFGS algorithm until a specified convergence threshold is satisfied. This approach has become a common training procedure which is adopted in several studies \cite{markidis2021old}. We also note that in its current formulation, the input data can potentially span different numerical scales. For example, the $x$ and $y$ coordinates as well as the length scale parameter $g$ depend on the dimension units of the problem, whereas the local strain is dimensionless, and typically exhibits values that are orders of magnitude below unity. Though not obligatory, this contrast between scales in many engineering applications requires some sort of input feature scaling in order to increase the network ability to learn, which is a common practice in the field of data science \cite{yin2022interfacing,datanormalization,moseley2021finite}. In the numerical examples presented in the next section, we normalize the local equivalent strain by a scaling factor in the form of $10^{c}$ ($c = 1$ or $2$), and subsequently unnormalize the output values by the same factor, as an initial attempt to increase the network learning speed.

It is useful to note here that the greater goal of this research effort is to design a single universal NN that can represent the local to non-local strain transformation regardless of problem-specific details, such as geometry, mesh discretization, boundary conditions, applied load, damage growth, etc. However, the development and training of such a network requires the abstraction of the spatial and temporal characteristics of non-local gradient damage evolution \cite{wang2021stressnet}. To the best of the author's knowledge, current PINN applications are always limited by a certain set of geometries, boundary conditions, and other model-specific details. {This is because PINNs are tied to the PDE they aim to approximate its solution, and the latter is sensitive to different configurations. Consequently, the state-of-the-art implementation of PINNs lacks the desired generalizability}. The achievement of this network requires further research on the design of the network as well as the pre-processing of the data that gets passed to the network, which is beyond the immediate objectives of this study and requires some advancements in the design of neural networks by the active research community in this field \cite{mishra2020,rao2020physics,markidis2021old}. Therefore, we bound the exploration space in most of the aforementioned parameters and mainly focus on the feasibility of integrating PINNs within a nonlinear finite element analysis, training a single PINN for a single combination of meshed geometry and applied load.

\subsection{FEM implementation}
\label{SubSec:FEM_implementation}

This section presents the mathematical formulation of the proposed non-local gradient-based damage model and details its implementation within a nonlinear finite element solver. We remind the reader that the goal of the proposed implementation is twofold: a) maintain the size of equations system as in the local damage framework, in other words avoid solving numerically for the non-local equivalent strain, and b) embed a non-local nature in the damage field variable, and calculate the element Jacobian and residual vector consistently with the non-local damage definition. At the end of this section we include Table \ref{Table_Methodologies_Comparison} which presents a comprehensive comparison between the local, non-local gradient and proposed methodology. This allows for a straightforward observation of the similarities and differences between the three methods, and confirms the advantages of the proposed model. For a complete derivation of the classical local \cite{kachanovbook} and non-local gradient \cite{peerlings1996gradient} damage methods, the reader is referred to \ref{Appendix:AppendixA_Local_Damage_FEM} and \ref{Appendix:AppendixB_NonLocal_Gradient_Damage_FEM} respectively.

We begin by formulating the strong form of the governing PDE, which is the equilibrium condition. Substituting Equation \eqref{Cauchy_stress} in \eqref{Equilibrium_Condition} leads to: 

\begin{equation}
\begin{split}
    \left[C_{ijkl}(d)\varepsilon_{kl}\right]_{,j} = 0 \; \; \; in \; \; \Omega
\end{split}
\label{FEM_Strong}
\end{equation}

\noindent where:

\begin{equation}
    C_{ijkl}(d) = (1-d) C_{ijkl} 
\label{FEM_Cijkl}
\end{equation}

Proceeding with the weak form of Equation \eqref{FEM_Strong}:

\begin{equation}
    {\bf{R}}(u) = \int_{\Omega} w^u \left[\left[C_{ijkl}(d)\varepsilon_{kl}\right]_{,j}\right] \; d\Omega
\label{FEM_WeakR}
\end{equation}

\noindent where $w^u$ are the displacement field weight functions. Integration by parts of Equation \eqref{FEM_WeakR} yields:  

\begin{equation}
    {\bf{R}}(u) = - \int_{\Gamma} \left[w^u t_i \right] d\Gamma + 
    \int_{\Omega} w^u _{,j} \left[C_{ijkl}(d)\varepsilon_{kl}\right] \; d\Omega
\label{FEM_WeakRexpand}    
\end{equation}

\noindent where $t_i$ corresponds to the external stress vector applied on the domain boundary. Denoting with $\bf{N}$ and $\bf{B}$ the displacement shape function matrix and its spatial derivatives respectively, the displacements and strains at the integration points are calculated as follows:

\begin{equation}
	{\bf{u}} = {\bf{N}} {\bf{\hat{u}}}; ~~~ {\bf{\varepsilon}}_{ij} = {\bf{B}} {\bf{\hat{u}}}
\label{FEM_IPdisps}
\end{equation}

\begin{equation}
	{\bf{w^u}} = {\bf{N}} {\bf{\hat{w}^u}}; ~~~ {\bf{w^u}}_{,j} = {\bf{B}} {\bf{\hat{w}^u}}
\label{FEM_IPweight}
\end{equation}

\noindent where the ($\hat{.}$) symbol is used to indicate nodal values. Substitution of Equations \eqref{FEM_IPdisps} and \eqref{FEM_IPweight} into Equation \eqref{FEM_WeakRexpand} results to the following expression of the residual vector:

\begin{equation}
    {\bf{R}}(u) = - \int_{\Gamma} \left[{\bf{N}} \hat{t}_i \right] d\Gamma + 
    \int_{\Omega} {\bf{B}}^{T} (1-d)C_{ijkl}{\bf{B}} {\bf{\hat{u}}} \; d\Omega
\label{FEM_WeakRexpand2}    
\end{equation}

In a quasi-static analysis, equilibrium at each load increment is achieved once the residual vector is sufficiently minimized. The problem non-linearity necessitates an incremental and iterative procedure in order to obtain the new equilibrium state, and appropriate nonlinear iterative numerical solvers such as the Newton-Raphson method or the arc-length method need to be implemented. In this study we adopt the Newton-Raphson approach, and a generic algorithm is provided in \ref{Sec:AppendixC}. In this scheme, denoting with $i$ the iteration counter, the new displacement vector $\bf{\hat{u}}_{i}$ can be computed as follows: 

\begin{equation}
    {\bf{\hat{u}}}_{i} = {\bf{\hat{u}}}_{i-1} + \delta{\bf{\hat{u}}}_{i-1}
\label{FEM_JduR_3}
\end{equation}

\noindent where ${\bf{\hat{u}}}_{i-1}$ is the displacement solution vector at the previous iteration, and $\delta{\bf{\hat{u}}}_{i-1}$ is the incremental change of the unknown degrees of freedom vector which is calculated as:

\begin{equation}
    {\bf{J}} \delta {\bf{\hat{u}}} = - {\bf{R}}(u)
\label{FEM_JduR_1}
\end{equation}

In Equation \eqref{FEM_JduR_1}, the term $\bf{J}$ is the Jacobian matrix and it is computed upon differentiation of Equation \eqref{FEM_WeakRexpand2} with respect to the nodal displacements:

\begin{equation}
    {\bf{J}} = \frac{\partial{\bf{R}}(u)}{\partial{\bf{\hat{u}}}}
\label{FEM_JduR_2}
\end{equation}

Numerical convergence within each increment is achieved once the following equation is satisfied:

\begin{equation}
    \frac{||\delta{\bf{\hat{u}}}_{i = last}||_{2}}{||\delta{\bf{\hat{u}}}_{i = 1}||_{2}} \leq tol
\label{FEM_JduR_4}
\end{equation}

\noindent where $\delta{\bf{\hat{u}}}_{i = last}$ and $\delta{\bf{\hat{u}}}_{i = 1}$ are the L2-norms of the displacement solution vector incremental changes at the last and first iteration respectively, and $tol$ is a numerical tolerance variable. 

We now focus on the derivation of the $\bf{J}$ and $\bf{R}$ terms, which constitute the novelty of the proposed methodology. Plugging Equation \eqref{FEM_WeakRexpand2} into Equation \eqref{FEM_JduR_2}, the expression for the global Jacobian matrix $\bf{J}$ reads:

\begin{equation}
    {\bf{J}} = {\bf{K}}(u) + \frac{\partial{\bf{K}}(u)}{\partial{\bf{\hat{u}}}}{\bf{\hat{u}}}
\label{FEM_Jexpr1}
\end{equation}

\noindent where: 

\begin{equation}
    {\bf{K}}(u) = \int_\Omega \! {\bf{B^{T}}} (1-d) {C_{ijkl}} {\bf{B}} \, \mathrm{d}\Omega
\label{FEM_K_prop}
\end{equation}

\begin{equation}
    \frac{\partial{\bf{K}}(u)}{\partial{\bf{\hat{u}}}} = \int_\Omega \! {\bf{B^{T}}} {C_{ijkl}} \, \left(-\frac{\partial{d}}{\partial {\bf{\hat{u}}}}\right) \, {\bf{B}} \, \mathrm{d}\Omega
\label{FEM_DKdu_prop}
\end{equation}

Following standard FEM discretization procedures, the integral formulation of Equations \eqref{FEM_K_prop} and \eqref{FEM_DKdu_prop} is replaced by a summation of these quantities computed on an element-level basis. Proceeding with element-level calculations, the damage variable $d$ in Equation \eqref{FEM_K_prop} is a function of the non-local strain PINN output, $d = d(\bar{\varepsilon}^{NN}_{eq})$, and it is calculated by substituting $\bar{\varepsilon}^{NN}_{eq}$ into the damage evolution law in Section \ref{Sec:ConstitutiveModeling}. Therefore, once the neural network is trained and it has learned the local to non-local strain transformation as explained in Section \ref{SubSec:MethodPINNs}, calculation of $d$ is straightforward. The term $\frac{\partial{d}}{\partial {\bf{\hat{u}}}}$ in Equation \eqref{FEM_DKdu_prop} reflects the dependency of $d$ to the nodal displacements ${\bf{\hat{u}}}$, which can be calculated as follows:

\begin{equation}
    \frac{\partial{d}}{\partial \hat{u}_{k}} = 
    \frac{\partial{d}}{\partial{{\bar{\varepsilon}}^{NN}_{eq}}}                \; 
    \frac{\partial{\bar{\varepsilon}^{NN}_{eq}}}{\partial{{\varepsilon}_{eq}}}       \; 
    \frac{\partial{\varepsilon_{eq}}}{\partial{\varepsilon_{ij}}}               \; 
    \frac{\partial{\varepsilon_{ij}}}{\partial{\hat{u}}_{k}}
\label{FEM_Dddu_prop}
\end{equation}

Equation \eqref{FEM_Dddu_prop} represents the complete dependence of $d$ to ${\hat{u}}_{k}$, accounting for all the intermediate transformation steps. Taking a closer look at each term in Equation \eqref{FEM_Dddu_prop}:

\begin{itemize}

\item the term $\frac{\partial{d}}{\partial{{\bar{\varepsilon}}^{NN}_{eq}}}$ reflects the dependence of damage to the non-local strain PINN output, and it is computed based on the selected damage evolution relationship. The damage evolution laws used in this study are described in Section \ref{Sec:ConstitutiveModeling}.

\item the term $\frac{\partial{\bar{\varepsilon}^{NN}_{eq}}}{\partial{{\varepsilon}_{eq}}}$ represents the differential relationship between between the PINN input ${\varepsilon}_{eq}$ and the PINN output $\bar{\varepsilon}^{NN}_{eq}$. This partial derivative is one of the two output variables of the neural network, and it is computed based on automatic differentiation. Its incorporation inside Equation \eqref{FEM_Dddu_prop} is the key step that affirms the consistent Jacobian matrix definition and ensures the sound performance of the Newton-Raphson algorithm.

\item the term $\frac{\partial{\varepsilon_{eq}}}{\partial{\varepsilon_{ij}}}$ depends on the definition of the local equivalent strain $\varepsilon_{eq}$ upon the tensorial strain components $\varepsilon_{ij}$. The definition of the equivalent strain used in this study is provided in Section \ref{Sec:ConstitutiveModeling}.

\item the term $\frac{\partial{\varepsilon_{ij}}}{\partial{\hat{u}}_{k}}$ is the $\bf{B}$ matrix components, representing the displacement-strain relationship at the integration points. 

\end{itemize}

This procedure computes the element-level Jacobian matrix ${\bf{J}}_{elem}$, and the global $\bf{J}$ is constructed upon the assembly of all ${\bf{J}}_{elem}$. As for the residual vector $\bf{R}$, the implementation of the shape function derivative matrix $\bf{B}$ in Equation \eqref{FEM_WeakR} yields:

\begin{equation}
    {\bf{R}}(u) = \int_{\Omega} {\bf{B}}^{T} (1 - d) C_{ijkl}\varepsilon_{kl} \; d\Omega
\label{FEM_WeakR_prop}
\end{equation}

The integral expression is evaluated at the element-level residuals ${\bf{R}}_{elem}$, which are then assembled to lead to the global $\bf{R}$ vector. Ultimately, substituting Equations \eqref{FEM_Jexpr1} and \eqref{FEM_WeakR_prop} into Equation \eqref{FEM_JduR_1} allows for the calculation of $\delta {\bf{\hat{u}}}$ at each iteration. The nonlinear iterative process continues until Equation \eqref{FEM_JduR_4} is satisfied and numerical convergence has been achieved. 

The workflow of the methodology is shown in Algorithm \ref{algorithm_hybrid_FEM_PINN}, which demonstrates how the trained neural network is integrated in the element-level computations and it is being utilized within a nonlinear, iterative procedure. The displacement and local strain fields are re-computed in each iteration, and the latter is used to update the non-local strain profile, jacobian matrix and residual vector. Therefore the PINN input and output are continuously updated within the Newton-Raphson procedure, which is the key element that allows for the residuals convergence. 

\begin{algorithm}
	\caption{Algorithm of the I-FENN framework in the case of non-local gradient damage}
	\label{algorithm_hybrid_FEM_PINN}
    \While {Equation $\eqref{FEM_JduR_4}$ is not satisfied}
    {
        \For{$\mathrm{each}$ $\mathrm{integration}$ $\mathrm{point}$ $and$ $\mathrm{each}$ $\mathrm{boundary}$ $\mathrm{node}$}
            {
            {Compute the shape functions $\bf{N}$ and their derivatives $\bf{B}$} \\
            Compute and extract coordinates, $g$ and $\varepsilon_{eq}$
            }
        Use the pre-trained PINN to predict $\bar{\varepsilon}^{NN}_{eq}$ and $\frac{\partial \bar{\varepsilon}^{NN}_{eq}}{{\partial \varepsilon}_{eq}}$ for all IPs \\
        \For{$\mathrm{each}$ $\mathrm{finite}$ $\mathrm{element}$}
        {
            \For{$\mathrm{each}$ $\mathrm{integration}$ $\mathrm{point}$}
            {
            Compute the shape functions $\bf{N}$ and their derivatives $\bf{B}$ \\
            Calculate $\frac{\partial \varepsilon_{eq}}{\partial \varepsilon_{ij}}$, based on the local equivalent strain definition \\
            Calculate $d$ and $\frac{\partial d}{\partial \bar{\varepsilon}^{NN}_{eq}}$, based on the governing damage law \\
            Compute the IP contribution to the element jacobian matrix and residual vector
            }
        }
        Assemble global Jacobian matrix $\bf{J}$, solve for the displacement vector $\bf{u}$, check convergence of residuals $\bf{R}$ 
    }
\end{algorithm}

Table \ref{Table_Methodologies_Comparison} provides a comparison across the local, non-local gradient and proposed framework with regards to the main computed variables. We highlight that the local damage method and the proposed framework share the same number of degrees of freedom, whereas the non-local gradient \cite{peerlings1996gradient} treats the non-local equivalent strain as an additional DOF. Specifically for a 2D finite element mesh with first order shape functions, the total number of DOFs for the local damage and the proposed method is $2n$ and for the gradient model it is $3n$. The residual expressions are provided in their weak form to allow for a more straightforward comparison across the three approaches, and we denote with $w^{\bar{\varepsilon}}$ the non-local strain weighting functions for the non-local gradient method. The mathematical derivation of the conventional local and non-local gradient methods can be found in \ref{Appendix:AppendixA_Local_Damage_FEM} and \ref{Appendix:AppendixB_NonLocal_Gradient_Damage_FEM} respectively. 

\begin{table}[H]
	\caption{Comparison between local, non-local gradient and proposed framework.}
	\label{Table_Methodologies_Comparison}\centering
	\begin{adjustbox}{width=\textwidth}
	\begin{tabular}{c||c|c|c}
		\toprule
		& Local & Non-local gradient & Proposed \\
		\midrule
        $\bf{\# Nodes}$ & $n$ & $n$ & $n$ \\

        $\bf{\# DOFs}$ (2D) & $2n$ & $3n$ & $2n$ \\
        
        \multirow{2}{*}{${\bf{R}}$ (Residual)} & \(\displaystyle
        R = \int_{\Omega} w^u \left[\left[C_{ijkl}(d)\varepsilon_{kl}\right]_{,j}\right] d\Omega \) & \(\displaystyle
        R^u = \int_{\Omega} w^u \left[\left[C_{ijkl}(d)\varepsilon_{kl}\right]_{,j}\right] d\Omega \) & \(\displaystyle
        R = \int_{\Omega} w^u \left[\left[C_{ijkl}(d)\varepsilon_{kl}\right]_{,j}\right] d\Omega \) \\
        &  & \(\displaystyle R^{\bar{\varepsilon}} = \int_{\Omega} w^{\bar{\varepsilon}} \left[\bar{\varepsilon}_{eq} - g \bar{\varepsilon}_{eq,ii}-\varepsilon_{eq}\right] d\Omega \)   \\
        
		${\bf{J}}$ (Jacobian) & $\begin{bmatrix} \frac{\partial{\bf{R}}}{\partial{\bf{\hat{u}}}} \end{bmatrix}$ & 
		$\begin{bmatrix}
        \frac{\partial{\bf{R}}^{u}}{\partial{\bf{\hat{u}}}} & \frac{\partial{\bf{R}}^{u}}{\partial{\bf{\hat{\bar{\varepsilon}}}}} \\
        \frac{\partial{\bf{R}}^{\bar{\varepsilon}}}{\partial{\bf{\hat{u}}}} & \frac{\partial{\bf{R}}^{\bar{\varepsilon}}}{\partial{\bf{\hat{\bar{\varepsilon}}}}}
        \end{bmatrix}$ & $\begin{bmatrix} \frac{\partial{\bf{R}}}{\partial{\bf{\hat{u}}}} \end{bmatrix}$  \\
		
		$\bf{Dependencies}$ & \(\displaystyle
        \frac{\partial{d}}{\partial \hat{u}_{k}} = 
        \frac{\partial{d}}{\partial{{\varepsilon}_{eq}}}       \; 
        \frac{\partial{\varepsilon_{eq}}}{\partial{\varepsilon_{ij}}}               \; 
        \frac{\partial{\varepsilon_{ij}}}{\partial{\hat{u}}_{k}}\) & 
        
        \(\displaystyle
        \frac{\partial{d}}{\partial \hat{\bar{\varepsilon}}_{eq}}, \frac{\partial{\varepsilon_{eq}}}{\partial{\hat{u}}_{k}} =  \frac{\partial{\varepsilon_{eq}}}{\partial{\varepsilon_{ij}}}               \; 
        \frac{\partial{\varepsilon_{ij}}}{\partial{\hat{u}}_{k}} \)
        
        & \(\displaystyle
        \frac{\partial{d}}{\partial \hat{u}_{k}} = 
        \frac{\partial{d}}{\partial{{\bar{\varepsilon}}^{NN}_{eq}}}                \; 
        \frac{\partial{\bar{\varepsilon}^{NN}_{eq}}}{\partial{{\varepsilon}_{eq}}}       \; 
        \frac{\partial{\varepsilon_{eq}}}{\partial{\varepsilon_{ij}}}               \; 
        \frac{\partial{\varepsilon_{ij}}}{\partial{\hat{u}}_{k}}\) \\
        \bottomrule
	\end{tabular}
	\end{adjustbox}
\end{table}

%%%%%%%%%%%%%%%%%%%%%%%

\section{Constitutive modeling}
\label{Sec:ConstitutiveModeling}

This section presents the relationships to compute the local equivalent strain and the associated damage variable. As for the scalar invariant measure of the deformation, $\varepsilon_{eq}$, we adopt two definitions. The first one follows Lemaitre's approach \cite{lemaitrebook}:

\begin{equation}
    \varepsilon_{eq}=\sqrt{\langle\varepsilon_{1}\rangle^{2} + \langle\varepsilon_{2}\rangle^{2} + \langle\varepsilon_{3}\rangle^{2}}
\label{StrainA}
\end{equation}

\noindent where $\varepsilon_{i}, i = 1, 2, 3$ are the principal strains and $\langle \, \cdot \: \rangle$ are the Macaulay brackets: $\langle . \rangle = \frac{1}{2}(. + |.|)$. Following this expression, $\varepsilon_{eq}$ depends evidently only in the positive principal strains. This is a common assumption in quasi-brittle materials, which exhibit significantly lower tensile strength than the compressive strength \cite{peerlings1998gradient}. The second definition of $\varepsilon_{eq}$ follows the modified von Mises strain definition and its expression is provided below \cite{de1995comparison}:

\begin{equation}
    \varepsilon_{eq} = \frac{k-1}{2k(1-2\nu)} + \frac{1}{2k} \sqrt{\frac{(k-1)^2}{(1-2\nu)^{2}}I_{1}^{2} + \frac{2k}{(1+\nu)^2}J_{2}}
\label{StrainB}
\end{equation}

\noindent where $I_{1}$ and $J_{2}$ are the strain invariants given by the following formulas ($\bf{\varepsilon}$ is the strain tensor):

\begin{equation}
    I_{1} = tr(\boldsymbol{\varepsilon})
\label{Invariant1}
\end{equation}

\begin{equation}
    J_{2} = 3tr(\boldsymbol{\varepsilon} \cdot \boldsymbol{\varepsilon}) - tr^{2}(\boldsymbol{\varepsilon})
\label{Invariant2}
\end{equation}

The damage evolution laws relate the value of the damage variable $d$ to the equivalent strain. For the sake of notation clarity, we note that $d = d(\varepsilon^{*}_{eq})$ where $\varepsilon^{*}_{eq}$ indicates any of the equivalent strains:

\begin{equation}
    \varepsilon^{*}_{eq} =
    \left\{
    	\begin{array}{lll}
    	\varepsilon_{eq} & \text {(local)}  \\
    	\bar{\varepsilon}_{eq} & \text {(non-local gradient)} \\
    	\bar{\varepsilon}^{NN}_{eq} & \text {(I-FENN)}
    	\end{array}
    \right.
\label{estardefinitions}
\end{equation}

In this study, we adopt two versions of an exponential-based damage model which are extensively used in the literature. The first version is the original model proposed by Mazars \cite{mazars1986description}, in which damage depends on the equivalent strain based on the following relationship:

\begin{equation}
    d(\varepsilon^{*}_{eq}) =
    \left\{
    	\begin{array}{ll}
    	0 & \text { if } \varepsilon^{*}_{eq} < \varepsilon_{D} \\
    	1 - \frac{\varepsilon_{D}(1-\alpha)}{\varepsilon^{*}_{eq}} - 
    	\frac{\alpha}{\exp(\beta(\varepsilon^{*}_{eq} - \varepsilon_{D}))} & \text { if } \varepsilon^{*}_{eq} \geq \varepsilon_{D}
    	\end{array}
    \right.
\label{DamageA}
\end{equation}

In the expression above, the variable $\varepsilon_{D}$ represents a strain threshold value which signals damage initiation, and it is a material property. The second damage model is a slightly modified version of the original formulation, in which damage is calculated as follows \cite{de2016gradient}:

\begin{equation}
    d(\varepsilon^{*}_{eq}) =
    \left\{
    	\begin{array}{ll}
    	0 & \text { if } \varepsilon^{*}_{eq} < \varepsilon_{D} \\
    	1 - \frac{\varepsilon_{D}}{\varepsilon^{*}_{eq}}\{ (1-\alpha) + \alpha \exp(\beta(\varepsilon_{D} - \varepsilon^{*}_{eq})) \}   
        & \text { if } \varepsilon^{*}_{eq} \geq \varepsilon_{D}
    	\end{array}
    \right.
\label{DamageB}
\end{equation}

\begin{figure}
	\centering
	\includegraphics[scale=0.7]{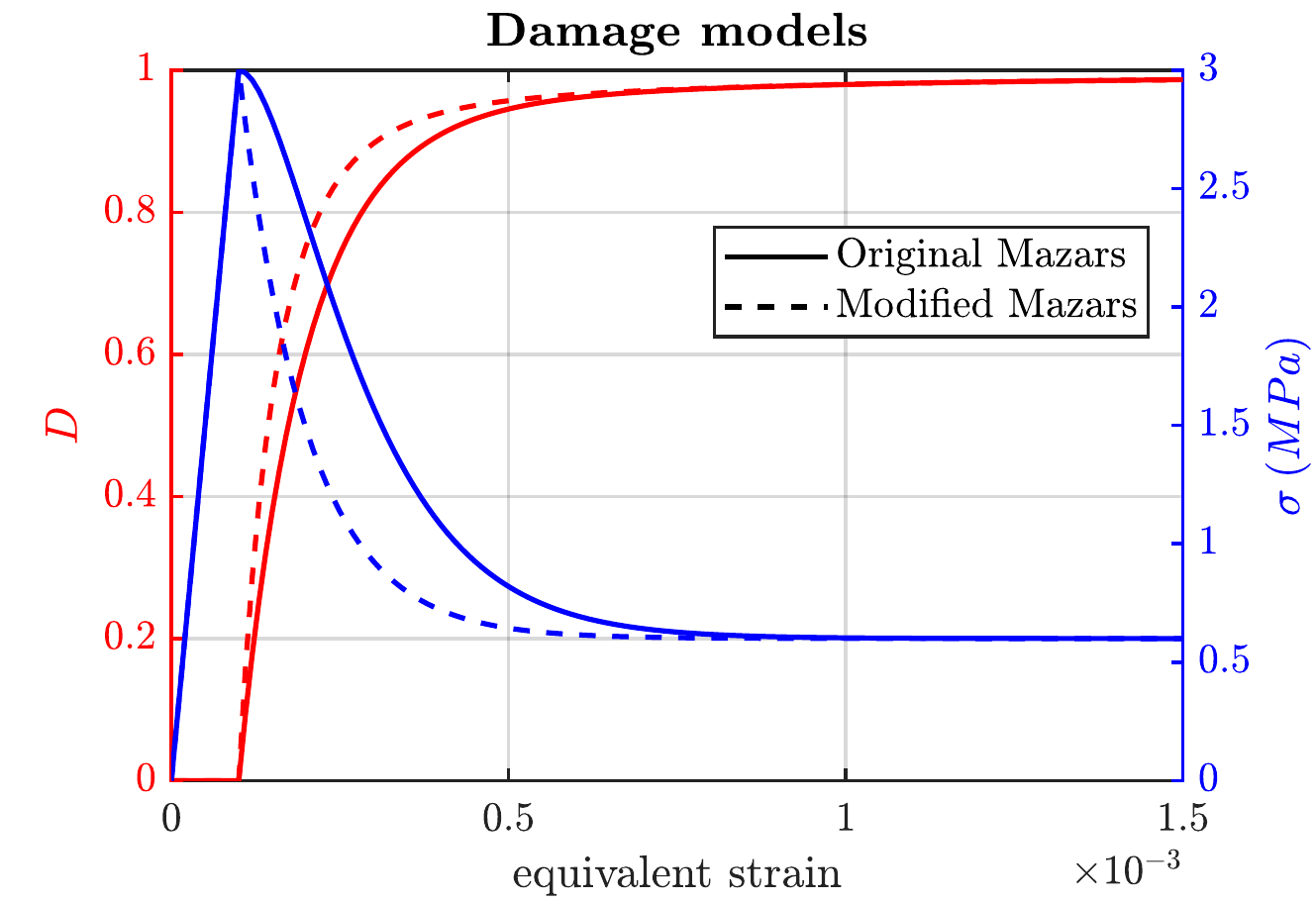}
	\caption{Graphical illustration of the damage evolution (red) and the associated stress-strain (blue) curves for uniaxial tension. The original Mazars model \cite{mazars1986description} is shown with continuous lines, and the modified version is depicted with dashed lines. In both cases $\alpha = 0.8$, $\beta = 10000$, $\varepsilon_{D} = 0.0001$, and the modulus of elasticity is $E = 30000 MPa$.}
	\label{Figure_Damage_models_two_in_one}
\end{figure}

In the definition of $d$ provided in Equations \eqref{DamageA} and \eqref{DamageB}, the parameter $\alpha$ is related to the residual strength in the material when the tensile strain approaches infinity, and the parameter $\beta$ is associated with the softening branch of the curve with higher values indicating a steeper post-peak slope and a more brittle behavior \cite{peerlings1998gradient}. A graphical illustration of the two versions is shown in Figure \ref{Figure_Damage_models_two_in_one}, along with the associated stress-strain curve in uniaxial tension for each model. Finally, we mention that maintaining thermodynamic consistency enforces damage irreversibility, and therefore the Karush-Kuhn-Tucker relationships need to be satisfied \cite{aifantis1986}. 

%%%%%%%%%%%%%%%%%%%%
\section{Numerical Examples}
\label{Sec:NumericalExamples}

This section presents the numerical implementation of the proposed methodology. Three distinct cases are analyzed, in order to examine the applicability and potential of I-FENN. First, in section \ref{SubSubSec:Single_notched_spec} we apply the method to a domain with a structured finite element mesh and a single crack. Then, in section \ref{SubSubSec:Double_notched_spec} we extend our investigation to a geometry with a structured mesh and two cracks, to examine whether the neural network can identify the presence of multiple strain localization regions. Finally, in section \ref{SubSubSec:Lshaped_spec} we implement I-FENN to a geometry with a single crack and an unstructured mesh, to verify that the choice of mesh discretization does not impact the learning ability of the neural network. In all cases we compare our results with the conventional non-local gradient method \cite{peerlings1996gradient}. The load application is displacement-controlled. For the remainder of this section the applied displacement/load is shown by means of a relative loadfactor value and it is denoted with the symbol $lf$. This variable depicts the applied displacement at the load increment of interest ${\bf{u}}^{app}$ normalized by the value of the total applied displacement $\bar{\bf{u}}$, therefore $lf = \frac{{\bf{u}}^{app}}{\bar{\bf{u}}}$. {Table {\ref{Table_PINNs_hyperparameters}} contains the values for the hyperparameters which are used for the training of the neural network of each test case. Specifically, we mention the number of layers, neurons per layer, epochs and learning rate.}

\begin{table}[H]
	\caption{List of hyperparameters used for the training of each test case.}
	\label{Table_PINNs_hyperparameters}\centering
    \begin{adjustbox}{width=\textwidth}
	\begin{tabular}{c||c|c|c||c|c||c|c|c}
		\toprule
		& \multicolumn{3}{c||}{$\bf{Single \ notch}$} & \multicolumn{2}{c||}{$\bf{Double \ notch}$} & \multicolumn{3}{c}{$\bf{L-shaped}$} \\
		
        & $lf = 0.42$ & $lf = 0.70$ & $lf = 0.82$ & $lf = 0.45$ & $lf = 0.70$ & $lf = 0.25$ & $lf = 0.50$ & $lf = 0.70$ \\ 
        
        \midrule

        $\bf{\# layers}$ & $10$ & $10$ & $10$ & $10$ & $10$ & $10$ & $10$ & $10$ \\

        $\bf{\# neurons/layer} $ & $50$ & $50$ & $50$ & $140$ & $120$ & $100$ & $100$ & $140$ \\
        
        $\bf{\# epochs} $ & $20000$ & $20000$ & $20000$ & $30000$ & $100000$ & $30000$ & $50000$ & $100000$ \\

        $\bf{learning \ rate} $ & $0.001$ & $0.001$ & $0.001$ & $0.001$ & $0.001$ & $0.0005$ & $0.0005$ & $0.0005$  \\
        
        \bottomrule
	\end{tabular}
 \end{adjustbox}
\end{table}

\subsection{Single-notched specimen}
\label{SubSubSec:Single_notched_spec}

\begin{figure}
	\centering
	\includegraphics[width=\linewidth]{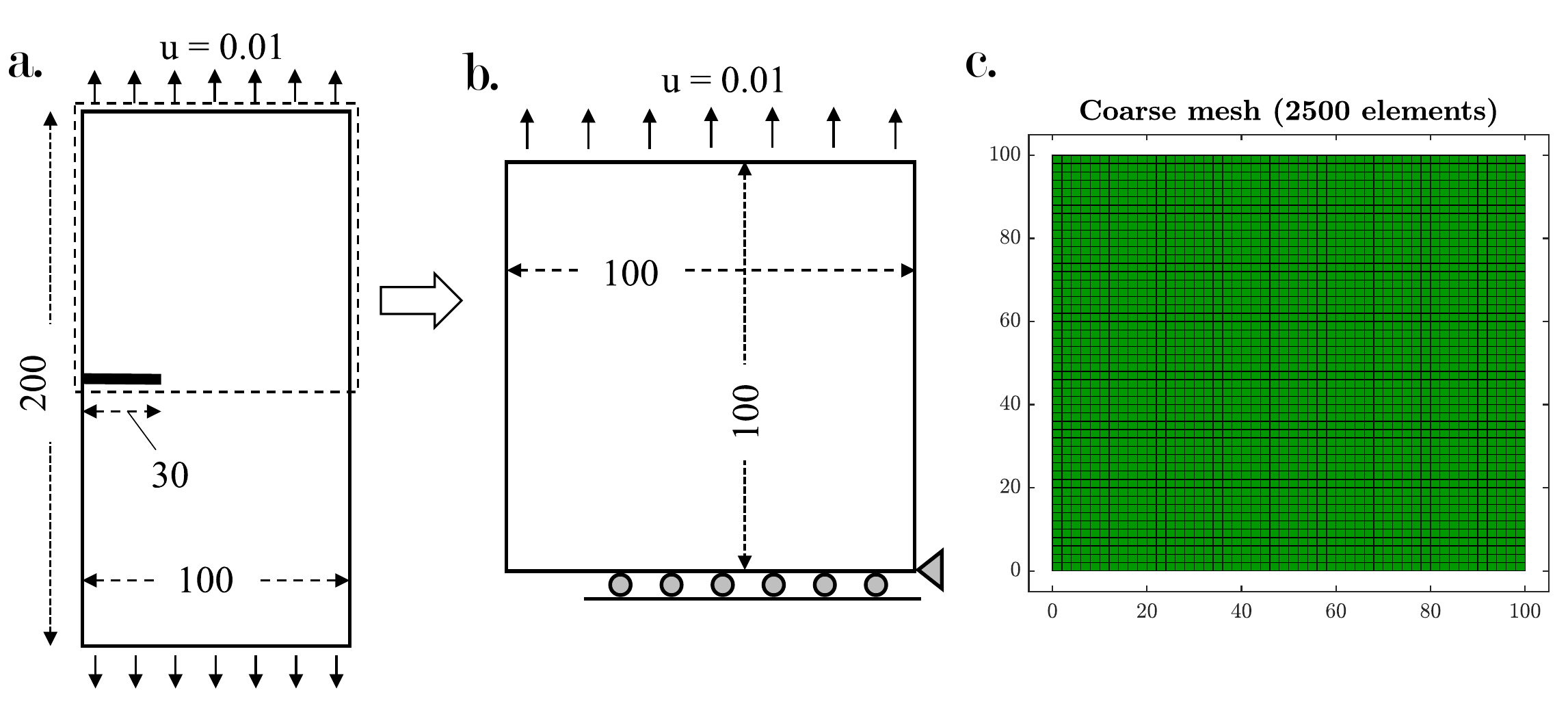}
	\caption{{\bf{a}}. Schematic illustration of the geometry, boundary conditions and loading for the single-notch case. {\bf{b}}. Due to symmetry, the upper half of the initial geometry is used in all numerical analyses. Domain continuity across the middle center-line is restored numerically through rollers. {\bf{c}}. The $Coarse$ finite element model of the domain, which is developed using a structured mesh of 2500 elements. All length units are in mm.}
	\label{Fig_Str100_Geometry_Mesh}
\end{figure}

The first numerical example is a mode-I loading case of a single-notch specimen. The geometry under consideration is shown in Figure \ref{Fig_Str100_Geometry_Mesh}. The symmetric nature of the problem allows the isolation and analysis of only the upper half of the domain, as shown in Figure \ref{Fig_Str100_Geometry_Mesh}b. A series of rollers is applied at the horizontal centerline of the intact domain, constraining the y-displacement of these nodes. The x-displacement of the bottom right node is also constrained. The shear modulus is taken as $G = 125000$ KPa and Poisson's ratio is $\nu$ $= 0.2$. Plane strain conditions are considered, and the convergence tolerance variable is set to $tol = 10^{-6}$.

In order to generate the training data, the FE model is first solved using the non-local gradient model described in \ref{Appendix:AppendixB_NonLocal_Gradient_Damage_FEM}. The domain is discretized using three structured meshes of varying element size. In the rest of this subsection, the three models are referred to as the \emph{Coarse}, \emph{Intermediate} and \emph{Fine}, and they each have 2500, 6400 and 10000 elements in total, respectively. Each side is discretized with 50, 80 and 100 elements and the finite element lengths for each case are $l_{elem} = 2 mm$, $l_{elem} = 1.25 mm$ and $l_{elem} = 1 mm$, respectively. In all cases the characteristic length is taken as $l_{c} = 4 mm$, thus ensuring that $l_{c}$ is at least twice the maximum element length. In this example we use the local strain definition of Equation \eqref{StrainA} and the damage law given in Equation \eqref{DamageA}, with $\alpha = 0.70$ and $\beta = 10000$.

\begin{figure}
	\centering
	\includegraphics[scale=0.6]{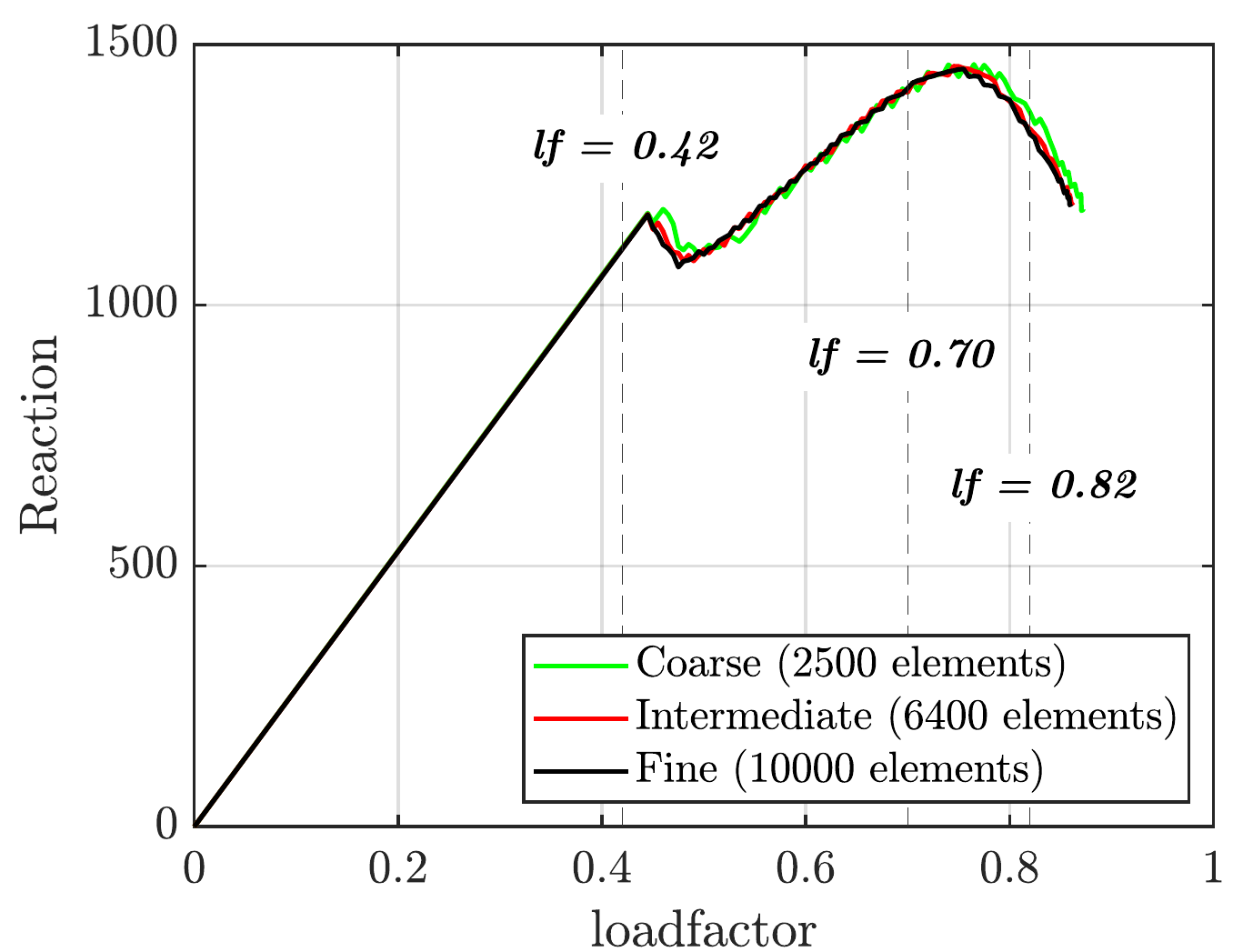}
	\caption{Reaction force-loadfactor curves for the single-notch case using the various mesh configurations.}
	\label{Figure_StrC100_ConvergenceStudy}
\end{figure}

\begin{figure}
	\centering
	\includegraphics[width=\linewidth]{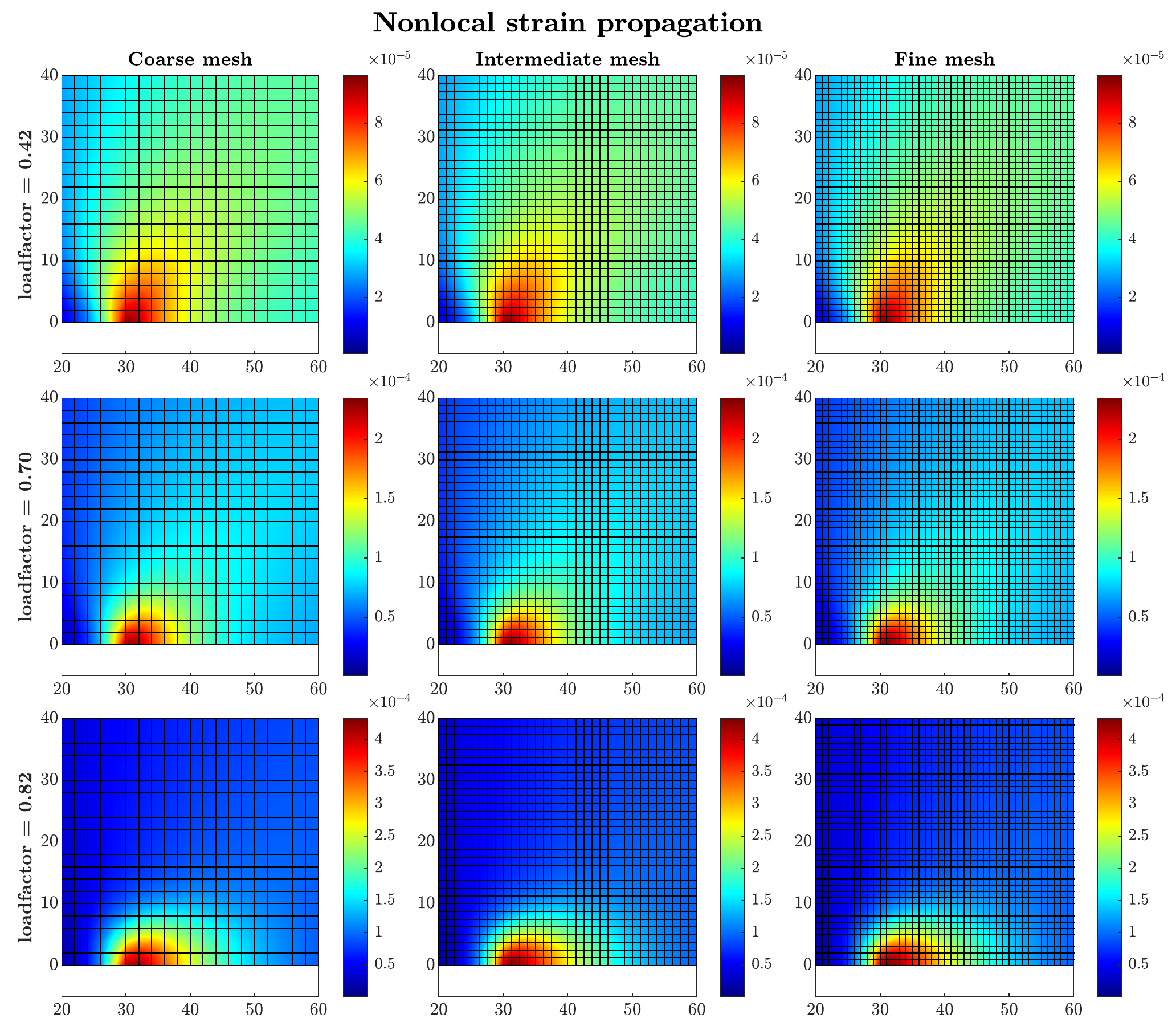}
	\caption{Snapshots of non-local equivalent strain ${\bar{\varepsilon}}_{eq}$ propagation across the Coarse, Intermediate and Fine meshes at loadfactor values $lf =  0.42, 0.70$ and $0.82$. For the sake of clarity, only a zoomed-in area around the crack tip is shown. The resulting contours demonstrate the mesh-independent nature of the non-local gradient solver.}
	\label{Figure_Str100_Strain_PropagationSnapshots_Zoomedin}
\end{figure}

\begin{figure}
	\centering
	\includegraphics[width=\linewidth]{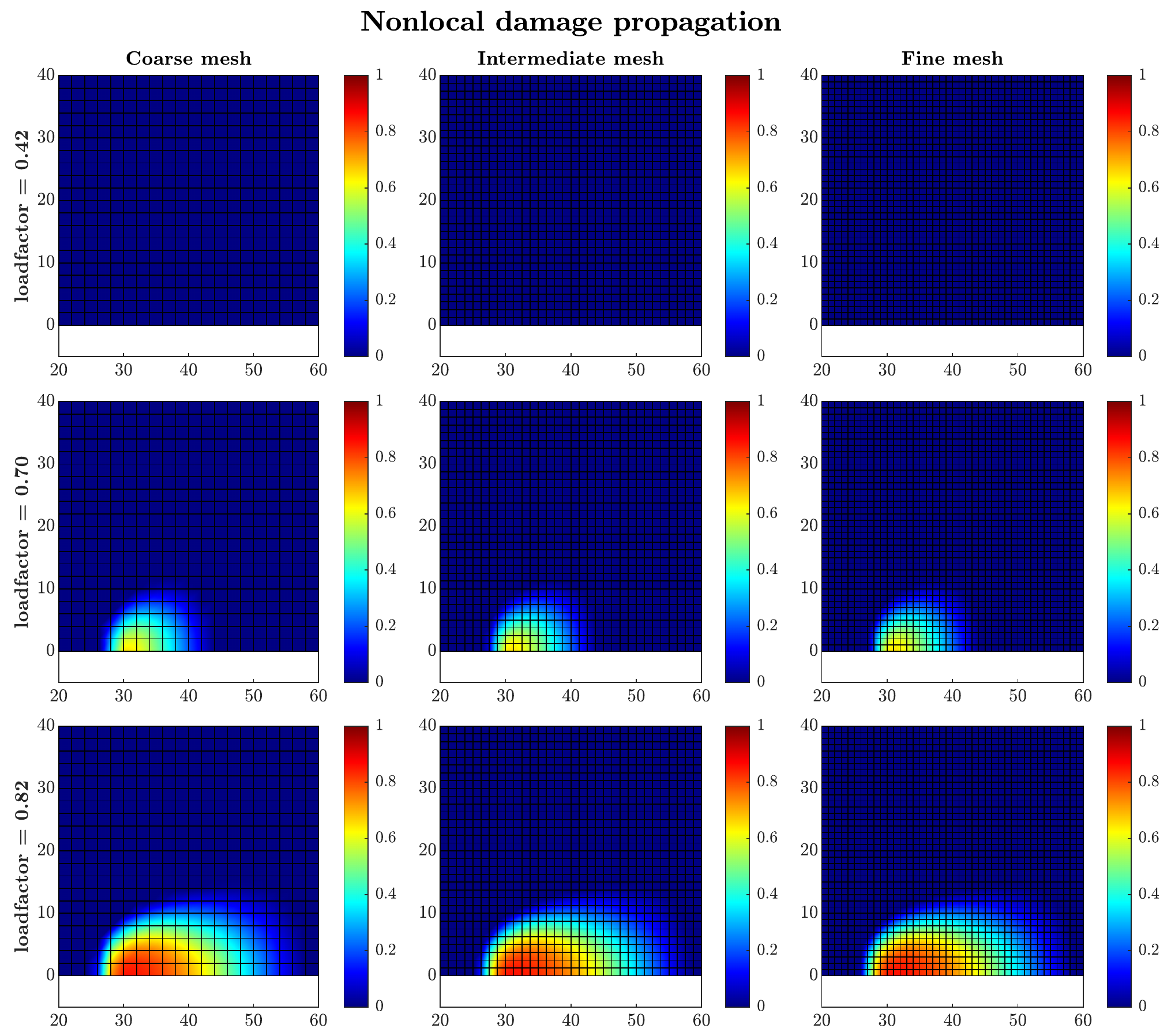}
	\caption{Snapshots of non-local damage $d$ propagation across the three different mesh idealizations at loadfactor values $lf = 0.42, 0.70$ and $0.82$. The contours are shown only for the region around the crack tip and they illustrate the mesh-objective character of the non-local gradient formulation.}
	\label{Figure_Str100_Damage_PropagationSnapshots_Zoomedin}
\end{figure}

First, we verify the mesh-independent character of the numerical solution. Figure \ref{Figure_StrC100_ConvergenceStudy} shows the reaction-loadfactor curves for the three cases. It can be observed that the resulting curves essentially coincide, which is the first indication of the solver mesh-independence. Additionally, Figure \ref{Figure_Str100_Strain_PropagationSnapshots_Zoomedin} shows a comparison of the non-local strain propagation across the three different meshes, with snapshots at the loadfactors marked on Figure \ref{Figure_StrC100_ConvergenceStudy}. These are $lf=0.42, \ 0.72, \ 0.82$. The first row of graphs corresponds to $lf = 0.42$, where the domains are still in the elastic regime. The second row corresponds to $lf = 0.70$, where damage has already initiated but the model is still in the hardening zone. The third row corresponds to $lf = 0.82$, which falls in the softening regime of the response. Following the same notation, Figure \ref{Figure_Str100_Damage_PropagationSnapshots_Zoomedin} demonstrates how damage evolves at these load increments. For the sake of a transparent cross-mesh comparison, only the crack tip neighbourhood $(20\leq x \leq 60, 0\leq y \leq 40)$ is shown, and the black grid in all subplots indicates the finite element edges. It is evident that both the non-local strain $\bar \varepsilon_{eq}$ and the damage variable $d$ share an almost identical profile across the three idealizations. This consistency further verifies that the non-local gradient method yields mesh-objective results throughout the entire load history.

\begin{figure}
	\centering
	\includegraphics[width=\linewidth]{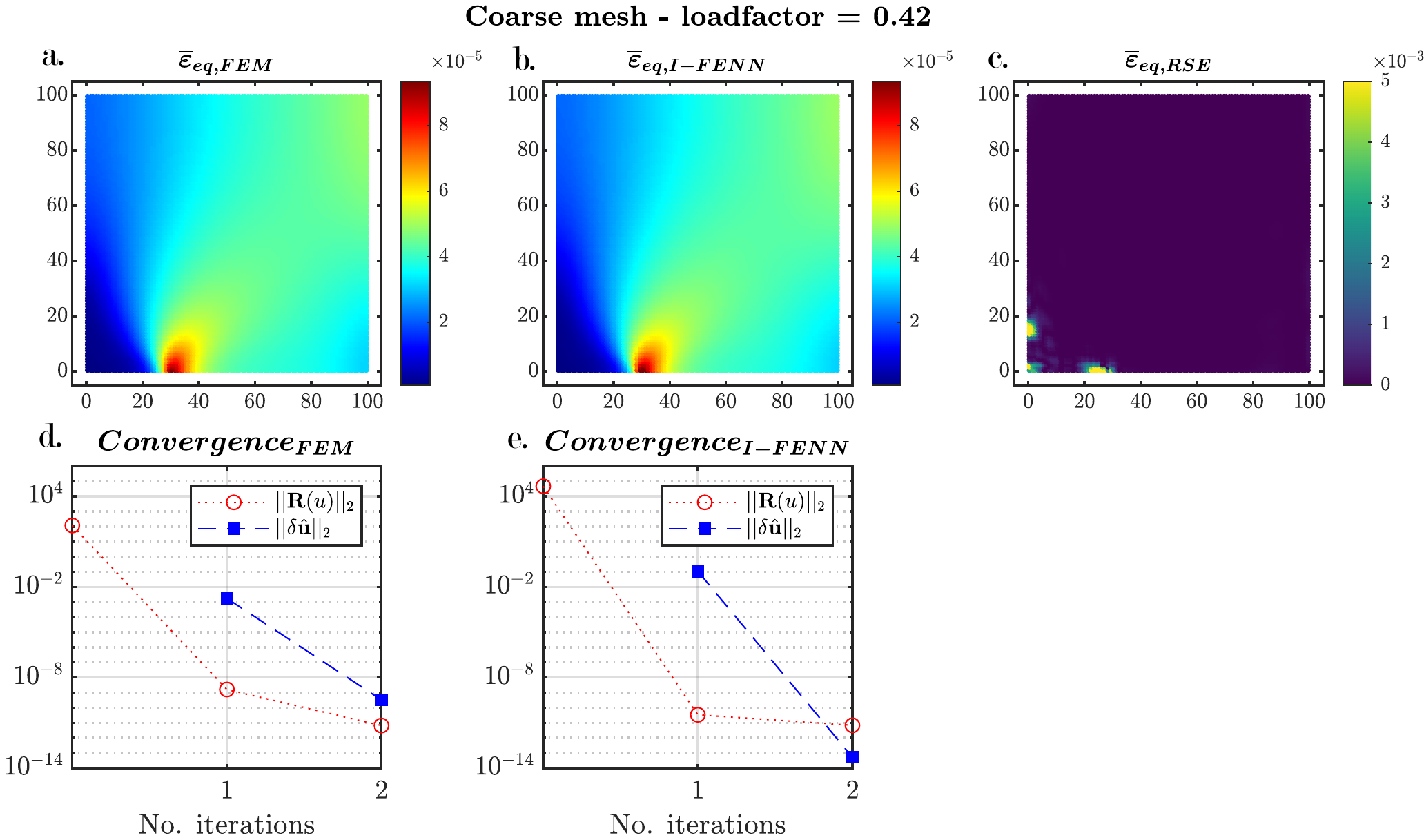}
	\caption{Comparison between FEM (left column) and I-FENN (middle column) for the Coarse model of the single-notch case, at a loadfactor value $lf = 0.42$. {\bf{a}}-{\bf{b}}. The non-local equivalent strain ${\bar{\varepsilon}}_{eq}$ maps for the two methods. Each filled circle in these plots indicates the value of ${\bar{\varepsilon}}_{eq}$ in the corresponding integration point. {\bf{c}}. The relative squared error of ${\bar{\varepsilon}}_{eq}$ calculated at each integration point using Equation \eqref{Eq_Metric_RSE}. With regards to the entire non-local strain vectors, the L2-norm of their differences is ${|| {\bar{\bf{\varepsilon}}}_{true} - {\bar{\bf{\varepsilon}}}_{pred} ||}_{2} = 1.88\times 10^{-5}$. {\bf{d}-\bf{e}}. Convergence performance of the two methods. The red line indicates the L2-norm of the residual vector ${\bf{R}}$, and the blue line shows the L2-norm of the displacement vector incremental change $\delta{\bf{\hat{u}}}$.}
	\label{Figure_StrC100_inc84_Method_Comparison}
\end{figure}

Figure \ref{Figure_StrC100_inc84_Method_Comparison} shows the comparison between FEM and I-FENN for the Coarse case at $lf = 0.42$. The domain is still in the elastic region and only the non-local strain profiles are shown. The first column of plots shows the \emph{FEM - true} solutions for the non-local strain (top) and converging residuals (bottom). Within the residuals subplot, the red line corresponds to the L2-norm of the internal stress residual vector ${\bf{R}}$ calculated based on Equation \eqref{FEM_WeakR_prop}. The blue line refers to the L2-norm of the displacement vector incremental change $\delta{\bf{\hat{u}}}$, and we remind the reader of the convergence criterion in Equation \eqref{FEM_JduR_4}. The second column of subplots depicts the \emph{I-FENN - predicted} solutions for the non-local strain (top) and convergence behavior (bottom). We observe a remarkable similarity between the predicted and targeted non-local strains. This similarity is both qualitative, captured by the overall contour shape, but also quantitative. The latter is verified by the top right subplot of Figure \ref{Figure_StrC100_inc84_Method_Comparison} which shows the Relative Squared Error (RSE) between the two. The formula for this error metric is shown in Equation \eqref{Eq_Metric_RSE}, and it is used to measure the relative error for both ${\bar{\varepsilon}}_{eq}$ and $d$:

\begin{subequations}
\begin{equation}
    {\bar{\varepsilon}}_{eq,RSE} = \frac{({\bar{\varepsilon}}_{eq,FEM} - {\bar{\varepsilon}}_{eq,I-FENN})^{2}}{({\bar{\varepsilon}}_{eq,FEM})^{2}}
\end{equation}

\begin{equation}
    d_{RSE} = \frac{(d_{FEM} - d_{I-FENN})^{2}}{(d_{FEM})^{2}}
\end{equation}
\label{Eq_Metric_RSE}
\end{subequations}

We underline that the plots in the first row of Figure \ref{Figure_StrC100_inc84_Method_Comparison} do not depict a continuous contour throughout the domain. Instead, each filled circle in these plots represents the value of the variable of interest in the corresponding integration point. This allows for a point-by-point comparison between the two methods, and ensures that the relative squared error is not smeared out due to plotting interpolation. A close observation of Figure \ref{Figure_StrC100_inc84_Method_Comparison}e reveals that the ${\bar{\varepsilon}}_{RSE}$ lies even below $0.1\%$ throughout the vast majority of the domain. This is an exceptional approximation of the target values, while there are only a few IPs where relative squared error gets higher. These points are located in the bottom left corner of the domain and specifically behind the crack tip. The elevated ${\bar{\varepsilon}}_{eq,RSE}$ at these points is attributed to the numerical issue of the range of $\bar{\varepsilon}_{eq}$ values. The non-local strain values in this area range between $10^{-6}$ and $10^{-5}$, whereas in the rest of the domain they range between $10^{-5}$ to $10^{-4}$. This difference in the order of magnitude yields more pronounced differences in the strain relative error, even if the absolute error is smaller in these locations as compared to the rest of the domain. This numerical feature remains present throughout the rest of the examined cases and it requires additional efforts on adapting a generalized PINN structure and training strategy to be alleviated, as discussed in Section \ref{SubSec:MethodPINNs}. For example, scalable approaches with individual subdomain normalization of the input data such as the FBPINNs framework \cite{moseley2021finite} could be a promising candidate to resolve this numerical issue. Nevertheless, we note that the area where the elevated values of $\bar{\varepsilon}_{eq,RSE}$ occur is sufficiently distanced from the damage process zone, and since the strain values are substantially low they do not affect the accuracy of damage predictions. This is shown more clearly in the next figures, where damage is also present. 

Another important observation from Figure \ref{Figure_StrC100_inc84_Method_Comparison} is that the L2-norms of ${\bf{R}}$ and $\delta{\bf{\hat{u}}}$ which are computed with I-FENN are continuously decreasing, until they have converged. This is the most crucial component of a successful non-linear finite element analysis, and proves the applicability of our proposed methodology. It is also important to note that our method converges with just two iterations, similar to the FEM solution, as we would expect for a load increment in the elastic regime. Based on these observations, we can conclude that Figure \ref{Figure_StrC100_inc84_Method_Comparison} provides evidence of both a) the successful training of the physics-informed neural network, and b) the feasibility of integrating it within the element stiffness function. 

\begin{figure}[H]
	\centering
	\includegraphics[width=\linewidth]{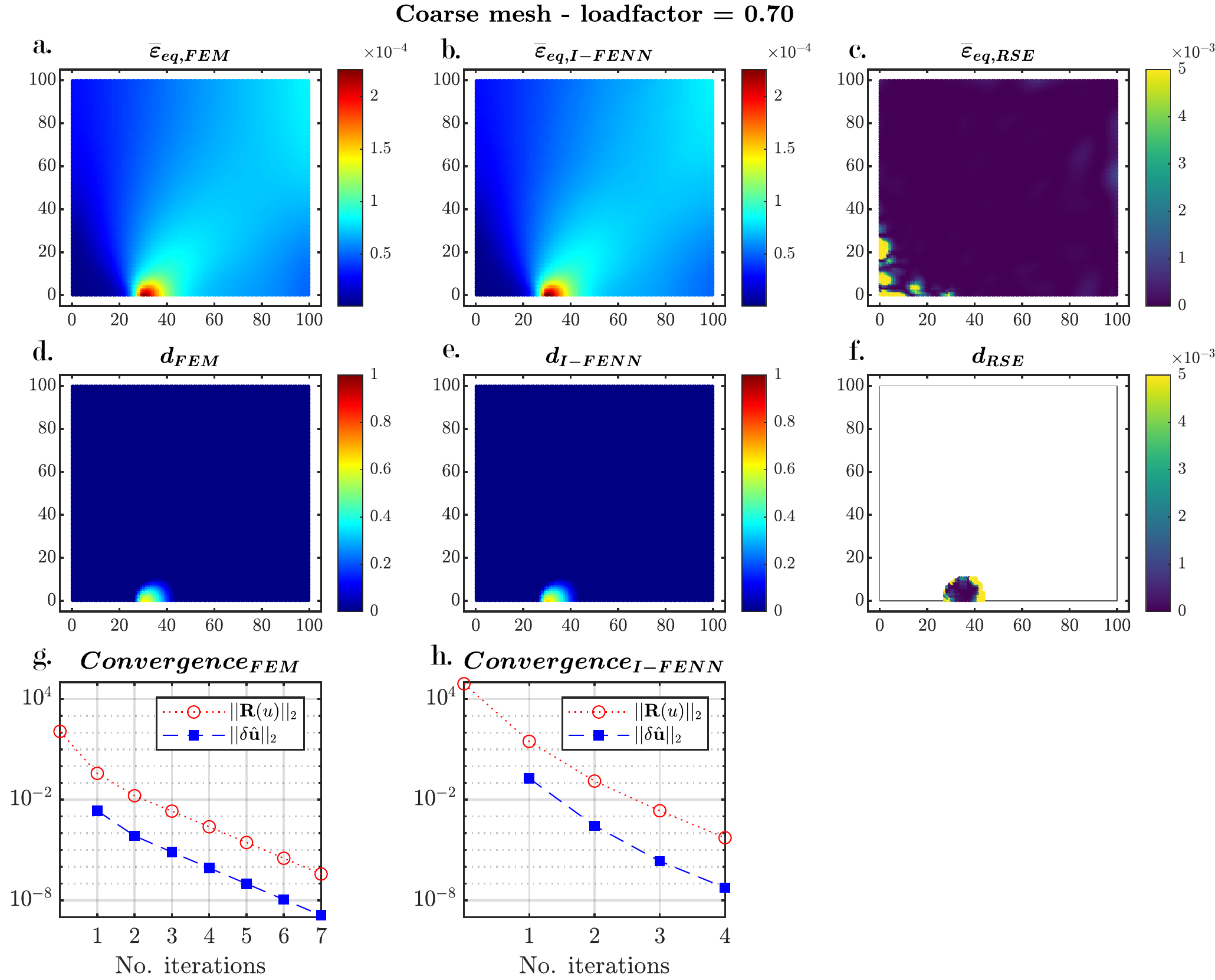}
	\caption{Comparison between FEM (left column) and I-FENN (middle column) for the Coarse model of the single-notch case, at loadfactor value $lf = 0.70$. The non-local equivalent strain and damage values at each integration point are shown in subplots {\bf{a}}-{\bf{b}} and {\bf{d}}-{\bf{e}} respectively. The relative squared error for the two variables is shown in subplots {\bf{c}} and {\bf{f}}, calculated using Equation \eqref{Eq_Metric_RSE}. The white space in subplot {\bf{c}} is due to the FEM value of $d = 0$, and therefore $d_{RSE}$ is not defined. The L2-norm of the strain vector is ${|| {\bar{\varepsilon}}_{FEM} - {\bar{\varepsilon}}_{I-FENN} ||}_{2} = 5.167 \times 10^{-5}$. {\bf{g}}-{\bf{h}}. Convergence of the two methods. The red line indicates the L2-norm of the residual vector ${\bf{R}}$, and the blue line shows the L2-norm of the displacement vector incremental change $\delta{\bf{\hat{u}}}$.}
	\label{Figure_StrC100_inc140_Method_Comparison}
\end{figure}

\begin{figure}[H]
	\centering
	\includegraphics[width=\linewidth]{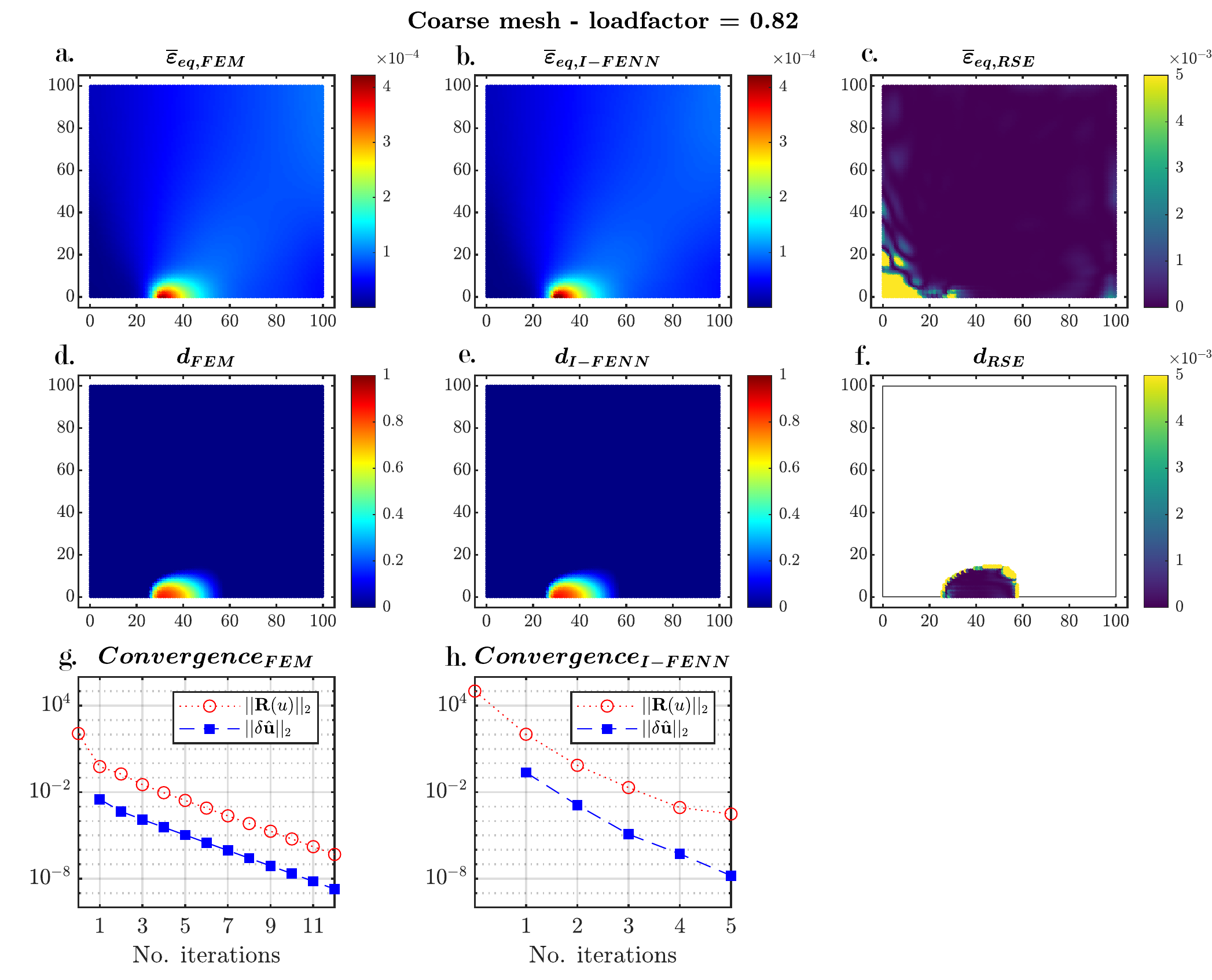}
	\caption{Comparison between FEM (left column) and I-FENN (middle column) for the Coarse model of the single-notch case at loadfactor value $lf = 0.82$. The non-local strain and damage values at each integration point, along with the relative squared error, are shown in the top and middle row respectively. The L2-norm of the ${\bar{\varepsilon}}_{eq}$ is ${|| {\bar{\varepsilon}}_{FEM} - {\bar{\varepsilon}}_{I-FENN} ||}_{2} = 1.296 \times 10^{-4}$. The bottom row of subplots illustrates the convergence performance of the two methods. The red line indicates the L2-norm of the residual vector ${\bf{R}}$, and the blue line shows the L2-norm of the displacement vector incremental change $\delta{\bf{\hat{u}}}$.}
	\label{Figure_StrC100_inc164_Method_Comparison}
\end{figure}

Figures \ref{Figure_StrC100_inc140_Method_Comparison} and \ref{Figure_StrC100_inc164_Method_Comparison} show a similar comparison for the Coarse case, at $lf = 0.70$ and $lf = 0.82$ respectively. The subplot layout is similar to Figure \ref{Figure_StrC100_inc84_Method_Comparison}, with the addition of the damage contours in the middle row of the figures. Even in the substantially more challenging damage-dominated zone, the evident conclusion is that I-FENN is capable of capturing with very good accuracy both the non-local strain and damage profiles. As far as the non-local strains are concerned, this observation is again subtly violated only at the area behind the crack tip, which is not critical for the crack propagation. With respect to the damage landscape, I-FENN also captures with very good accuracy its spatial extent and magnitude. We note that the white space in Figures \ref{Figure_StrC100_inc140_Method_Comparison}h and \ref{Figure_StrC100_inc164_Method_Comparison}h is due to the FEM $d = 0$, and therefore $d_{RSE}$ is mathematically undefined. Finally, we emphasize the continuously decreasing L2-norms of ${\bf{R}}$ and $\delta{\bf{\hat{u}}}$ in Figures \ref{Figure_StrC100_inc140_Method_Comparison}g and \ref{Figure_StrC100_inc164_Method_Comparison}h, which further verify the feasibility of our method in the presence of damage. 

\begin{figure}[H]
	\centering
	\includegraphics[scale=0.6]{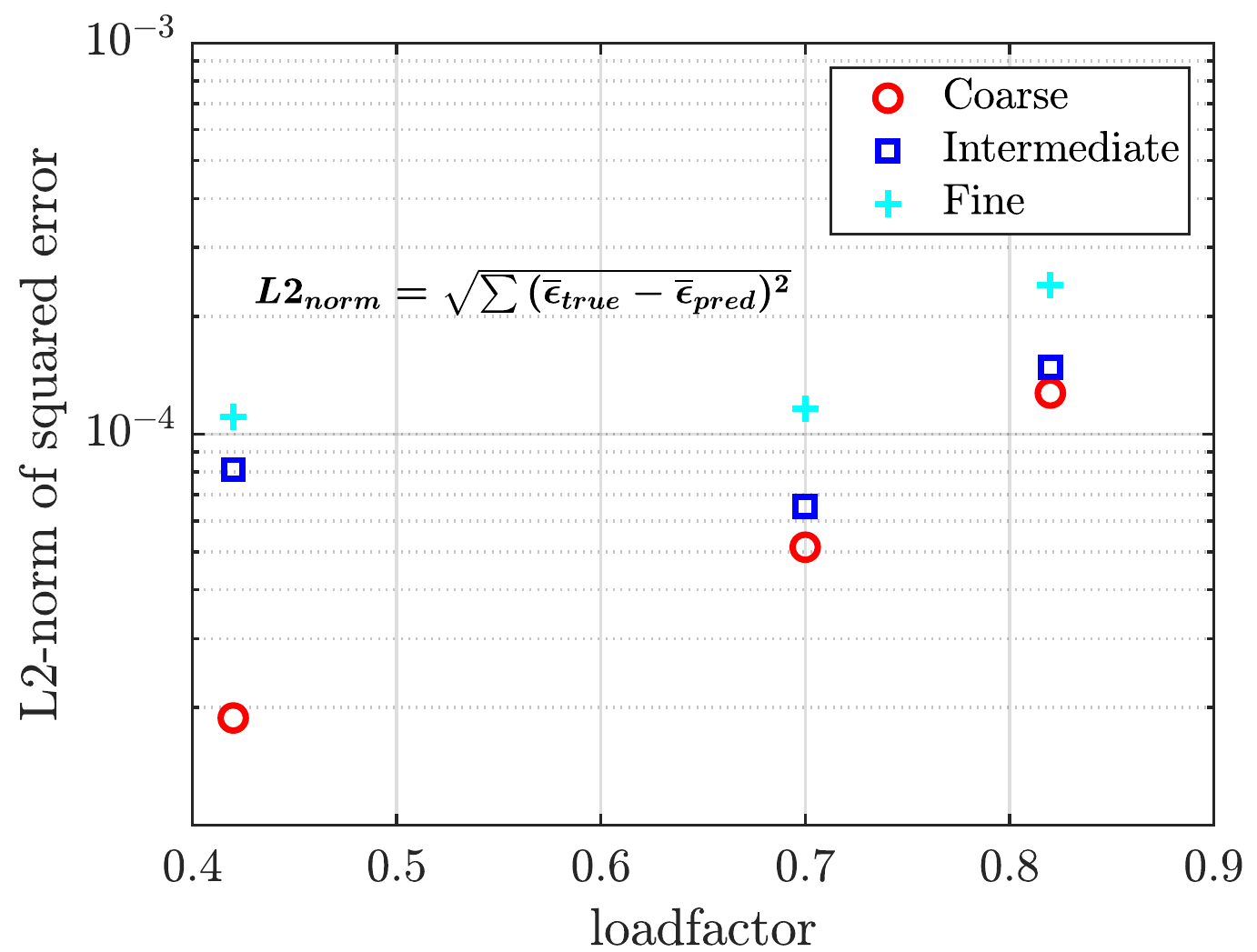}
	\caption{The L2-norms of the non-local equivalent strain errors when the PINN is trained and tested in the same mesh resolution. The formula for the L2-norm of the strain vector absolute error reads: $||.||_{2} = \sqrt{\sum {({\bar{\varepsilon}}_{eq,FEM} - {\bar{\varepsilon}}_{eq,I-FENN})^{2}}}$}
	\label{Figure_StrainErrors_selftest}
\end{figure}

Additionally, we provide in Figure \ref{Figure_StrainErrors_selftest} a concise way to monitor the error metrics. Figure \ref{Figure_StrainErrors_selftest} shows the L2-norms of the non-local strain vector absolute error between FEM and I-FENN. In this graph the PINNs are trained and tested against the same mesh resolution. Overall we observe values of the L2-norm in the order of $10^{-5}$ to $10^{-4}$, while there are two general trends: a) the error value increases monotonically as the mesh resolution is refined, and b) the error value tends to increase as damage grows and the domain enters the softening stage. Both trends are rather anticipated, the first due to already discussed presence of more integration points contributing to the error accumulation, and the second due to the higher absolute values in the non-local strain profile as damage increases.

\begin{figure}
	\centering
	\includegraphics[width=\linewidth]{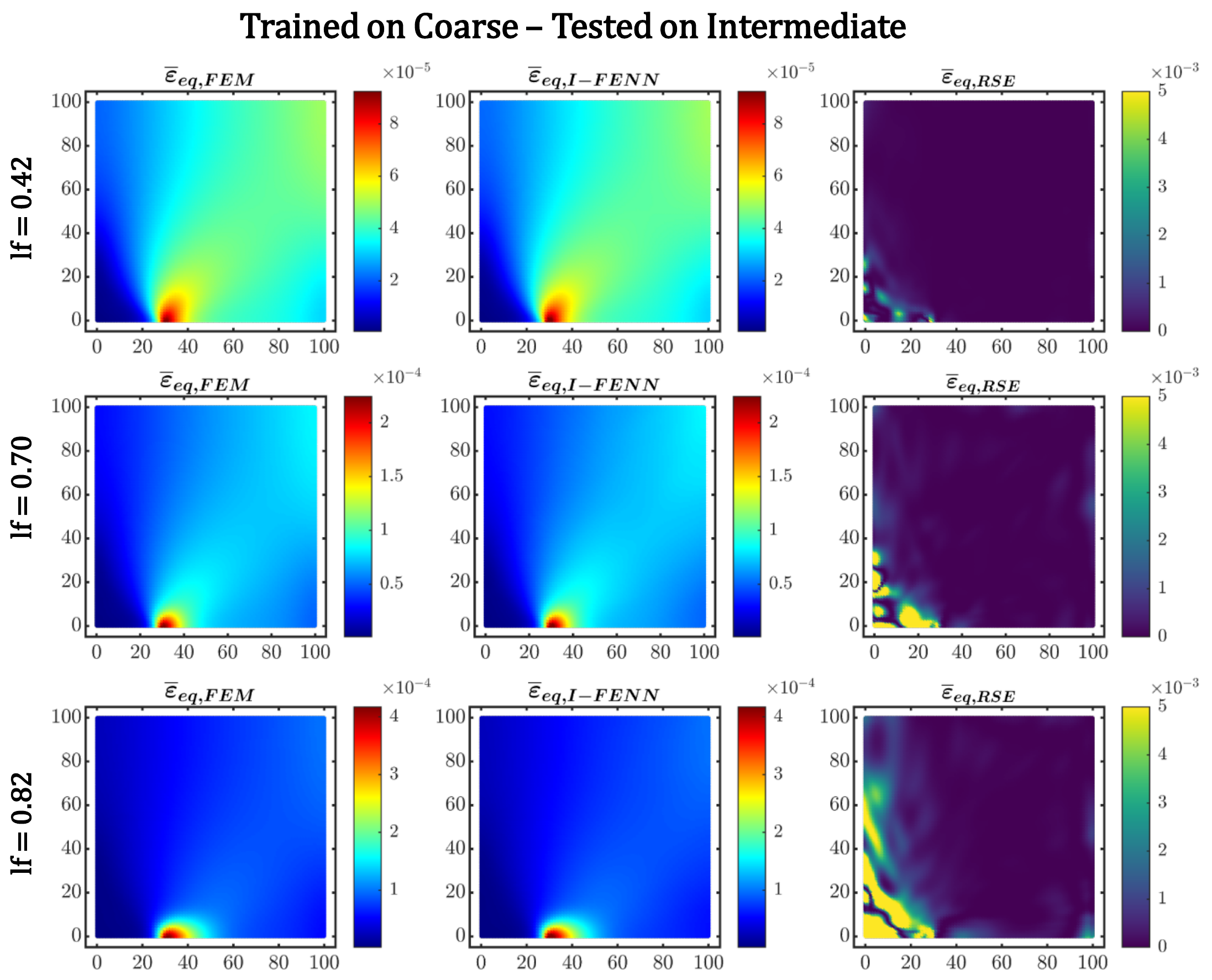}
	\caption{Cross-mesh generalization: The PINN is trained on the Coarse mesh data of the single-notch case and it is used to predict the non-local strain ${\bar{\varepsilon}}_{eq}$ profile in the Intermediate mesh model. The left column shows the FEM results of ${\bar{\varepsilon}}_{eq}$ in the Intermediate mesh, the middle column illustrates the I-FENN predictions, and the right column depicting the relative squared error. Each row of subplots corresponds to a different loadfactor value.}
	\label{Figure_Str100_CrossMesh_CtoI}
\end{figure}

Next, we examine a critical aspect of our methodology: the ability of the neural network to generalize across different mesh densities. Evidently, the gain from training a PINN on a coarsely discretized geometry and utilizing it for predictions in substantially finer meshes is a crucial feature from a computational efficiency standpoint, opening another pathway through which I-FENN could decrease the numerical cost of numerical simulations. Here we touch upon this path as well, by testing the three PINNs which are trained on the Coarse mesh against the data from the other two finer mesh discretizations at the same loadfactor values. {From a computational execution standpoint, this is feasible because we maintain the same number for the input matrix (4 variables: x, y, g, and {${\varepsilon}_{eq}$}) and output vector (1 variable: {${\bar{\varepsilon}}_{eq}$}) variables, regardless the mesh density. This ensures the consistency between the inner dimensions of the weight matrices multiplication inside the network, and therefore only the outer dimensions of the input matrix and output vector change, which correspond to the number of collocation points.} We also underline that in this comparative study, we evaluate the trained PINN itself, without integrating it inside the nonlinear solver. In other words, we feed the input from the Intermediate/Fine mesh to the Coarse-trained PINN, and implement only one forward propagation pass to get the output predictions without further refining the predictions using the Newton-Raphson solver. 

\begin{figure}
	\centering
	\includegraphics[width=\linewidth]{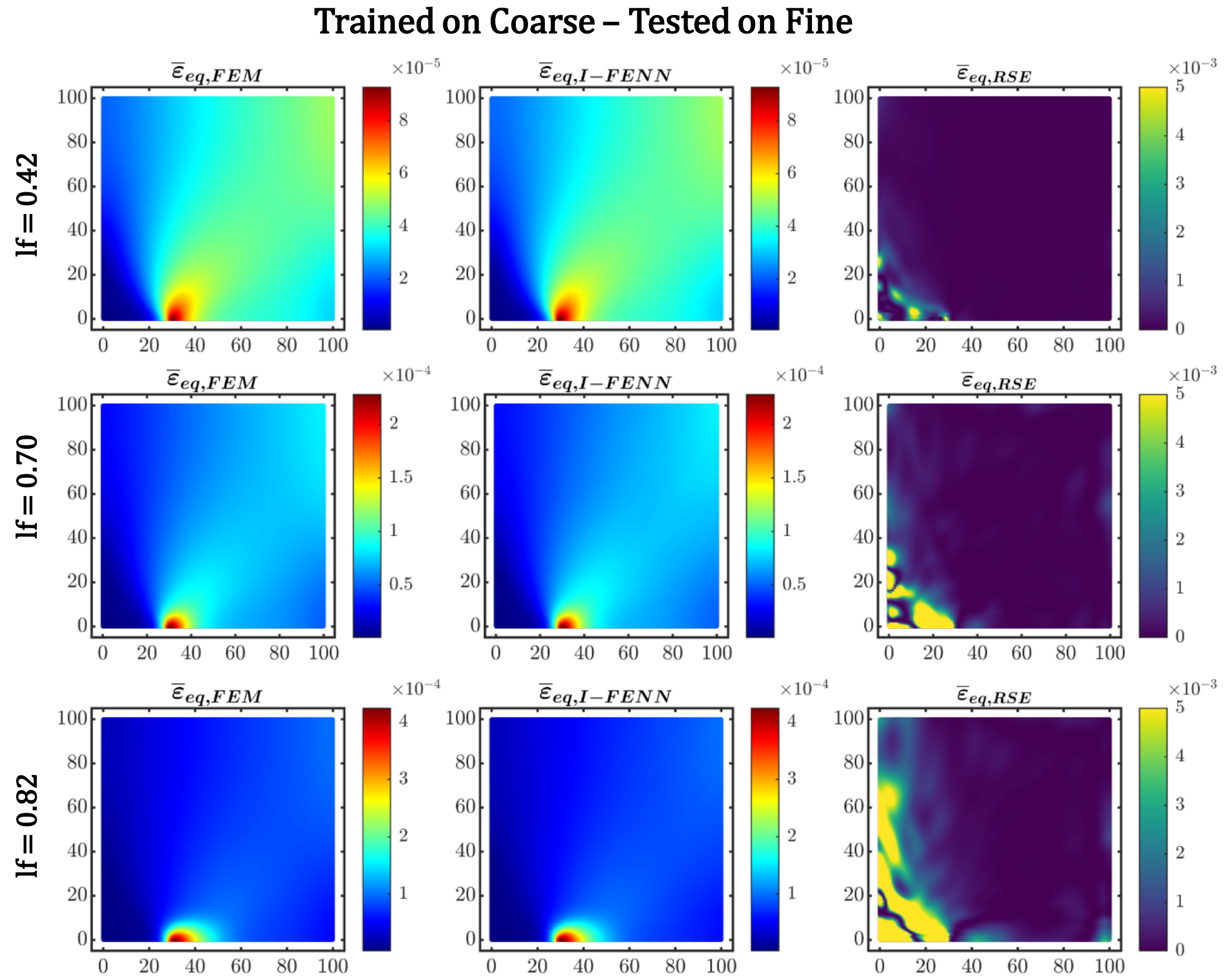}
	\caption{Cross-mesh generalization: The PINN which is trained on the Coarse mesh data of the single-notch case is used to predict the non-local strain ${\bar{\varepsilon}}_{eq}$ profile in the Fine mesh model. The left column shows the FEM results of ${\bar{\varepsilon}}_{eq}$ in the Intermediate mesh, the middle column illustrates the I-FENN predictions, and the right column depicting the relative squared error. Each row of subplots corresponds to a different loadfactor value.}
	\label{Figure_Str100_CrossMesh_CtoF}
\end{figure}

The results of this study are shown in Figures \ref{Figure_Str100_CrossMesh_CtoI} and \ref{Figure_Str100_CrossMesh_CtoF}, for the Intermediate and Fine models respectively, and each row of plots corresponds to a different loadfactor value. Across the entire domain, we observe a remarkable resemblance of the non-local strain profile between the FEM and I-FENN calculations for all three loadfactor values. The relative squared error is particularly low when the comparison is conducted in the elastic regime, which corresponds to the first row of plots in Figures \ref{Figure_Str100_CrossMesh_CtoI} and \ref{Figure_Str100_CrossMesh_CtoF}. It can be observed that the error metric performance gradually deteriorates as we move deeper into the presence of damage; however, this is an artifact of the growing difference between the maximum and minimum strain value range. The spread in the $\bar{\varepsilon}_{eq,RSE}$ is observed to increase in the regions where the strains are much lower than the damage initiation threshold strain $\varepsilon_{D}$ and do not contribute to damage evolution. Additionally, a closer observation in the RSE plots between Figures \ref{Figure_Str100_CrossMesh_CtoI} and \ref{Figure_Str100_CrossMesh_CtoF} reveals that the Coarse-trained network performs slightly better on the Intermediate mesh as compared to the Fine mesh. This is anticipated, as the larger number of integration points in the Fine mesh is expected to dilate the accumulated error; nevertheless the non-local strain profile is still captured with very good accuracy as shown in the second column of plots in Figure \ref{Figure_Str100_CrossMesh_CtoF}.

\begin{figure}
	\centering
	\includegraphics[width=\linewidth]{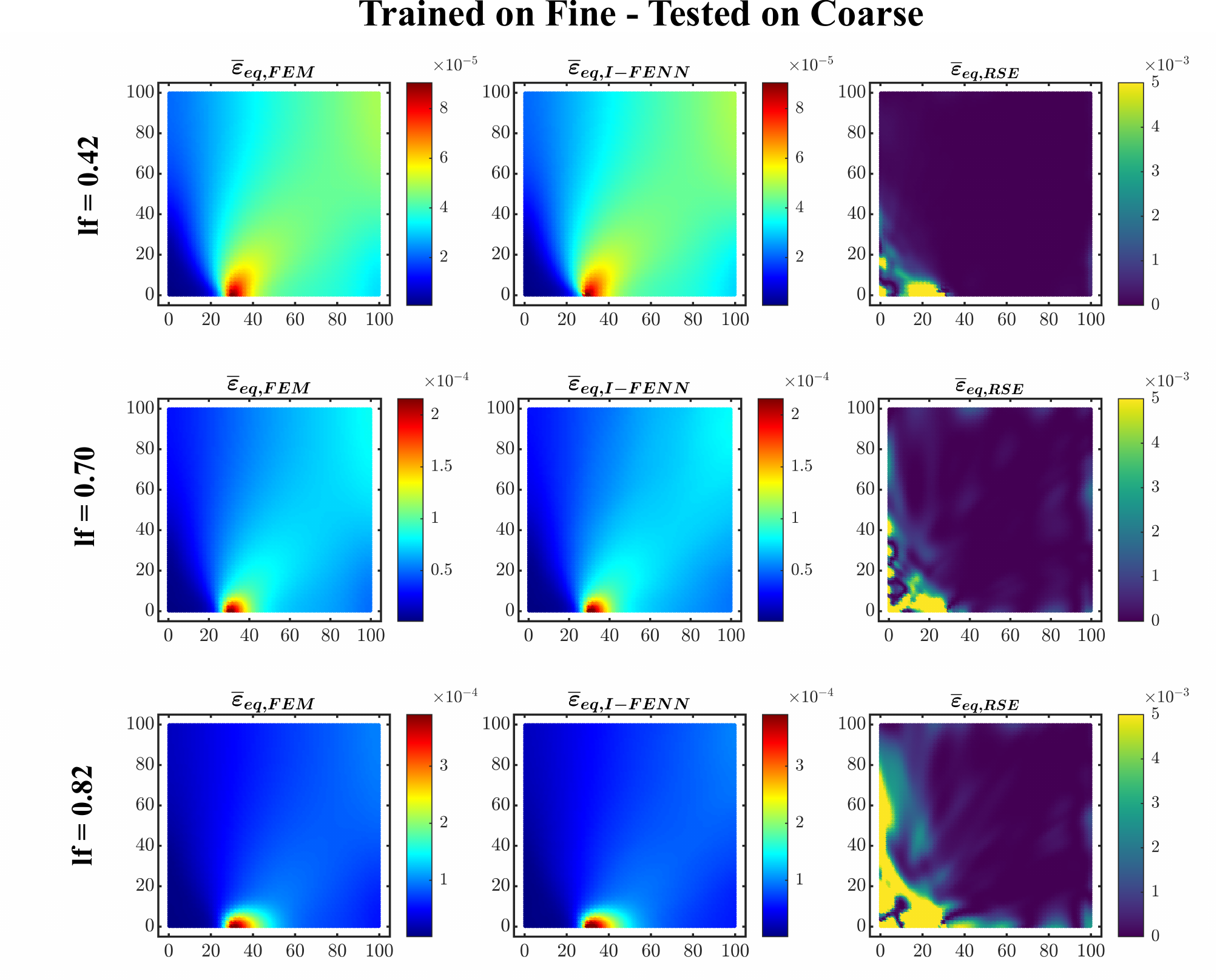}
	\caption{Cross-mesh generalization: The PINN which is trained on the Fine mesh data of the single-notch case is used to predict the non-local strain ${\bar{\varepsilon}}_{eq}$ profile in the Coarse mesh model. The left column shows the FEM results of ${\bar{\varepsilon}}_{eq}$ in the Intermediate mesh, the middle column illustrates the I-FENN predictions, and the right column depicting the relative squared error. Each row of subplots corresponds to a different loadfactor value.}
	\label{Figure_Str100_CrossMesh_FtoC}
\end{figure}

{The converse response is also investigated and the results are shown in Figure {\ref{Figure_Str100_CrossMesh_CtoI}}. A PINN which is trained in the Fine mesh discretization is tested against the data generated with the Coarse mesh, at the loadfactor values of interest. The network is capable of predicting the nonlocal strain profile with good accuracy. The most accurate results are obtained when training and testing occurs at identical mesh refinement levels, but} overall, this parametric study shows promising results on the feasibility of training and testing a PINN at different mesh discretizations. This is expected to have a significant impact in accelerating the numerical analysis of densely meshed geometries in multi-scale and multi-physics simulations.

\begin{figure}[H]
	\centering
	\includegraphics[width=\linewidth]{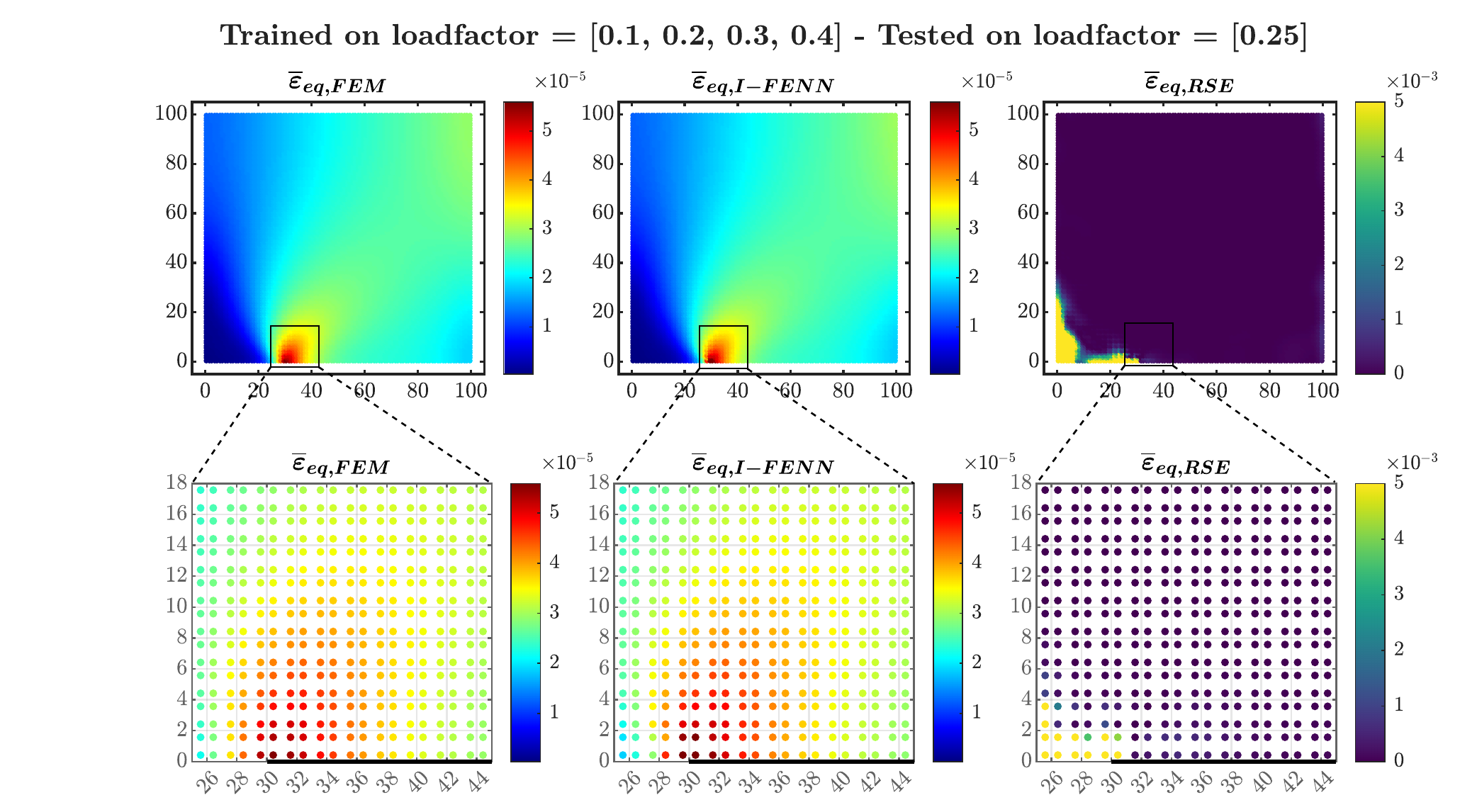}
	\caption{Cross-load history generalization: The PINN is trained on input data from four elastic load increments with $lf = 0.1, 0.2, 0.3$ and $0.4$, and then tested against the data corresponding to $lf = 0.25$. The top row of subplots shows the FEM (left) and I-FENN (middle) results for the non-local strain ${\bar{\varepsilon}}_{eq}$, and the relative squared error is shown in the right. The bottom row of graphs show a point-by-point comparison in a zoomed-in region around the crack tip, further demonstrating the predicting accuracy of I-FENN.}
	\label{Figure_StrC100_Multi_lf_elastic}
\end{figure}

Thus far, we have illustrated how a neural network which is trained and tested at the same load increment is capable of predicting the non-local strain profile. Next we examine how this approach can be generalized across multiple load increments. First, we confine our interest in the elastic regime and train a PINN with input data from four load increments, namely at $lf  = 0.1, 0.2, 0.3$ and $0.4$. We then test the ability of the PINN against unseen input data generated at loadfactor $lf = 0.25$. {We also note that for this numerical study, where the PINN is trained and tested at different loadfactor values, these values are used as an additional input variable into the network.} The results of this parametric study are shown in Figure \ref{Figure_StrC100_Multi_lf_elastic}. The top series of graphs shows a comparison between FEM and I-FENN non-local strain values throughout the entire domain. The bottom series of plots zooms on the crack tip neighbourhood, allowing for an even more clear point-by-point comparison in the region of high strains concentration. We observe a very good correlation between the target and predicted values, both in the vicinity of the crack tip and also across the full domain.

\begin{figure}[H]
	\centering
	\includegraphics[width=\linewidth]{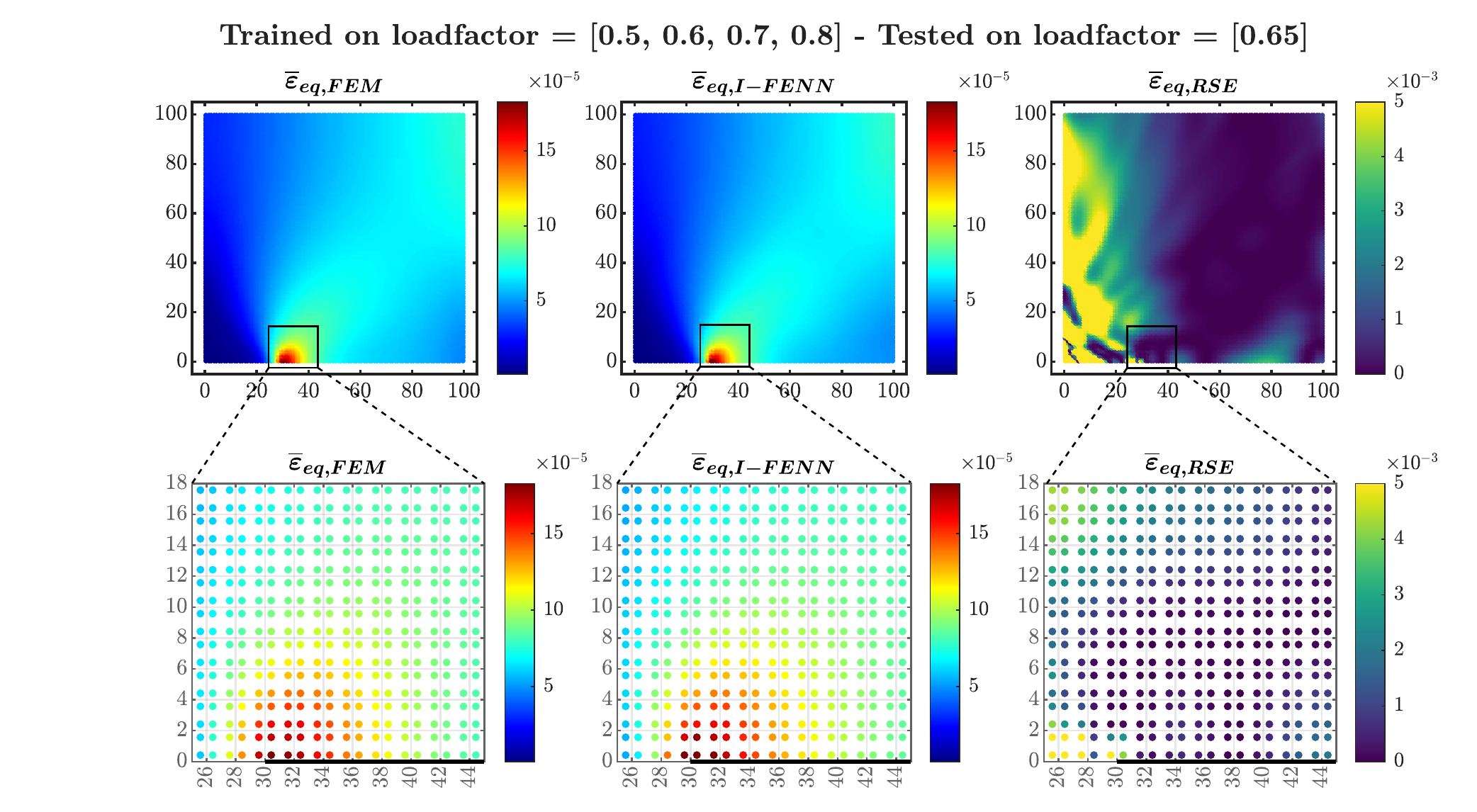}
	\caption{Cross-load history generalization: The PINN is trained on input data from four inelastic load increments with $lf = 0.5, 0.6, 0.7$ and $0.8$, and it is used to make predictions at $lf = 0.65$. The top row of subplots shows the FEM (left) and I-FENN (middle) results for the non-local strain ${\bar{\varepsilon}}_{eq}$, and the relative squared error is shown in the right. I-FENN is capable of capturing with sufficient accuracy the strain landscape, particularly around the critical crack-tip area, as it is shown in the zoomed-in subplots in the bottom row of the figure.}
	\label{Figure_StrC100_Multi_lf_inelastic}
\end{figure}

We then follow a similar approach for the inelastic zone, training the PINN with data from $lf = 0.5, 0.6, 0.7$ and $0.8$ and testing against $lf = 0.65$. The results of this parametric study are shown in Figure \ref{Figure_StrC100_Multi_lf_inelastic}. As shown in the top row of graphs, even though the network shows a slightly deteriorated performance when compared to the study conducted in the elastic regime, it is still capable to generalize with adequate accuracy both overall and particularly in the zone where damage is expected to propagate. This is more clearly demonstrated in the bottom series of zoomed-in graphs, where the clarity of these plots enables a straightforward comparison of the critical integration points behavior.  

\begin{figure}[H]
	\centering
	\includegraphics[scale=0.7]{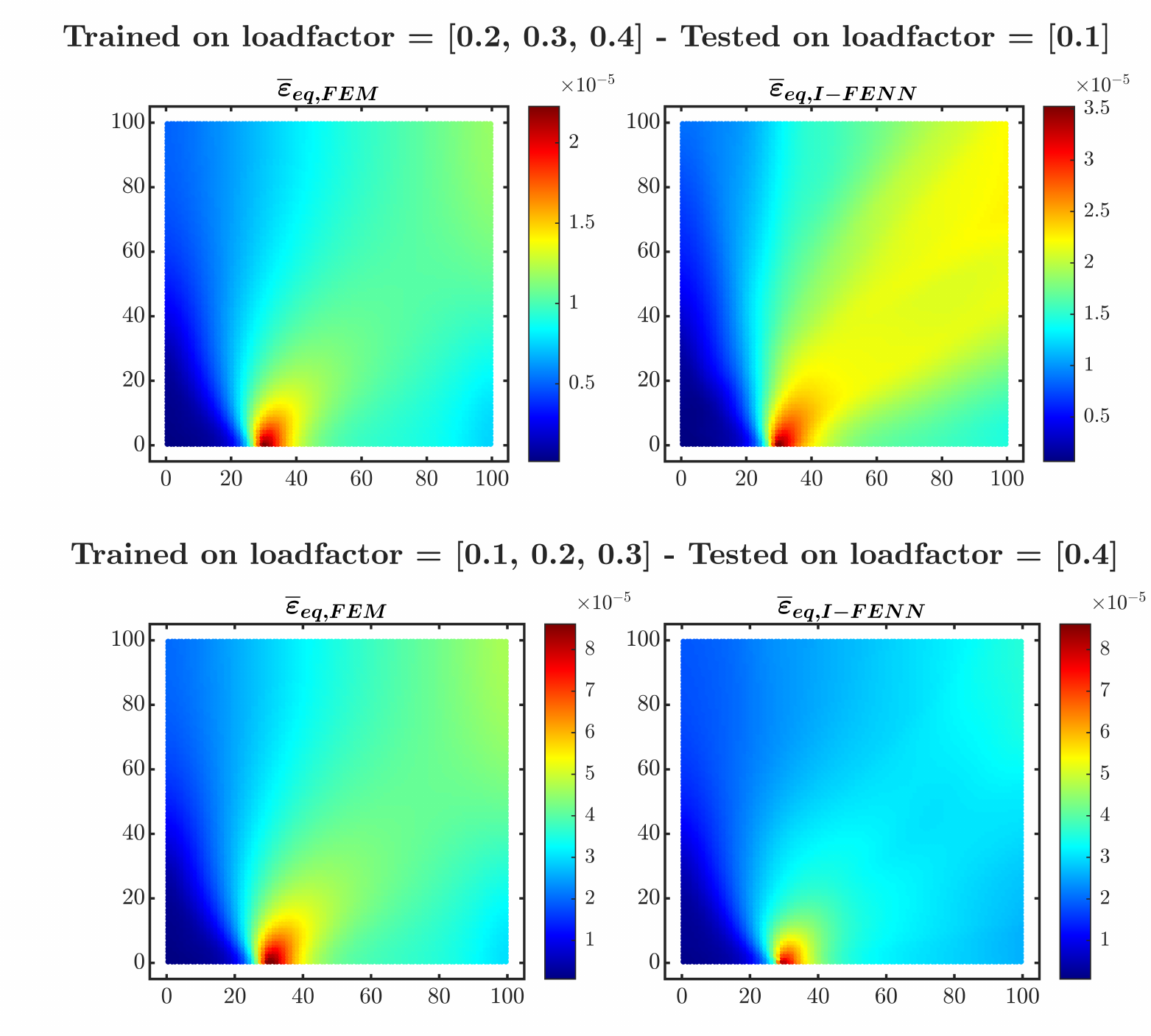}
	\caption{Cross-load history generalization: The PINN is trained on input data from three load increments and tested against an unseen load value which lies outside the range of the training dataset. The top row of subplots shows the case where the training dataset corresponds to $lf = 0.1$, $lf = 0.2$, $lf = 0.3$ and the test dataset to $lf = 0.4$ (extrapolation to a higher loadfactor value). The bottom row of subplots shows the case where the training dataset corresponds to $lf = 0.2$, $lf = 0.3$, $lf = 0.4$ and the test dataset to $lf = 0.1$ (extrapolation to a lower loadfactor value). The left column shows the FEM results and the right the I-FENN results for the non-local strain ${\bar{\varepsilon}}_{eq}$. The PINN approximates qualitatively the non-local strain profile in both cases, but fails to capture accurately its magnitude.}
	\label{Figure_Extrapolation_Elastic}
\end{figure}

{Until this point, generalization of the PINN predictive capability has been attempted in an \textit{interpolating} sense, since the unseen test dataset corresponds to a loadfactor which lies in between the training ones. We also attempt to examine the \textit{extrapolating} capability of the trained PINN, by conducting two additional numerical studies. We have trained a PINN on loadfactor values $lf = 0.1, 0.2, 0.3$ and tested on $lf = 0.4$, as well as training on $lf = 0.2, 0.3, 0.4$ and testing on $lf = 0.1$. These two parametric studies aim to shed light on how the PINN performs when the unseen test dataset lies either above or below the training dataset, and the results of this investigation are shown in Figure {\ref{Figure_Extrapolation_Elastic}}. The PINN approximates qualitatively the non-local strain profile in both cases, but fails to capture correctly its magnitude. This is rather anticipated, since PINNs in their standard form are well-known for their poor extrapolation performance, which is also confirmed in this case.}

Overall, the results presented at Figures {\ref{Figure_StrC100_Multi_lf_elastic}} {,} {\ref{Figure_StrC100_Multi_lf_inelastic}} {and {\ref{Figure_Extrapolation_Elastic}}} shed some initial yet promising light towards the feasibility of the PINN to generalize beyond the input data it has been trained upon. {Even in its current vanilla formulation, the PINN demonstrates a good interpolating capability. This is a crucial dimension of I-FENN, since it potentially allows to carefully select few training datasets that correspond to the loadfactor range boundaries, as well as a few in-between steps, and based on the interpolation capability of the PINN make predictions in the intermediate range. Yet, we underline that the purpose of Figures {\ref{Figure_StrC100_Multi_lf_elastic}} - {\ref{Figure_Extrapolation_Elastic}} is only to provide some light on the generalization aspect of the PINN across the load history, and not to establish rigorous guidance on how this generalization should be executed. For this purpose, more sophisticated network architectures such as GRUs or LSTMs may prove more suitable candidates.} 

To summarize, in this section we have implemented the new non-local damage model to a geometry with mode-I loading conditions and a structured finite element mesh. Three different discretizations of varying mesh density were used. After demonstrating the non-local character of the non-local gradient solver across the three models, we illustrated the following key points:

\begin{itemize}

\item The pre-trained PINN is capable of learning accurately the local to non-local strain transformation at the offline training stage. This is shown in the first row of subplots in Figures \ref{Figure_StrC100_inc84_Method_Comparison} - \ref{Figure_StrC100_inc164_Method_Comparison}, where the non-local strain profile from I-FENN shows very good resemblance to the FEM solution.

\item The PINN is integrated within the element-level stiffness function and it is repeatedly executed within the iterative Newton-Raphson method. Its input and output variables are continuously updated as it acts jointly with the finite element interpolation functions. As a result, a damage profile with a non-local character is generated, which matches with remarkable accuracy the damage profile from FEM. This is illustrated in the second row of subplots in Figures \ref{Figure_StrC100_inc84_Method_Comparison} - \ref{Figure_StrC100_inc164_Method_Comparison} 

\item The robust convergence of the Newton-Raphson method in I-FENN is demonstrated in the bottom row of subplots in Figures \ref{Figure_StrC100_inc84_Method_Comparison} - \ref{Figure_StrC100_inc164_Method_Comparison}. The L2-norms of the residual force vector and incremental displacement change are continuously decreased, until the convergence criterion of Equation \ref{FEM_JduR_3} is satisfied.

\item We tested the neural network which is trained on the Coarse mesh against the other two mesh models, across all three loadfactor values. The results are shown in Figures \ref{Figure_Str100_CrossMesh_CtoI} - \ref{Figure_Str100_CrossMesh_CtoF}. We observe a very good match between the FEM and I-FENN predictions, which shows evidence of a cross-mesh generalization capability of the proposed methodology.

\item We trained a neural network on four elastic load increments and tested against the unseen data from another load increment, and we repeated this study in the inelastic zone. The results of this study are shown in Figures \ref{Figure_StrC100_Multi_lf_elastic} - \ref{Figure_StrC100_Multi_lf_inelastic}. In both cases we observe very good predictions of I-FENN, a trend which is even more profound in the elastic load case. 

\end{itemize}

This section has a) illustrated the feasibility of I-FENN and b) it has shed some initial light across potential generalization pathways. In the next two subsections we are mainly focusing on the first aspect, and we demonstrate the applicability of I-FENN on more challenging geometries by extending our investigation to: a) a geometry with two cracks and a structured mesh, and b) a geometry with one crack and unstructured mesh.

\subsection{Double-notched specimen}
\label{SubSubSec:Double_notched_spec}

\begin{figure}[H]
	\centering
	\includegraphics[width=\linewidth]{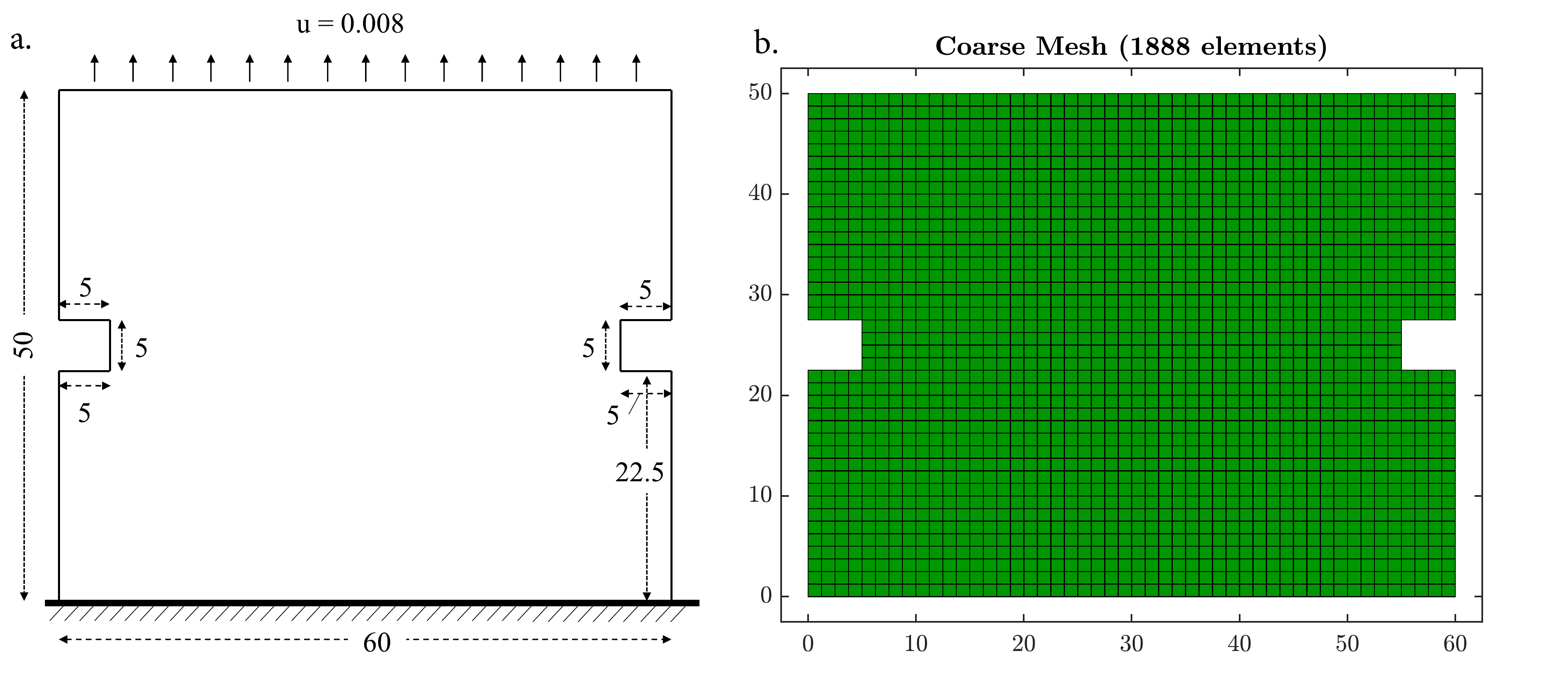}
	\caption{Geometry, loading, boundary conditions and the structured coarse mesh discretization of the double-notch domain.}
	\label{Figure_DoubleNotch_Geometry_Mesh}
\end{figure}

The next numerical example is a direct tension test on a double-notched geometry that has been previously modeled in \cite{peerlings1998gradient}, and the purpose of this implementation is to examine the ability of the network to recognize the presence of more than one crack locations. The geometric details of the domain are shown in Figure \ref{Figure_DoubleNotch_Geometry_Mesh}a. The specimen is fixed on the bottom edge and a vertical upward displacement load is applied at the top nodes. The shear modulus is $G = 125000$ KPa and Poisson's ratio is $\nu = 0.20$. In this case we adopt the Equation \eqref{StrainB} for the local strain definition and the damage law from Equation \eqref{DamageB}, with $\alpha = 0.99$ and $\beta = 400$. The numerical tolerance for convergence is $tol = 10^{-4}$.

\begin{figure}[H]
	\centering
	\includegraphics[scale=0.7]{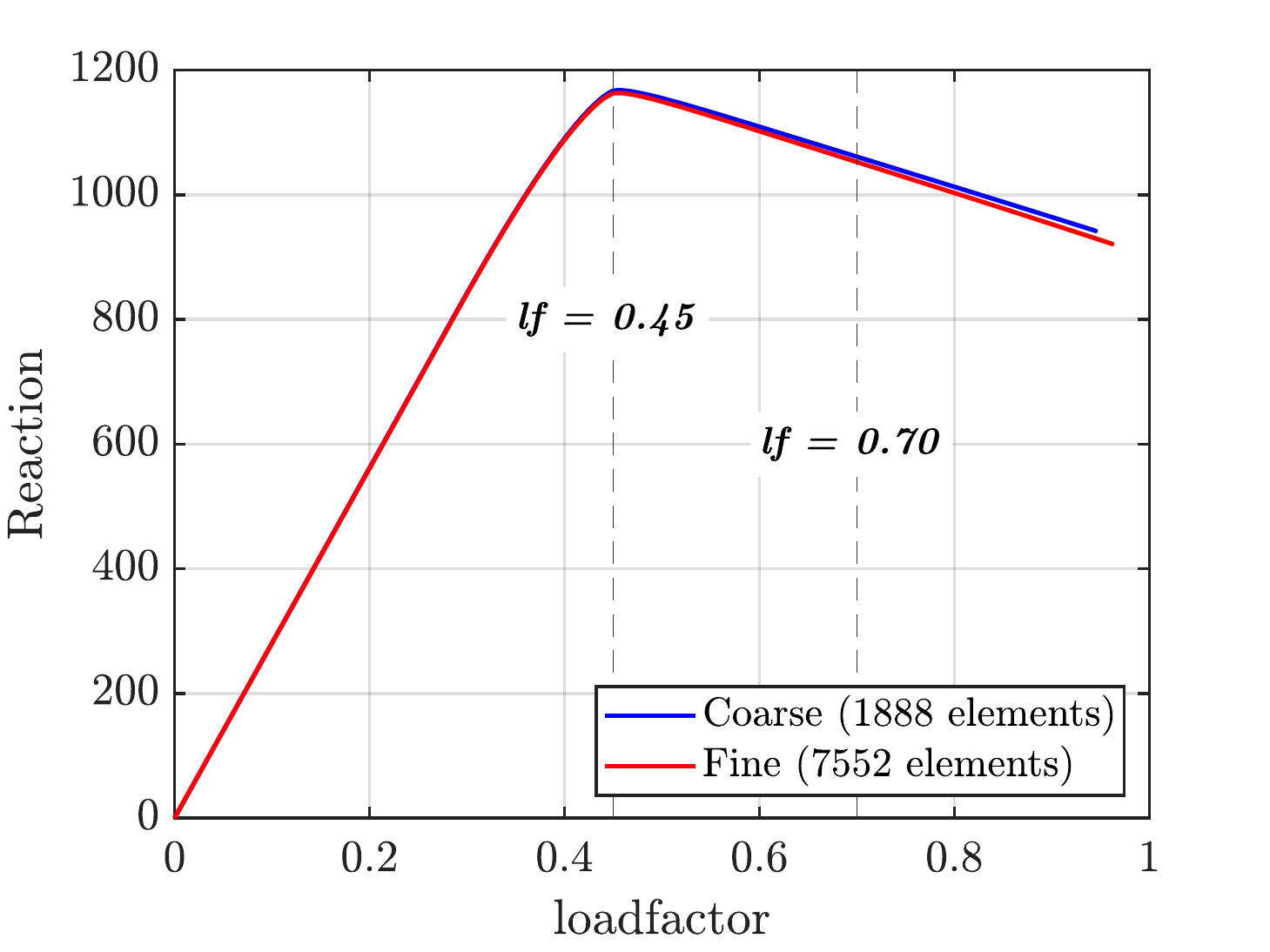}
	\caption{Reaction - loadfactor curves for the two mesh idealizations of the double-notch specimen.}
	\label{Figure_DoubleNotch_ConvergenceStudy}
\end{figure}

For this geometry we apply a structured finite element mesh discretization with two mesh densities. The two resulting models are termed $Coarse$ and $Fine$, with element sizes of $1.25$mm and $0.625$mm respectively. The characteristic element length is $l_{c} = 2$mm. Figure \ref{Figure_DoubleNotch_Geometry_Mesh}b shows the Coarse model, which employs $1888$ elements. Similar to the single-notch problem, we first execute the numerical simulation using the conventional non-local gradient method, and we demonstrate once again the mesh-independence of the solver by comparing the overlapping reaction-loadfactor curves of the two models as shown in Figure \ref{Figure_DoubleNotch_ConvergenceStudy}.

\begin{figure}[H]
	\centering
	\includegraphics[scale=0.7]{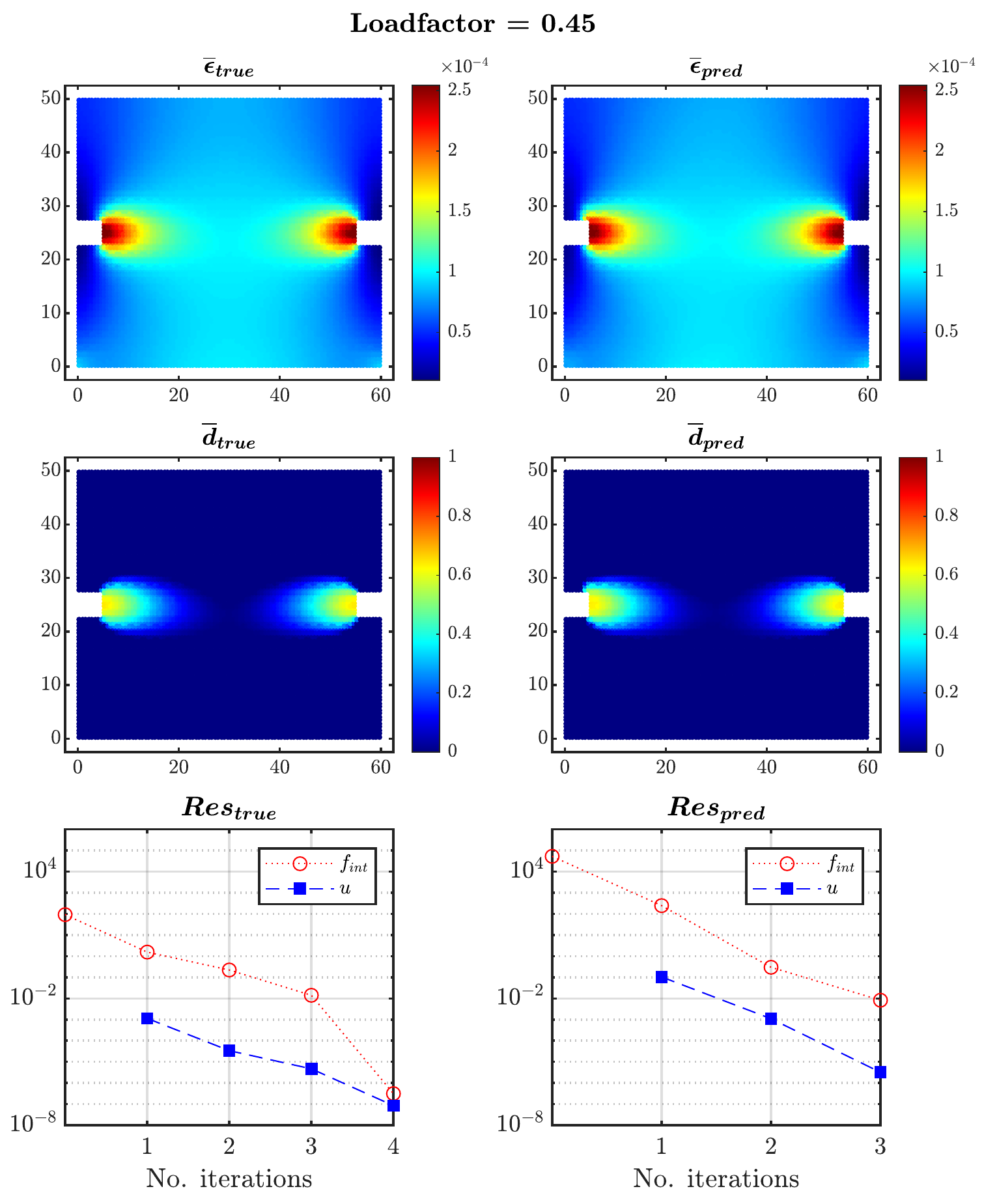}
	\caption{Comparison between FEM (left column) and I-FENN (right column) for the Coarse model of the double-notch case, at a loadfactor $lf = 0.45$. The first row of plots shows the non-local equivalent strain ${\bar{\varepsilon}}_{eq}$ profile, the second row depicts damage $d$, and the third row reports the calculated residuals. The results demonstrate the capability of I-FENN to identify the presence of more than one strain localization regions, and therefore to detect the presence of multiple crack locations in the domain. The L2-norm of the strain mismatch vector is ${|| {\bar{\varepsilon}}_{FEM} - {\bar{\varepsilon}}_{I-FENN} ||}_{2} = 2.2778 \times 10^{-4}$.}
	\label{Figure_DoubleNotch_inc90_MethodComparison}
\end{figure}

We compare the results of IFENN against the FEM simulations at two load increments. The first case is at $lf = 0.45$, where the reaction curve has reached its peak value and damage has already initiated around the crack tips. The results of this comparison are shown in Figure \ref{Figure_DoubleNotch_inc90_MethodComparison}. The first column of plots are the target values generated with FEM, and the second column stems from the I-FENN implementation. The first row of plots depicts the non-local equivalent strains ${\bar{\varepsilon}}_{eq}$, the second row displays the corresponding damage profile $d$, and the third row of plots shows the convergence performance of the two methods. Similar to the previous case, the red line in Figures \ref{Figure_DoubleNotch_inc90_MethodComparison}e and \ref{Figure_DoubleNotch_inc90_MethodComparison}f corresponds to the L2-norm of the residual stress vector ${\bf{R}}$, and the blue line indicates the L2-norm values of the displacement vector incremental change $\delta{\bf{\hat{u}}}$. Comparison of the non-local strain profiles in \ref{Figure_DoubleNotch_inc90_MethodComparison}a and \ref{Figure_DoubleNotch_inc90_MethodComparison}b shows the apparent capability of the trained network to identify the presence of more than one region where strain is localized. This is a very important observation which shows the flexibility of the training algorithm, and provides preliminary evidence of the network's capability to adapt to more complex cases than a single crack. The network's ability to identify both regions results to very similar damage $d$ profiles between the two methods, as shown in \ref{Figure_DoubleNotch_inc90_MethodComparison}c and \ref{Figure_DoubleNotch_inc90_MethodComparison}d. Finally, we emphasize the excellent convergence response of I-FENN shown in \ref{Figure_DoubleNotch_inc90_MethodComparison}f, where the norms of both ${\bf{R}}$ and $\delta{\bf{\hat{u}}}$ decrease at an almost constant rate until the convergence criterion of Equation \eqref{FEM_JduR_3} is satisfied. 

\begin{figure}[H]
	\centering
	\includegraphics[scale=0.7]{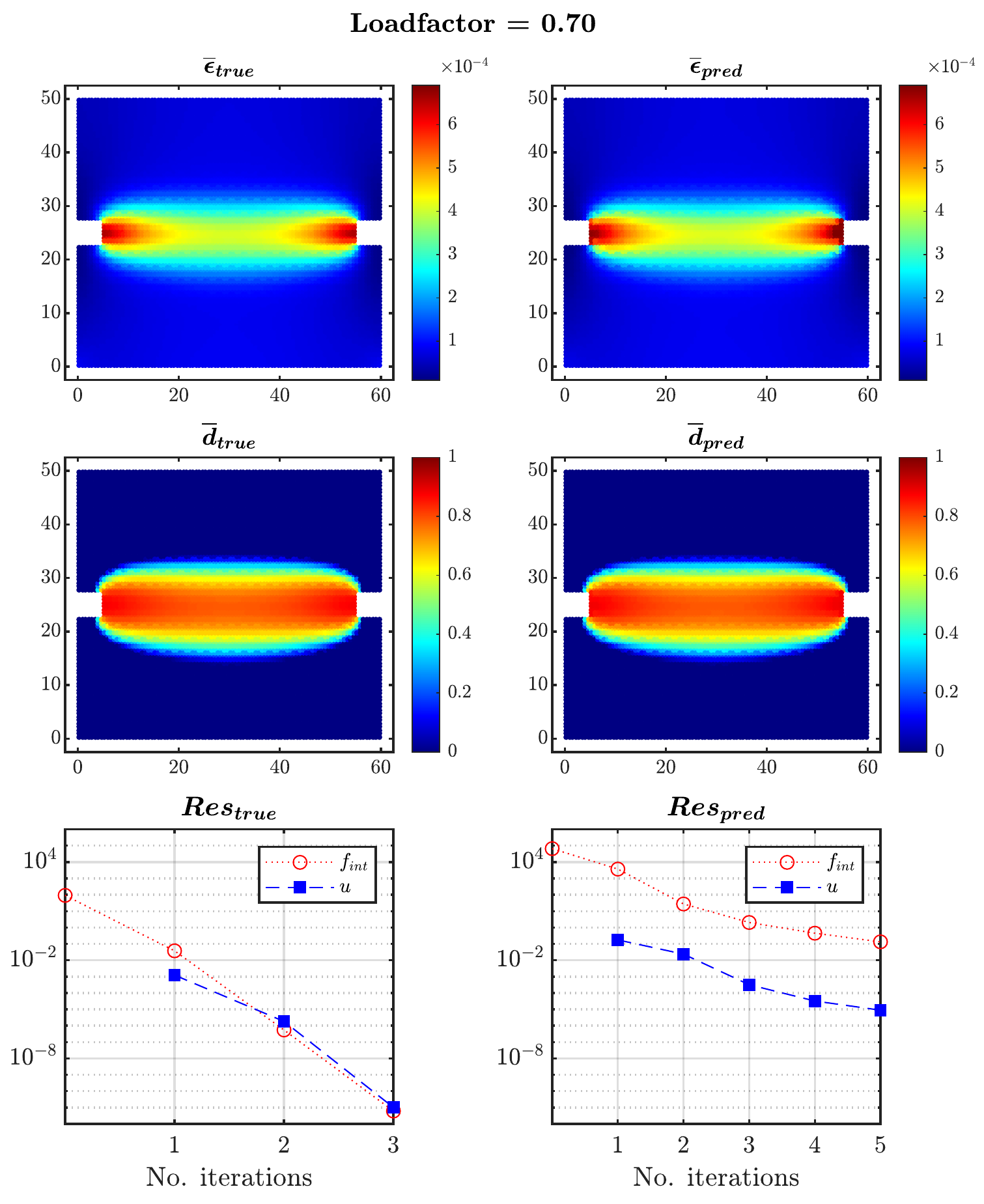}
	\caption{Comparison between FEM (left column) and I-FENN (right column) for the Coarse model of the double-notch case, at a loadfactor $lf = 0.70$. The first row of plots shows the non-local equivalent strain ${\bar{\varepsilon}}_{eq}$ profile, the second row depicts damage $d$, and the third row reports the calculated residuals. The trained PINN captures with very good accuracy the non-local strain landscape and subsequently I-FENN calculates a damage profile which is very similar to the target one. The L2-norm of the strain differences is ${|| {\bar{\varepsilon}}_{FEM} - {\bar{\varepsilon}}_{I-FENN} ||}_{2} = 8.418 \times 10^{-4}$.}
	\label{Figure_DoubleNotch_inc140_MethodComparison}
\end{figure}

We examine one additional case for the double-notch specimen, at $lf = 0.70$. At this load increment the domain has entered the softening regime and damage is widespread across a much wider zone. Figure \ref{Figure_DoubleNotch_inc140_MethodComparison} depicts the comparative study between FEM and I-FENN, where the subplots follow a similar layout as in Figure \ref{Figure_DoubleNotch_inc90_MethodComparison}. These graphs clearly indicate the excellent performance of the proposed methodology, since both the predicted non-local strain and damage profiles resemble with sufficient accuracy the target values. Additionally, the numerical solution converges within a few iterations and the residuals norms follow a monotonically decreased trajectory. Overall, the results of this example establish further confidence in the feasibility of the proposed methodology, and demonstrate its successful implementation in geometries with more than one strain localization regions.

\subsection{L-shaped specimen}
\label{SubSubSec:Lshaped_spec}

\begin{figure}[H]
	\centering
	\includegraphics[width=\textwidth]{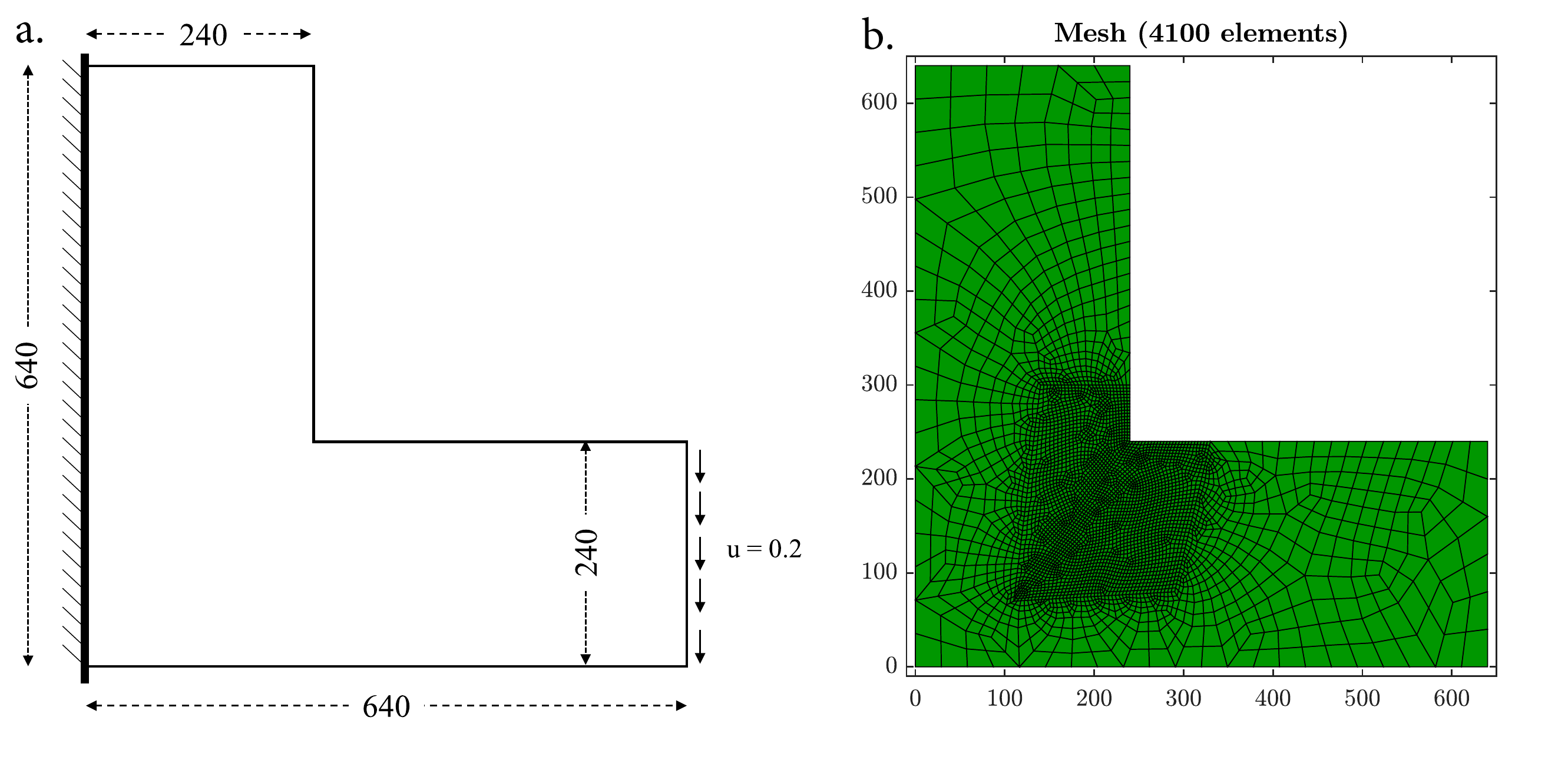}
	\caption{{\bf{a}}. Geometric and loading details for the L-shaped specimen. {\bf{b}}. Finite element discretization of the domain using an unstructured mesh of $4100$ elements in total.}
	\label{Figure_Lshaped_Geometry_Mesh}
\end{figure}

Thus far, we have validated the proposed methodology in geometries which are discretized using a structured finite element mesh. In this example, which is an L-shaped geometry, we utilize an unstructured mesh and investigate whether this modeling choice has an impact in the solution performance. The geometric and loading/boundary conditions of this example are shown in Figure \ref{Figure_Lshaped_Geometry_Mesh}a. The L-shaped specimen is fixed along its left side, and a downward displacement-controlled load is applied across the right edge. The shear modulus is $G = 125000$ KPa and Poisson's ratio is $\nu = 0.20$. For this example we use Equation \eqref{StrainB} for the local strain definition and the damage law from Equation \eqref{DamageB}, with $\alpha = 0.99$ and $\beta = 350$. The numerical tolerance for convergence is set to $tol = 10^{-4}$. Figure \ref{Figure_Lshaped_Geometry_Mesh}b shows the resulting finite element discretization, using a mesh of $4100$ elements in total. In the fine-mesh zone, the element length is $l_{elem} \approx 2.5$mm. The characteristic element length is $l_{c} = 5$mm. We anticipate damage initiation at the inner corner of the domain, and therefore we refine the mesh in that region as well as along the expected damage path. 

\begin{figure}[H]
	\centering
	\includegraphics[scale = 0.9]{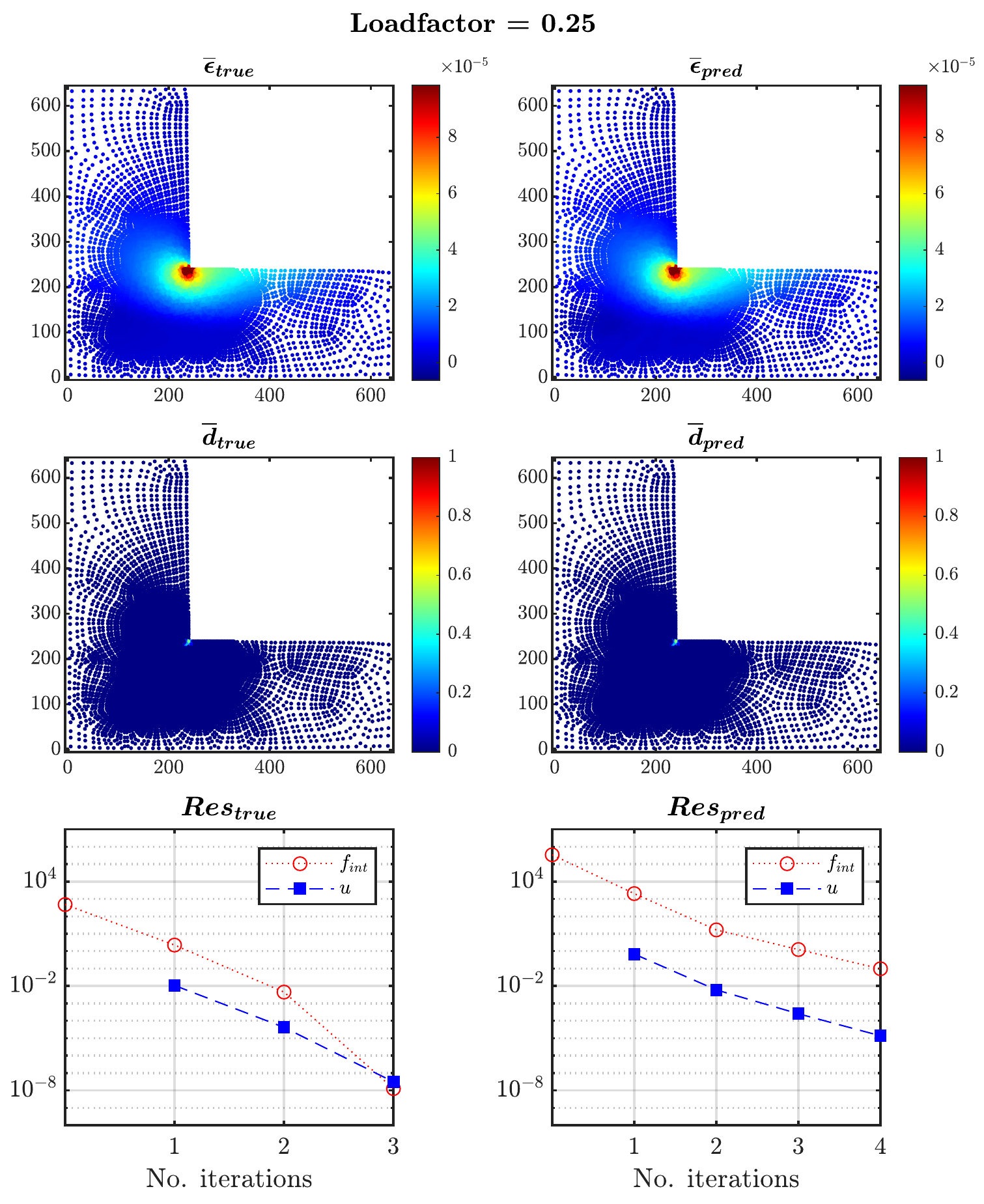}
	\caption{Comparison between FEM (left column) and I-FENN (right column) for the L-shaped specimen at loadfactor $lf = 0.25$. The first row shows the non-local equivalent strain ${\bar{\varepsilon}}_{eq}$, the second row depicts the damage $d$ profile, and the third reports the residuals during this load increment. The L2-norm of the strain differences vector is ${|| {\bar{\varepsilon}}_{FEM} - {\bar{\varepsilon}}_{I-FENN} ||}_{2} = 6.834 \times 10^{-5}$.}
	\label{Figure_LshapedCoarse_inc50_Method_Comparison}
\end{figure}

\begin{figure}[H]
	\centering
	\includegraphics[scale=0.9]{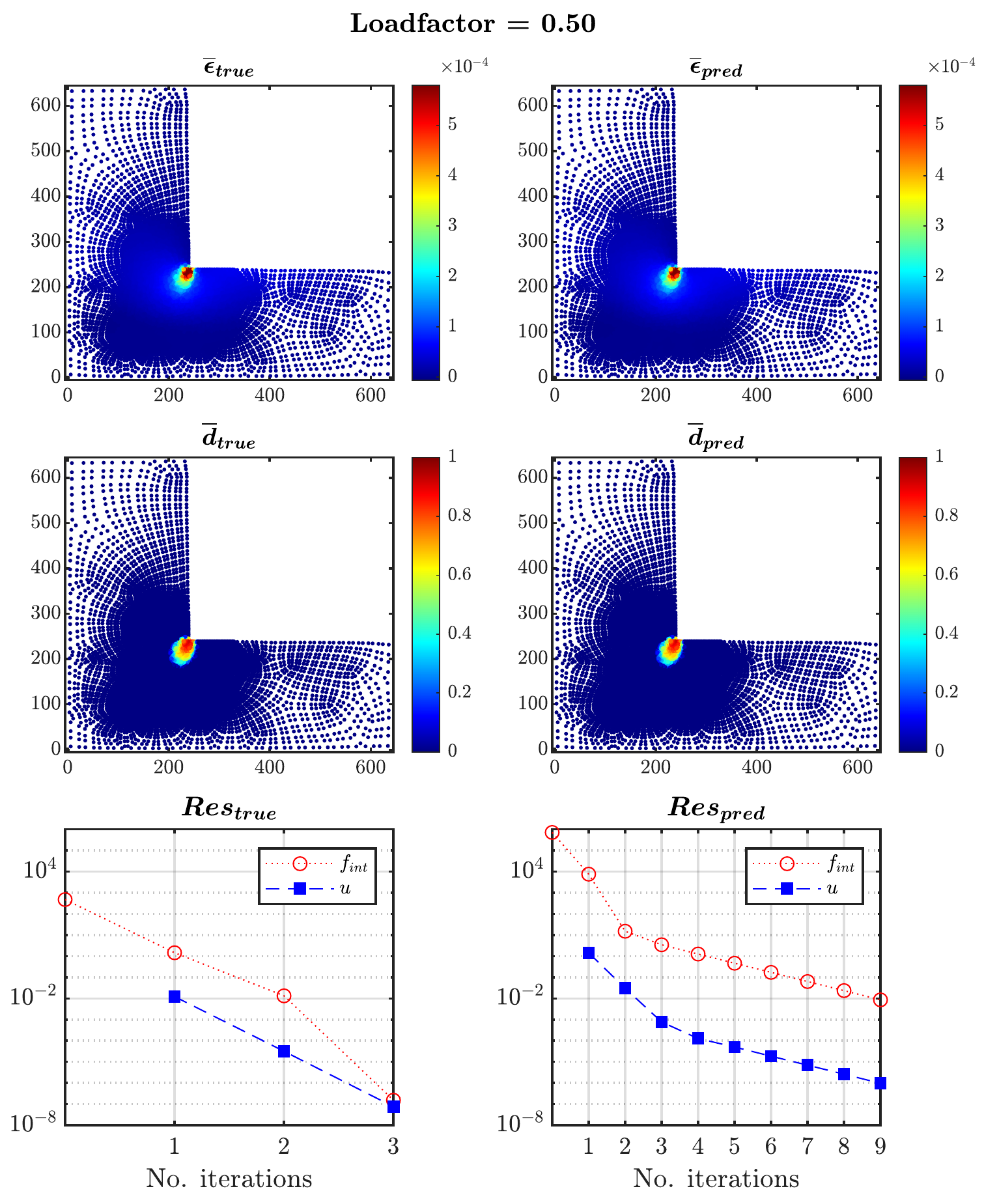}
	\caption{Comparison between FEM (left column) and I-FENN (right column) for the L-shaped specimen at loadfactor $lf = 0.50$. The first row shows the non-local equivalent strain ${\bar{\varepsilon}}_{eq}$, the second row depicts the damage $d$ profile, and the third reports the residuals during this load increment. The L2-norm of the strain mismatch vector is ${|| {\bar{\varepsilon}}_{FEM} - {\bar{\varepsilon}}_{I-FENN} ||}_{2} = 6.797 \times 10^{-4}$.}
	\label{Figure_LshapedCoarse_inc100_Method_Comparison}
\end{figure}

\begin{figure}[H]
	\centering
	\includegraphics[scale=0.9]{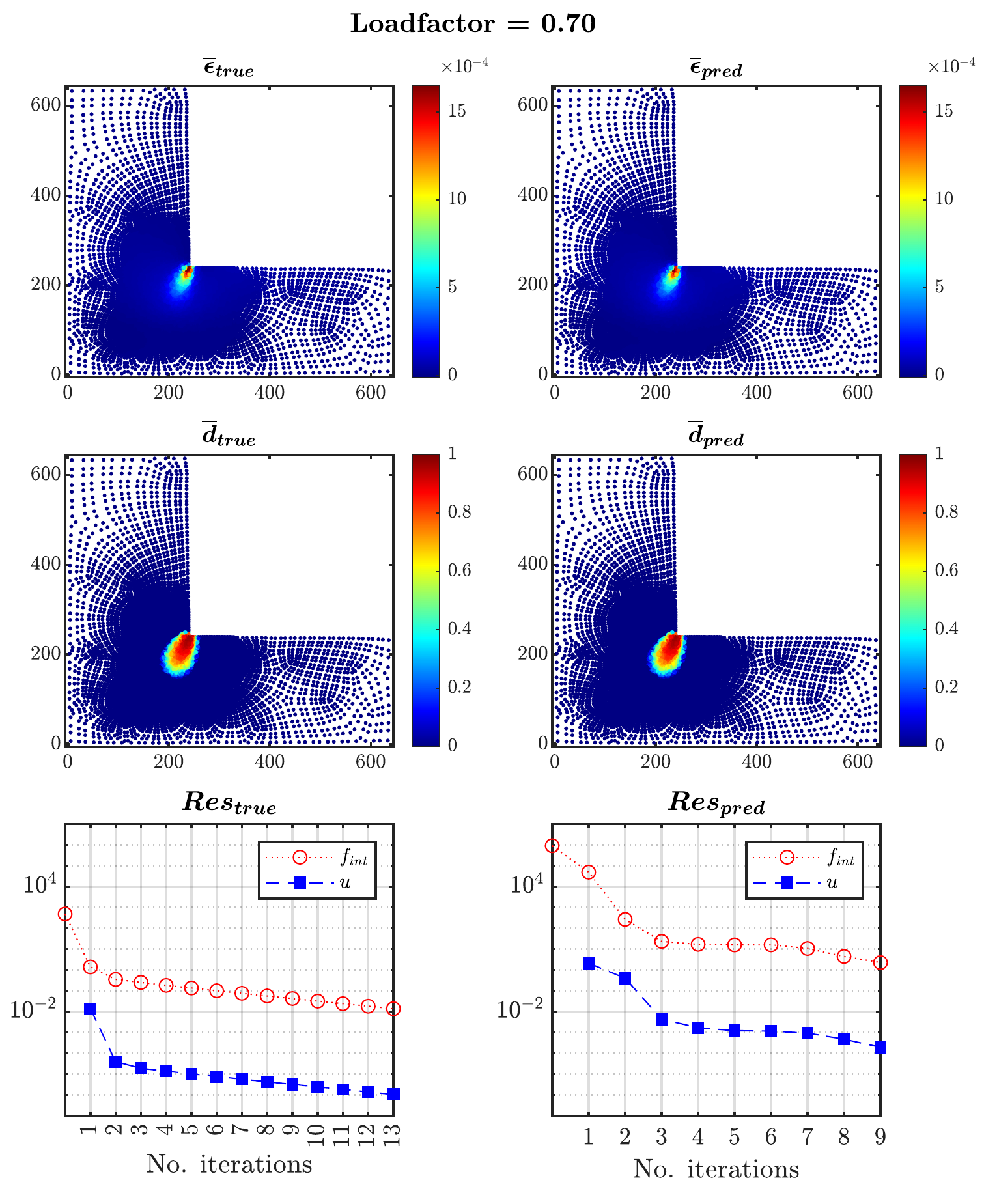}
	\caption{Comparison between FEM (left column) and I-FENN (right column) for the L-shaped specimen at loadfactor $lf = 0.75$. The first row shows the non-local equivalent strain ${\bar{\varepsilon}}_{eq}$, the second row depicts the damage $d$ profile, and the third reports the residuals during this load increment. The L2-norm of the strain error vector is ${|| {\bar{\varepsilon}}_{FEM} - {\bar{\varepsilon}}_{I-FENN} ||}_{2} = 4.441 \times 10^{-3}$.}
	\label{Figure_LshapedCoarse_inc140_Method_Comparison}
\end{figure}

Figures \ref{Figure_LshapedCoarse_inc50_Method_Comparison} - \ref{Figure_LshapedCoarse_inc140_Method_Comparison} present the results of the method implementation in the L-shaped geometry, for loadfactor values $lf = 0.25, 0.50$ and $0.75$. All three cases belong to the inelastic zone, but they represent distinctively different levels of damage spread. The subplots of each figure follow a similar layout to the previous examples, where the FEM results of strain, damage and convergence behavior are shown across the first column of the subplots, and the corresponding results using I-FENN are illustrated across the second column. Overall, similar trends as in the other two numerical examples are observed, and I-FENN is capable of producing results which are almost indistinguishable from FEM. The results alignment is clear across all three load increments and holds true with respect to both variables of interest, namely the non-local strain and damage. Additionally, as shown in the bottom row of plots in these figures, the algorithm converges within a reasonable number of iterations while showing a constantly decreasing trend. The results of this example indicate that the developed framework is insensitive to the choice of mesh orientation, structure and refinement and provide further evidence of the generalizability of I-FENN.

\subsection{Discussion on computational efficiency}
\label{SubSubSec:ComputationalTime}

{To this end, we evaluate the computational efficiency of I-FENN by comparing its simulation runtime vs the conventional FEM solver. We perform this comparison across all the investigated geometries and at the load increments of interest, and we report these values in Table {\ref{Table_Runtime_Comparison}}. We also note that we utilize a monolithic and not a staggered solver for the FEM, since despite their advantages the latter are well-known for the increased cost and difficulty of capturing the softening regime {\cite{bharali2022robust}}. We observe that I-FENN outperforms the conventional FEM implementation of the non-local gradient damage model in all but one cases. This case, which corresponds to the L-shaped geometry at $lf = 0.50$ is dominated by the fact that I-FENN needs 9 iterations to converge, whereas the conventional solver takes 3. In all other cases the performance is significantly improved, with a runtime acceleration ratio between 2-5 times. Since this comparison is conducted only at specific load increments, Table {\ref{Table_Runtime_Comparison}} provides just an initial insight on the computational benefit of I-FENN, and we note that the full potential of the proposed framework should be evaluated once the entire load history is analyzed. We also note that in the computational time of I-FENN we do not account for the offline PINN training time. This is because once the PINN is trained, it acts as a standalone nonlinear function which can be used for an arbitrary number of FEM simulations. Therefore, a fair comparison between the two methods should account for just the numerical simulation part, which is a common practice in the literature} \cite{mozaffar2019pathplasticity,liang2018deep}

\begin{table}[H]
	\caption{Comparison of simulation runtimes between I-FENN and FEM. All numbers correspond to seconds.}
	\label{Table_Runtime_Comparison}\centering
	\begin{adjustbox}{width=\textwidth}
	\begin{tabular}{c|c|c|c|c|c||c|c|c|c||c|c|c|c|c|c}
		\toprule

        \multicolumn{6}{c||}{$\bf{Single \ notch}$} & \multicolumn{4}{c||}{$\bf{Double \ notch}$} & \multicolumn{6}{c}{$\bf{L-shaped}$} \\

        \midrule
        
        \multicolumn{2}{c|}{$\bf{lf = 0.42}$} & \multicolumn{2}{c|}{$\bf{lf = 0.70}$} & \multicolumn{2}{c||}{$\bf{lf = 0.82}$} & 
        \multicolumn{2}{c|}{$\bf{lf = 0.45}$} & \multicolumn{2}{c||}{$\bf{lf = 0.70}$} & 
        \multicolumn{2}{c|}{$\bf{lf = 0.25}$} & \multicolumn{2}{c|}{$\bf{lf = 0.50}$} & \multicolumn{2}{c}{$\bf{lf = 0.70}$} \\
        
        $I-FENN$ & $FEM$ & $I-FENN$ & $FEM$ & $I-FENN$ & $FEM$ & $I-FENN$ & $FEM$ & $I-FENN$ & $FEM$ & $I-FENN$ & $FEM$ & $I-FENN$ & $FEM$ & $I-FENN$ & $FEM$  \\

        $4.727$ & $6.821$ & $5.941$ & $20.914$ & $7.523$ & $35.413$ & $2.796$ & $7.786$ & $4.078$ & $5.317$ & $16.784$ & $31.645$ & $37.923$ & $32.766$ & $35.591$ & $131.264$ \\
	
        \bottomrule
	\end{tabular}
	\end{adjustbox}
\end{table}

%%%%%%%%%%%%%%%%%%%%%%%%%

\section{Summary and Concluding remarks}
\label{Sec:Conclusions}

In this paper we developed a conceptually new approach of utilizing machine-learning techniques in order to reduce the computational expense of nonlinear computational solid mechanics problems. We introduced an Integrated Finite Element Neural Network framework (I-FENN), where pre-trained neural networks are directly deployed into the finite element stiffness function and act cooperatively with element interpolation functions to compute element-level variables. The trained network is used to compute the state variables, and their derivatives; and hence, it contributes to the calculation of the Jacobian matrix and residual vector utilized in the non-linear FEM analysis. The nonlinear and iterative nature of the solution algorithm is preserved, which allows for the continuous update of the trained network input and output variables until numerical convergence is achieved. 

%We derived the complete mathematical formulation of the model, and illustrated how the pre-trained neural network is incorporated in the element stiffness function for this specific problem.

We presented the implementation of the I-FENN framework in the development of a new continuum damage mechanics model. The proposed non-local damage model operates at the lower computational cost of the local damage method, while providing a fully consistent non-local damage description. The key point is the transformation of the local strain to a non-local strain at each material point, a transformation which is operated by the trained neural network. A PINN is developed and trained on the gradient non-local model that is commonly used in non-local damage modeling. The two PINN outputs, i.e. the non-local strain and its derivative with respect to the local, are directly used in the computation of the element Jacobian matrix and residual vector. We showcased through a series of examples the feasibility of the proposed non-local damage model in three distinctly different cases: a) structured mesh with a single crack, b) structured mesh with two cracks, and c) unstructured mesh with a single crack. Through these comparative studies we provided sufficient evidence of the results accuracy and the robustness of numerical convergence. We also illustrated potential generalization pathways of the method, by either a) training the PINN on coarsely meshed geometries and using it for predictions in finer discretizations, or b) by training the PINN on few selected load increments and applying it against data from unseen time steps. 

To the authors' best knowledge, this is the first effort towards a full integration of trained neural network within the framework of the iterative non-linear finite element modeling. The sound implementation and numerical examples show the potential of the I-FENN approach to aid the efforts of improving the computational efficiency of the solution of non-linear problems in mechanics. The deployment of the I-FENN approach within additional mechanics problems may lead to a paradigm shift in the way that machine learning tools are utilized and developed for computational mechanics modeling. The concepts which are proposed in this new benchmark implementation can be tailored and further utilized in many other cases where the need for numerical solution of additional PDEs increases the computational cost to prohibitive levels, such as multi-physics, multi-scale and/or multiple length-scale problems. 

As we illustrate the potential of the proposed I-FENN framework, we also mention the outstanding challenges that persist. {PINNs are inherently tied to the underlying target problem: the partial differential equation whose solution they approximate. The cost of solving a PDE is notoriously amplified when different scenarios are investigated, such as varying initial and boundary conditions, different domain geometries, different inputs, etc {\cite{wang2021learning}}. Consequently, a well-known drawback of PINNs is that once they are trained, they lack adaptiveness to these variations. Our current setup is based on PINNs, and hence it is bounded by its inherent limitations. To this end, machine learning models with a more versatile nature could be even better candidates to approximate the solution of parametric PDEs. This extension could aid to address variations in the geometry, loading setup and boundary conditions, and alleviate the need for repetitive PINN training at each configuration. Additionally, incorporation of the time-dependent behavior of the nonlinear problem requires a more in-depth study of the network design and architecture.}

%%%%%%%%%%%%%%%%%%%%%%%%%

\section{Data Availability}

{The neural network training code, as well as several excel files with training datasets that correspond to the numerical examples, have been made publicly available on Google Drive at:

https://drive.google.com/drive/u/0/folders/1mM35pXk7gzyx27NbIF3flqQxehr1tebh}

%%%%%%%%%%%%%%%%%%%%%%%%%

\newpage
\appendix
\section{FEM discretization for the local damage method}
\label{Appendix:AppendixA_Local_Damage_FEM}

This appendix presents the mathematical derivation of the numerical solution of the local damage method. Indicial notation is mostly used throughout this section. 

%%%%%%%%%%%%%%%%%%%%%%%%%%%%%%%%%%%%%%%%%%%%%%%%%%%%%%%%%%%%%%%%%%%%%%%%%%%%%%%%
\textbf{Elasticity tensor}:

\begin{equation}
	C_{ijkl} = \left[K \delta_{ij}\delta_{kl}+\mu \left[\delta_{ik}\delta_{jl}+\delta_{il}\delta_{jk}-\frac{2}{3}\delta_{ij}\delta_{kl}\right]\right]
\label{C_tensor1}
\end{equation}

\noindent where $K$ is the bulk modulus, $\mu$ is the shear modulus, and $\delta$ is the Kronecker delta. 

%%%%%%%%%%%%%%%%%%%%%%%%%%%%%%%%%%%%%%%%%%%%%%%%%%%%%%%%%%%%%%%%%%%%%%%%%%%%%%%%
\textbf{Compatibility equation}:

Assuming small deformations, the compatibility equation reads: 
\begin{equation}
	\varepsilon_{ij} = \frac{1}{2}\left[u_{i,j}+u_{j,i}\right] 
\label{e_tensor1}
\end{equation}

%%%%%%%%%%%%%%%%%%%%%%%%%%%%%%%%%%%%%%%%%%%%%%%%%%%%%%%%%%%%%%%%%%%%%%%%%%%%%%%%
\textbf{Strong form of governing equations}:

The equilibrium equation reads:
\begin{equation}
	\sigma_{ij,j} = 0   
\label{Equil_PDE_local}
\end{equation}

%%%%%%%%%%%%%%%%%%%%%%%%%%%%%%%%%%%%%%%%%%%%%%%%%%%%%%%%%%%%%%%%%%%%%%%%%%%%%%%%
\textbf{Weak form}:

The weak form of Equation \ref{Equil_PDE_local} reads:

\begin{equation}
    \boldsymbol{R}(u) = \int_{\Omega} w^u \left[\left[C_{ijkl}(d)\varepsilon_{kl}\right]_{,j}\right] \; d\Omega
\label{WeakR_displ_local}
\end{equation}

\noindent where $w^u$ are the displacement field weight functions and $C_{ijkl}(d) = (1-d) C_{ijkl}$. Integration by parts yields:  

\begin{equation}
    \boldsymbol{R}(u) = - \int_{\Gamma} \left[w^u t_i \right] d\Gamma + 
    \int_{\Omega} w^u _{,j} \left[C_{ijkl}(d)\varepsilon_{kl}\right] \; d\Omega
\label{WeakR_displ_expanded_local}    
\end{equation}

\noindent where $t_i$ corresponds to the external stress vector applied on the domain boundary.

%%%%%%%%%%%%%%%%%%%%%%%%%%%%%%%%%%%%%%%%%%%%%%%%%%%%%%%%%%%%%%%%%%%%%%%%%%%%%%%%
\textbf{FEM discretization}:

The displacements and strains at the integration points are calculated as follows:

\begin{equation}
	u = \boldsymbol{N} \hat{u}; ~~~ \varepsilon_{ij} = \boldsymbol{B} \hat{u}
\label{displacements_strains_local}
\end{equation}

\begin{equation}
	w^u = \boldsymbol{N} \hat{w}^u; ~~~ w^u_{,j} = \boldsymbol{B} \hat{w}^u
\label{weight_functions_local}
\end{equation}

\noindent where $\boldsymbol{N}$ and $\boldsymbol{B}$ are the displacement shape function matrix and its derivatives respectively. Nodal values are denoted with the ($\hat{.}$) symbol. Substituting equations \ref{displacements_strains_local} and \ref{weight_functions_local} into \ref{WeakR_displ_expanded_local} results to:

\begin{equation}
    \boldsymbol{R}(u) = - \int_{\Gamma} \left[\boldsymbol{N} \hat{t}_i \right] d\Gamma + 
    \int_{\Omega} \boldsymbol{B}^{T} (1-d)C_{ijkl}\boldsymbol{B} \hat{u} \; d\Omega
\label{WeakR_displ_expanded_local_2}    
\end{equation}

In a nonlinear iterative solver, the system of equations to be solved is:

\begin{equation}
	\boldsymbol{J} \delta \hat{u} = - \boldsymbol{R}(u) 
\label{system_equations_local1}
\end{equation}

\noindent where $\boldsymbol{J}$ is the Jacobian matrix and it is calculated as follows:

\begin{equation}
	\boldsymbol{J} = - \frac{\partial \boldsymbol{R}(u)}{\partial \hat{u}} = \boldsymbol{K}(u) \boldsymbol{I} \; + \; \frac{\partial \boldsymbol{K}(u)}{\partial \hat{u}} \hat{u}
\label{system_equations_local2}
\end{equation}

\noindent where $\boldsymbol{I} = \frac{\partial \hat{u}}{\partial \hat{u}}$ is the identity matrix and the expressions for the stiffness matrix $\boldsymbol{K}$ and its derivative with respect to the nodal displacements are provided below (matrix notation is used for the elasticity tensor):

\begin{equation}
	\boldsymbol{K}(u) = \int_{\Omega} \boldsymbol{B}^{T} (1-\boldsymbol{d}) \boldsymbol{C}\boldsymbol{B}^u d\Omega
\label{appendix_damageK_formula}
\end{equation}

\begin{equation}
	\frac{\partial \boldsymbol{K}(u)}{\partial \hat{u}} = - \int_{\Omega} \boldsymbol{B} \boldsymbol{C}\boldsymbol{B} \frac{\partial \boldsymbol{d}}{\partial \hat{u}} d\Omega
\label{appendix_partialK_partialu}
\end{equation}

The derivative of damage with respect to the nodal displacements at each Gauss point is calculated using the chain rule:

\begin{equation}
	\frac{\partial d}{\partial \hat{u}_{k}} = \frac{\partial d}{\partial \varepsilon_{eq}} \, \frac{\partial \varepsilon_{eq}}{\partial \varepsilon_{ij}} \, \frac{\partial \varepsilon_{ij}}{\partial \hat{u}_{k}}
\label{appendix_partiald_partialu}
\end{equation}

\noindent where the first term is given by the governing damage law, the second term depends on the relationship between the local equivalent strain and the tensor strain, and the third term is the shape function derivatives (Equation \ref{displacements_strains_local}).

\newpage
\section{ FEM discretization for the non-local gradient damage method}
\label{Appendix:AppendixB_NonLocal_Gradient_Damage_FEM}

This appendix presents the mathematical derivation of the numerical solution of the non-local gradient damage method. Indicial notation is mostly used throughout this section. 

%%%%%%%%%%%%%%%%%%%%%%%%%%%%%%%%%%%%%%%%%%%%%%%%%%%%%%%%%%%%%%%%%%%%%%%%%%%%%%%%
\textbf{Elasticity tensor}:

\begin{equation}
	C_{ijkl} = \left[K \delta_{ij}\delta_{kl}+\mu \left[\delta_{ik}\delta_{jl}+\delta_{il}\delta_{jk}-\frac{2}{3}\delta_{ij}\delta_{kl}\right]\right]
\label{C_tensor2}
\end{equation}

\noindent where $K$ is the bulk modulus, $\mu$ is the shear modulus, and $\delta$ is the Kronecker delta. 

%%%%%%%%%%%%%%%%%%%%%%%%%%%%%%%%%%%%%%%%%%%%%%%%%%%%%%%%%%%%%%%%%%%%%%%%%%%%%%%%
\textbf{Compatibility equation}:

Assuming small deformations, the compatibility equation reads: 
\begin{equation}
	\varepsilon_{ij} = \frac{1}{2}\left[u_{i,j}+u_{j,i}\right] 
\label{e_tensor2}
\end{equation}

%%%%%%%%%%%%%%%%%%%%%%%%%%%%%%%%%%%%%%%%
\textbf{Strong form of governing equations}:

The strong form of the equilibrium equation and the non-local equivalent strain PDE is provided below:
\begin{equation}
	\sigma_{ij,j} = 0   
\label{Equil_PDE_nonlocal}
\end{equation}

\begin{equation}
	\bar{\varepsilon}_{eq} - g \bar{\varepsilon}_{eq,ii} = \varepsilon_{eq}
\label{GradStrain_PDE}
\end{equation}

\noindent where $\varepsilon_{eq} = \varepsilon_{eq}(\varepsilon_{ij})$ is the local equivalent strain and $\bar{\varepsilon}_{eq}$ is the non-local equivalent strain.

%%%%%%%%%%%%%%%%%%%%%%%%%%%%%%%%%%%%%%%%
\textbf{Weak form}:

The weak form of Equation \ref{Equil_PDE_nonlocal} reads:

\begin{equation}
    R^u_i = \int_{\Omega} w^u \left[\left[C_{ijkl}(d)\varepsilon_{kl}\right]_{,j}\right] \; d\Omega
\label{WeakR_displ_nonlocal}
\end{equation}

\noindent where $w^u$ are the displacement field weight functions and $C_{ijkl}(d) = (1-d)C_{ijkl}$. Upon integration by parts, Equation \ref{WeakR_displ_nonlocal} yields:  

\begin{equation}
    R^u_i = - \int_{\Gamma} \left[ w^u t_i \right] d\Gamma + 
    \int_{\Omega} w^u _{,j} \left[C_{ijkl}(d)\varepsilon_{kl}\right] \; d\Omega
\label{WeakR_displ_expanded_nonlocal}    
\end{equation}

\noindent where $t_i$ corresponds to the external stress vector applied on the domain boundary. The weak form of the gradient equation \ref{GradStrain_PDE} is:

\begin{equation}
    R^{\varepsilon} = \int_{\Omega} w^{\varepsilon} \left[\bar{\varepsilon}_{eq} - g \bar{\varepsilon}_{eq,ii}-\varepsilon_{eq}\right] \; d\Omega 
\label{WeakR_strain}
\end{equation}

\noindent where $w^{\varepsilon}$ are the non-local strain field weight functions. Upon integration by parts, Equation \ref{WeakR_strain} yields:

\begin{equation}
	R^\varepsilon = \int_{\Gamma} \left[w^\varepsilon  \bar{\varepsilon}_{eq,i} n_i\right] d\Gamma - \int_{\Omega} w^\varepsilon_{,i}\left[-g \bar{\varepsilon}_{eq,i}\right] \; d\Omega + \int_{\Omega} w^\varepsilon\left[\bar{\varepsilon}_{eq}-\varepsilon_{eq}\right] \; d\Omega
\label{WeakR_strain_expanded}
\end{equation}

The first term corresponds to the boundary condition expression as given in Equation \ref{NonlocalGradientPDE_BCs}, and therefore vanishes to zero. Upon rearrangement, Equation \ref{WeakR_strain_expanded} reads:

\begin{equation}
	R^\varepsilon = \int_{\Omega} w^\varepsilon_{,i}\left[g \bar{\varepsilon}_{eq,i}\right] +  w^\varepsilon\left[\bar{\varepsilon}_{eq} - \varepsilon_{eq}\right] \; d\Omega 
\label{WeakR_strain_expanded_2}
\end{equation}

%%%%%%%%%%%%%%%%%%%%%%%%%%%%%%%%%%%%%%%%
\textbf{FEM discretization}:

The displacement and non-local strain values at the integration points are calculated based on the following expressions:

\begin{equation}
	u = \boldsymbol{N}^u \hat{u}; ~~~ \bar{\varepsilon} = \boldsymbol{N}^\varepsilon \hat{\bar{\varepsilon}}
\end{equation}

\begin{equation}
	w^u = \boldsymbol{N}^u \hat{w}^u; ~~~ w^\varepsilon = \boldsymbol{N}^\varepsilon \hat{w}^\varepsilon
\label{weight_functions}
\end{equation}

\noindent where $\boldsymbol{N}^{u}$ and $\boldsymbol{N}^{\varepsilon}$ are the interpolation functions for displacements and strains respectively, and ($\hat{.}$) denotes the corresponding nodal values. As for the local strains and the derivatives of the non-local strains at the integration points, the following equations are used:

\begin{equation}
    \varepsilon_{ij} = \boldsymbol{B}^u \hat{u}; ~~~ \bar{\varepsilon}_{,j} = \boldsymbol{B}^\varepsilon \hat{\bar{\varepsilon}}
\end{equation}
 
\begin{equation}
    w^u_{,j} = \boldsymbol{B}^u \hat{w}^u; ~~~ w^\varepsilon_{,j} = \boldsymbol{B}^\varepsilon \hat{w}^\varepsilon
\label{weight_functions_derivatives}
\end{equation}

\noindent where $\boldsymbol{B}^u$ and $\boldsymbol{B}^\varepsilon$ are the shape function derivatives for the displacements and non-local strains respectively. 

Using Equations \ref{weight_functions} and \ref{weight_functions_derivatives}, we rewrite Equations \ref{WeakR_displ_expanded_nonlocal} and \ref{WeakR_strain_expanded} as follows: 

\begin{equation}
    R^u_i = - \int_{\Gamma} {\boldsymbol{N}^u}^T t_i \; d\Gamma + \int_{\Omega}{\boldsymbol{B}^u}^TC_{ijkl}(d)\varepsilon_{kl} \; d\Omega  
\end{equation}

\begin{equation}
    R^\varepsilon_i = \int_{\Omega}{\boldsymbol{B}^\varepsilon}^T g \bar{\varepsilon}_{eq,i} + {\boldsymbol{N}^\varepsilon}^T \bar{\varepsilon}_{eq} - {\boldsymbol{N}^\varepsilon}^T \varepsilon_{eq} \; d\Omega
\end{equation}

Differentiating the residuals with respect to the displacement and non-local strain fields, results to the following terms for the Jacobian matrix components (matrix notation is used for the elasticity tensor):

\begin{equation}
      {\boldsymbol{J}}^{uu} = \frac{\partial R^u}{\partial u} = \int_{\Omega} {\boldsymbol{B}^{uT}} (1 - \boldsymbol{d}) \boldsymbol{C} \boldsymbol{B}^{u} \; d\Omega
\label{Eqn:Juu}
\end{equation}

\begin{equation}
      {\boldsymbol{J}}^{u\varepsilon} = \frac{\partial R^u}{\partial \bar{\varepsilon}} = - \int_{\Omega} {\boldsymbol{B}^{uT} \boldsymbol{C}} \frac{\partial \boldsymbol{d}}{\partial \bar{\varepsilon}} \varepsilon_{kl} {\boldsymbol{N}}^{\varepsilon} \; d\Omega
\label{Eqn:Jue}
\end{equation}

\begin{equation}
\begin{split}
      {\boldsymbol{J}}^{\varepsilon u} = \frac{\partial R^\varepsilon}{\partial u} = - \int_{\Omega} {\boldsymbol{N}^{\varepsilon T}} \frac{\partial \varepsilon_{eq}}{\partial \varepsilon_{ij}} {\boldsymbol{B}^{u}} \; d\Omega
\end{split}
\label{Eqn:Jeu}
\end{equation}

\begin{equation}
\begin{split}
      {\boldsymbol{J}}^{\varepsilon \varepsilon} = \frac{\partial R^\varepsilon}{\partial \bar{\varepsilon}} = \int_{\Omega} ({\boldsymbol{N}^{\varepsilon T} \boldsymbol{N}^{\varepsilon} + \boldsymbol{B}^{\varepsilon T}} g {\boldsymbol{B}^{\varepsilon}}) \; d\Omega
\end{split}
\label{Eqn:Jee}
\end{equation}

The discretization of the nodal internal and external forces reads:

\begin{equation}
\begin{split}
      \hat{\boldsymbol{f}}^{u}_{int} = \int_{\Omega} {\boldsymbol{B}^{uT}} \hat{\sigma} \; d\Omega
\end{split}
\end{equation}

\begin{equation}
\begin{split}
      \hat{\boldsymbol{f}}^{u}_{ext} = \int_{\Gamma} {\boldsymbol{N}^{uT}} \hat{t} \; d\Omega
\end{split}
\end{equation}

\begin{equation}
\begin{split}
      \hat{\boldsymbol{f}}^{\bar{\varepsilon}}_{int} = {\boldsymbol{J}}^{\bar{\varepsilon} \bar{\varepsilon}} \hat{\bar{\varepsilon}}_{eq}
\end{split}
\end{equation}

\begin{equation}
\begin{split}
      \hat{\boldsymbol{f}}^{\bar{\varepsilon}}_{ext} = \int_{\Omega} {\boldsymbol{N}^{\bar{\varepsilon} T}} \varepsilon_{eq} \, d\Omega
\end{split}
\end{equation}

\noindent where $\hat{t}$ are the nodal boundary forces. Ultimately, the matrix equation system for the non-local gradient damage method reads:

\begin{equation}
\begin{array}{cccccc}
\begin{bmatrix}
  {\bf{J}}^{uu}                    & {\bf{J}}^{u\bar{\varepsilon}} \\
  {\bf{J}}^{\bar{\varepsilon} u}   & {\bf{J}}^{\bar{\varepsilon} \bar{\varepsilon}}
\end{bmatrix} 
&
\begin{bmatrix}
  \delta \hat{{\bf{u}}} \\
  \delta \hat{{\bar{{\boldsymbol{\bar{\varepsilon}}}}}}_{eq} 
\end{bmatrix}
 = 
& -
\begin{bmatrix}
  \hat{\bf{f}}^{u}_{int} \\
  \hat{\bf{f}}^{\bar{\varepsilon}}_{int}
\end{bmatrix}
& + 
& 
\begin{bmatrix}
  \hat{\bf{f}}^{u}_{ext} \\
  \hat{\bf{f}}^{\bar{\varepsilon}}_{ext} 
\end{bmatrix}
\end{array}
\label{NonlocalGradientPDE_Equations_Appendix}
\end{equation}

\newpage
\section{Appendix C: Newton-Raphson algorithm}
\label{Sec:AppendixC}

The algorithm below presents the implementation details of the nonlinear iterative Newton-Raphson solver. We note that $\delta \hat{\boldsymbol{u}}_{F}$ denotes the incremental change of the nodal displacements at the free (Neumann) boundary. Also, the $\textit{tolerance}$ is a sufficiently small number dictating numerical convergence and it is specified in the order of $10^{-4}$ to $10^{-6}$.

\begin{algorithm}
  \caption{Newton-Raphson algorithm}
	\label{algorithm_classic_NewtonRaphson}
	Initialize all variables, set $\mathrm{flag}$ as $\mathrm{true}$ \\
	\While{$loadfactor<1$}
	    {$iter = 1$ \\ 
        Evaluate global Jacobian matrix $\boldsymbol{J}$, residual forces vector $\boldsymbol{R}$ and history variable matrix \\  
        \While{$\mathrm{flag}$ $\mathrm{is}$ $\mathrm{true}$}
            {Solve $\boldsymbol{J} \boldsymbol{\delta \hat{u}} = -\boldsymbol{R}$; update $\boldsymbol{\hat{u}}$ \\
	        Evaluate global Jacobian $\boldsymbol{J}$, residual forces vector $\boldsymbol{R}$ and history variable matrix \\  
	        {\uIf {$\frac{||\boldsymbol{\delta \hat{u}}_{F,iter}||_{2}}{||\boldsymbol{\delta \hat{u}}_{F,1}||_{2}} < \textit{tolerance}$ and $iter < max\_iter$}
	            {Update history variable; calculate reactions; increase loadfactor; set $\mathrm{flag}$ as $\mathrm{false}$ \\}
            \Else 
                {Discard the updated history variable matrix; $iter = iter + 1$ \\
        		{\uIf{$iter = max\_iter$}
        		{Set flag as false; decrease the applied load increment}
        		}}}}}    		
\end{algorithm}

\newpage
\section{Appendix D: Element Jacobian formulation for the local and non-local gradient damage methods}
\label{Sec:AppendixD}

The following algorithms describe the element Jacobian formulation $\boldsymbol{J_{elem}}$ for the conventional local and non-local gradient methods:
`
\begin{algorithm}
	\caption{Element Jacobian formulation for the local damage method}
	\label{algorithm_elstif_local}
	Initialize variables \\
    {
       \For{$\mathrm{each}$ $\mathrm{finite}$ $\mathrm{element}$}
        {
        Compute number, positions and weights of integration points \\
        \For{$\mathrm{each}$ $\mathrm{integration}$ $\mathrm{point}$}
            {
            Compute the shape functions $\boldsymbol{N}$ and their derivatives $\boldsymbol{B}$ \\
            Calculate the local strain vector $\boldsymbol{\epsilon} = \boldsymbol{B} \times \boldsymbol{d}$ \\
            Compute the local equivalent strain $\varepsilon_{eq}$ and its derivative w.r.t. the strain vector $\frac{\partial{\varepsilon_{eq}}}{\partial{\varepsilon_{ij}}}$ \\
            Calculate the local damage variable $D$ and keep $max(D, D_{stored})$ \\
            Evaluate the stress vector $\boldsymbol{\sigma} = (1 - D) \times {\boldsymbol{C}} \times \boldsymbol{\varepsilon}$ \\
            Compute the point contribution to the $\boldsymbol{R}_{elem} = \boldsymbol{R}_{elem} + \boldsymbol{B} \times \boldsymbol{\sigma} $ \\
            Compute the point contribution to the $\boldsymbol{J}_{elem}$ based on Equations \ref{appendix_damageK_formula}, \ref{appendix_partialK_partialu} and \ref{appendix_partiald_partialu}.
            }}}
\end{algorithm}

\begin{algorithm}[H]
	\caption{Element Jacobian formulation for the non-local gradient damage method}
	\label{algorithm_elstif_nonlocal_gradient}
	Initialize variables \\
    {
       \For{$\mathrm{each}$ $\mathrm{finite}$ $\mathrm{element}$}
        {
        Identify the number, positions and weights of integration points \\
        \For{$\mathrm{each}$ $\mathrm{integration}$ $\mathrm{point}$}
            {
            Compute the shape functions $\boldsymbol{N}$, $\bar{\boldsymbol{N}}$ and their derivatives $\boldsymbol{B}$, $\bar{\boldsymbol{B}}$ for $\boldsymbol{d}$ and $\bar{\boldsymbol{\varepsilon}}$ respectively \\
            Calculate the local strain vector $\boldsymbol{\varepsilon} = \boldsymbol{B} \times \boldsymbol{d}$ \\
            Calculate the local equivalent strain $\varepsilon_{eq}$ and its derivative w.r.t. the strain vector $\frac{\partial{\varepsilon_{eq}}}{\partial{\varepsilon_{ij}}}$  \\
            Compute the non-local equivalent strain $\bar{\varepsilon}$ $=$ $\bar{\boldsymbol{N}}$ $\times$ $\hat{\bar{\varepsilon}}$ and history variable \\
            Evaluate the stress vector $\boldsymbol{\sigma} = (1 - \bar{D}) \times {\boldsymbol{C}} \times \boldsymbol{\varepsilon}$ \\
            Compute the point contribution to the $\boldsymbol{R}_{elem} = \boldsymbol{R}_{elem} + \boldsymbol{B} \times \boldsymbol{\sigma} $ \\
            Compute the point contribution to $\boldsymbol{J}_{elem}$ based on Equations \ref{Eqn:Juu}, \ref{Eqn:Jue}, \ref{Eqn:Jeu} and \ref{Eqn:Jee}.
            }}}
\end{algorithm}

\newpage
\bibliography{bibliography}

\end{document}